%% file: main.tex
\documentclass[11pt]{article}
\input{preamble}

\author{
Vladimir Braverman\thanks{Johns Hopkins University, Google, Rice University, and UT Health. Email: \texttt{vbraverman@google.com}.}
\and
Prathamesh Dharangutte\thanks{University of Copenhagen. Email: \texttt{prathamesh.dharangutte@di.ku.dk}. Significant part of the work was completed when the author was affiliated with Rutgers University and the author would like to acknowledge support from NSF IIS-2229876, DMS-2220271, DMS-2311064, CCF-2208663, CCF-
2118953 and CNS-2515159 during that period.}
\and
Chenglin Fan\thanks{Seoul National University. Email: \texttt{fanchenglin@gmail.com}.
The author was supported in part by the National Research Foundation of Korea (NRF) under Grant No. RS-2026-25484051.}
\and
Chen Wang\thanks{Rensselaer Polytechnic Institute. Email: \texttt{wangc33@rpi.edu}.}
\and
Shenghao Xie\thanks{Carnegie Mellon University. Email: \texttt{shenghax@andrew.cmu.edu}. The author is supported in part by NSF CCF-2335411, and in part by Office of Naval Research award number N000142112647 and a Simons Investigator Award. Part of the work was completed when the author was affiliated with Texas A\&M University.}
\and
Samson Zhou\thanks{Texas A\&M University. Email: \texttt{samsonzhou@gmail.com}. The author is supported in part by NSF CCF-2335411 and gratefully acknowledges funding provided by the Oak Ridge Associated Universities (ORAU) Ralph E. Powe Junior Faculty Enhancement Award.}
}

\title{Fully Dynamic Correlation Clustering with Vertex Updates in Polylogarithmic Update Time}

\date{}

\begin{document}

\maketitle

\begin{abstract}
We present a fully dynamic algorithm for correlation clustering that supports \emph{vertex} insertions and deletions while maintaining an $O(1)$ approximation in polylogarithmic amortized update time. This resolves an open problem of vertex deletions posed by a recent line of work on dynamic correlation clustering.

We adopt the min-disagreement objective for correlation clustering, i.e., given a signed complete graph $G=(V,E^+\cup E^-)$, we want to minimize the total number of $\pl$ edge crossing clusters and $\nl$ edges within clusters. The input is given in the adjacency list model, where each update inserts or deletes a node along with its full set of $\pl$ neighbors. Our main technical contribution is a fault-tolerant sparse-dense decomposition framework which unlike the existing decomposition-based frameworks allows for structural inconsistencies as long as the induced cost is bounded. We also show a separation between explicitly maintaining the (stronger) sparse-dense decomposition and maintaining a constant-factor approximation to correlation clustering, which may be of independent interest.
\end{abstract}

\section{Introduction}
\label{sec:intro}
Correlation clustering is a fundamental problem in unsupervised learning that formalizes the goal of grouping similar elements together and separating dissimilar elements. 
Given a complete graph $G = (V, E^+ \cup E^-)$ with edges partitioned into positive $\pl$ edges $E^+$ and negative $\nl$ edges $E^-$, correlation clustering seeks a partition of $V$ minimizing the number of disagreements, i.e., $\pl$ edges with endpoints in different clusters and $\nl$ edges with endpoints in the same cluster. 
Correlation clustering was first introduced by Bansal, Blum, and Chawla~\cite{bansal2004correlation} and has since been the subject of extensive theoretical study (see, e.g.,~\cite{GiotisG06,BonchiGK13,BhaskaraDV18,AhmadianE0M20,Cohen-AddadLMNP21,Assadi022,BehnezhadCMTFOCS22,Cohen-AddadFLMN22,Veldt22,BehnezhadCMT23,Cohen-AddadLMP24,CaoCL0NV24,DalirrooyfardMM24,braverman2025fully,CaoCL0LNTVY025,SorianoS26}, and references therein), as well as a wide range of practical applications, including automated labeling~\cite{AgrawalHKMT09,ChakrabartiKP08}, community mining~\cite{ChenSX12,levorato2017evaluating,VeldtWG17,VeldtGW18,ShiDELM21,AADGW22}, document clustering~\cite{bansal2004correlation,zhang2008correlation,BonchiGK13}, duplicate detection~\cite{ArasuRS09}, and many others. 

While the classical formulation of correlation clustering assumes a fixed signed complete graph $G = (V, E^+ \cup E^-)$, many real-world applications involve data that evolves over time. 
For example, in a social network, individuals can join and leave the network, causing the friendly and hostile relationships to evolve, reflected as the changes in $\pl$ and $\nl$ edges.
More generally, input labels can fluctuate due to noisy data sources, and a vertex might be temporarily dropped from the network due to reliability issues. 
In such scenarios, the most straightforward way to maintain correlation clustering solutions is to recompute an optimal clustering from scratch after every update.
However, this approach is computationally expensive and impractical, which motivates the design of efficient and high-quality correlation clustering algorithms in the \emph{dynamic setting}.

As a result, there has recently been considerable research studying dynamic algorithms for correlation clustering~\cite{BehnezhadDHSS19, ChechikZ19, BehnezhadCMT23, DalirrooyfardMM24, Cohen-AddadLMP24, BCCGM24, braverman2025fully,CCLLLNTVYZdynamicCC}. 
One line of work assumes that the input is a signed complete graph in which all edges are initially negative, and each update consists of flipping the label of a single edge, either from negative to positive or vice versa. 
The goal is to maintain a clustering that approximately minimizes disagreement after each such update. 
Efficiency is typically measured by the \emph{amortized update time}, defined as the total time spent over a sequence of updates divided by the number of updates performed. 
Achieving an update time of $\mathrm{polylog}(n)$ is a natural target, as it almost matches the cost of reading a single edge label. 
Initial breakthroughs by~\cite{BehnezhadDHSS19, ChechikZ19} gave dynamic algorithms with $3$-approximation guarantees and $\mathrm{polylog}(n)$ amortized update time. 
Subsequent work improved the efficiency further:~\cite{BehnezhadCMT23, DalirrooyfardMM24} achieved constant update time for $O(1)$ approximation, and 
~\cite{BCCGM24} improved the approximation ratio to $3 - \Omega(1)$ while retaining $\mathrm{polylog}(n)$ update time. 
Most recently, \cite{CCLLLNTVYZdynamicCC} designed a dynamic algorithm that almost matches the best offline approximation factor using $\mathrm{polylog}(n)$ update time, as long as the clustering is represented with a special data structure.

\paragraph{Vertex insertions/deletions.} 
While much of the prior work focuses on edge label updates in a fixed vertex set, many practical scenarios involve more fundamental structural changes, namely, the insertion or deletion of vertices along with their incident edges. 
For instance, in evolving social or communication networks, users may join or leave the system, causing a cascade of updates to the pairwise relationships encoded in the graph. 
This naturally motivates the study of fully dynamic correlation clustering algorithms that can efficiently handle vertex updates. 

Formally, the input is an evolving signed complete graph on a fixed set of at most $n$ nodes. 
The algorithm operates in the adjacency list model, where it has direct access to the current set of $\pl$ neighbors for each node. 
We assume that adjacency lists are maintained and updated externally by users or an upstream system, and are provided to the algorithm in fully updated form at each step. 
This reflects common practical settings, such as social or communication networks, where users independently report their connections, and such updates can be performed in parallel or asynchronously outside the algorithm's control. 
As a result, the algorithm is not charged for assembling or modifying adjacency lists.

The graph is implicitly defined by these lists: a pair of nodes shares a $\pl$ edge if they appear in each other’s lists, and a $\nl$ edge otherwise. 
Each update consists of either the insertion or deletion of a single node's adjacency list. 
In an insertion, a new node $v$ arrives along with its full set of $\pl$ neighbors, which implicitly defines the labels of all edges incident to $v$. 
In a deletion, a node $v$ and all of its incident edges are removed from the graph. 
The algorithm must maintain, at all times, a clustering that achieves a constant-factor approximation to the minimum number of disagreements. 
We remark that while it is natural to aim for $\polylog(n)$ update time for each \emph{edge} insertions and deletions, it is significantly less clear what is achievable for \emph{vertex} insertions and deletions. 
In particular, a single vertex update can simultaneously affect up to $\Theta(n)$ edge labels. 
Therefore, the following question is a major open problem in dynamic algorithms for correlation clustering:
\begin{quote}
\centering
\emph{Is it possible to maintain a constant-factor approximation for correlation clustering using $\polylog(n)$ update time in the vertex insertion/deletion model?}
\end{quote}

\subsection{Our Contributions}
In this paper, we answer the above question in the affirmative, presenting the first \emph{fully dynamic algorithm} for correlation clustering that maintains an $O(1)$-approximation under a worst-case fixed ordering of vertex insertions and deletions, with polylogarithmic amortized update time.
A key distinction between our work and recent dynamic correlation clustering algorithms~\cite{Cohen-AddadLMP24,dalirrooyfard2025sparse} lies in the treatment of deletions.
While all of these works, including ours, support arbitrary order of vertex insertions, prior algorithms are limited to \emph{random-order deletions}, where deleted vertices are sampled at random from the current vertex set. 
In contrast, our algorithm is the \emph{first to support arbitrary order of vertex deletions} while maintaining an $O(1)$-approximation and polylogarithmic amortized update time. 
This makes our method substantially more robust in fully dynamic settings.


En route to our efficient dynamic algorithm, we encountered and resolved several major technical challenges, and the ideas in our solutions can be of independent interest (see \Cref{subsec:tech-overview} for an overview).
First, explicitly maintaining the standard sparse-dense decomposition needs $\Omega(\sqrt{n})$ amortized update time (\Cref{lem:vanilla-sdd-lower-bound}). We bypass this bottleneck by introducing a ``partially correct tolerant decomposition'' framework. This allows for controlled structural errors by partitioning the graph into singletons, tolerant clusters, and isolated clusters, which makes it possible to maintain efficiently. We form each cluster as a standard almost-clique, but maintain it only under the relaxed definition (\Cref{lem:old-sdd-is-valid-for-new}).
Second, we extend the charging scheme of \cite{Assadi022,AssadiSW23} to bound the costs of tolerant and isolated clusters, and we charge misclassified singletons, level by level, to the truly sparse singletons (\Cref{thm:approximation-partial-correct-sdd}).
Third, to achieve polylogarithmic update time, we combine three mechanisms:
(i) \emph{lazy refreshment}: a cluster $K$ is rebuilt only after $\eps\card{K}$ newly inserted vertices have joined it, or $\eps\card{K}$ of its past members have been deleted, where $\card{K}$ is its size at formation (\Cref{alg:vertex-dynamic-alg} and \Cref{alg:refresh-ac});
(ii) \emph{identification}: when a cluster forms, the outside vertices adjacent to at least half of it are moved to the singleton set, so that vertices which a later deletion may make dense with the new cluster are already tracked (\Cref{alg:new-ac});
(iii) \emph{approximate degree buckets}: every singleton enters all $O(\log n)$ degree lists, and periodically, sampling decides whether level $i$ must be scanned fully, or only until $2^i\polylog(n)$ scanned vertices test sparse (\Cref{alg:deletion-clique-forming} and \Cref{alg:check-and-try-cluster}).
Finally, for the running time, a forward-charging budget scheme lets each new cluster pay for its formation from the budgets of past updates and pass fresh budgets to the vertices it identifies (\Cref{sec:charging:budgeting}).
Together, these techniques give the guarantees of \Cref{thm:master-theorem-CC}.

\begin{restatable}{theorem}{MainTheorem}
\label{thm:master-theorem-CC}
For the correlation clustering problem on a signed complete graph $G = (V, E^+ \cup E^-)$,
there exists a randomized fully dynamic vertex-update algorithm that, given the positive subgraph
$G^+ = (V, E^+)$ and a sequence of $\poly(|V|)$ worst-case vertex insertions and deletions, together with all
incident $\pl$ edges, maintains an $O(1)$-approximation to the optimal correlation clustering with high probability in
$\polylog{(n)}$ amortized update time.
\end{restatable}

Adversarial robustness is an important goal for dynamic and streaming algorithms whose future inputs may depend on their past behavior.
Instead of having an input stream fixed before the algorithm starts, in the standard \emph{black-box} model, the adversary chooses each update after seeing the algorithm's past outputs.
This matters in practice whenever the output feeds back into the input, for instance, on social platforms, users join, leave, or change connections in response to the communities they are shown, and a dynamic clustering algorithm used as a subroutine to determine the communities receives updates that depend on its earlier answers~\cite{Wajc20}.
Therefore, adversarial robustness has been studied extensively for streaming problems~\cite{Ben-EliezerJWY20,HassidimKMMS20}, and it has also drawn more attention in dynamic graph algorithms~\cite{ChuzhoyKhanna19,Wajc20,BernsteinEtAl22,braverman2025fully}.
Essentially, our algorithm in \Cref{thm:master-theorem-CC} is robust against the more powerful \emph{white-box} model~\cite{AjtaiBJSSWZ22}, where a computationally unbounded adversary may inspect the entire maintained state and all previously drawn random bits before choosing the next vertex insertion or deletion.
Thus we achieve robustness while preserving the $O(1)$-approximation and the $\polylog(n)$ amortized update time guarantees (see \Cref{thm:white-box}).



\section{Technical Overview}
\label{subsec:tech-overview}
Our goal is to maintain an $O(1)$-approximate solution for min-disagreement correlation clustering under vertex insertions and deletions, while using only polylogarithmic amortized time per update. 
The main difficulty is that a single vertex update can modify $\Theta(n)$ edges, which may drastically change the structural role of many remaining vertices. 
In particular, under vertex deletions, dense structure can emerge after a carefully chosen blocking set disappears, and an algorithm that search for local neighborhood changes may fail to detect this change with polylogarithmic time. 
This makes the model much more challenging than the comparable \emph{edge dynamic} settings in \cite{BehnezhadCMT23,DalirrooyfardMM24,braverman2025fully,CCLLLNTVYZdynamicCC}.

\subsection{Our framework: partially correct tolerant sparse-dense decomposition}
Roughly speaking, there are two popular ways to obtain $O(1)$-approximation for correlation clustering, pivot algorithm (see, e.g.~\cite{ailon2008aggregating,BehnezhadCMTFOCS22,BehnezhadCMT23,CambusKLPU24,DalirrooyfardMM24,dalirrooyfard2025sparse}) and sparse-dense decomposition (see, e.g.~\cite{Cohen-AddadLMNP21,Assadi022,AssadiSW23,Cohen-AddadLMP24})\footnote{More recent algorithms for state-of-the-art approximation correlation clustering uses LP-based techniques (e.g.~\cite{Cohen-AddadLNFOCS2022,Cohen-AddadLLNFOCS2023,CaoCL0NV24,CaoCL0LNTVY025}) or local search (\cite{Cohen-AddadLPTYZSTOC2024}). However, since they run sparse-dense decomposition as a pre-clustering step, most of these results still need techniques in sparse-dense decomposition.}.

\paragraph{Difficulty of pivot-based algorithms.}
Pivot-based algorithms iteratively select an unassigned vertex as a ``pivot'' and form a cluster around it, based on the similarity relationships with the remaining vertices. 
They often achieve strong approximation ratios, but although \cite{dalirrooyfard2025sparse} adapted them to handle vertex \emph{insertions}. 
However, the pivot algorithm faces some inherent challenges with vertex \emph{deletions}. 
To see the issue, suppose a pivot is selected and its cluster is formed, and then the pivot is deleted. 
Then this can invalidate a large number of cluster assignments, requiring a large number of updates that seem incompatible with both efficiency and maintaining a good approximation guarantee. 

\paragraph{The sparse-dense decomposition technique.}
Another popular technique for correlation clustering is the sparse-dense decomposition (SDD) technique (also called ``preclustering'' in the literature), which partitions the graph into sparse singleton vertices and dense almost-cliques. 
Roughly speaking, we define a \emph{sparse edge} as one whose endpoints have significantly different neighborhoods and a \emph{sparse vertex} if many of its incident edges are sparse; intuitively, these vertices act as bridges connecting unrelated parts of the graph. 
Removing them leaves a graph that decomposes into ``almost-cliques'': groups of vertices whose sizes are similar to their degrees, and whose numbers of non-neighbors inside and neighbors outside are small. 
In the offline setting, this decomposition gives a constant-factor approximation by clustering each almost-clique together and keeping each sparse vertex as singletons (see, e.g.~\cite{Cohen-AddadLMNP21,Assadi022,AssadiSW23}). 

A notable advantage of the sparse-dense decomposition technique is that, when randomization is allowed, we can efficiently use \emph{samples} to determine whether a vertex should join a cluster. 
As such, it is quite natural to use this technique in the vertex dynamic setting, where sampling is perhaps the most natural operation.
Indeed, building on the decomposition techniques, \cite{Cohen-AddadLMP24} gave the first dynamic algorithm that achieves $O(1)$-approximation with $\polylog{n}$ update time that supports \emph{arbitrary-order insertion} and \emph{random deletion}.  
Indeed, their analysis crucially relies on the assumption that only a random vertex can be deleted, as this randomness ensures that the graph structure only change gradually, thereby ruling out worst-case scenarios in which the removal of a carefully chosen set of ``blocking'' vertices suddenly force the formation of a dense cluster.

\paragraph{Sparse-dense decomposition cannot work.}
We would naturally wonder whether a simple variant of the algorithm in \cite{Cohen-AddadLMP24} can support worst-case vertex deletion.
Alas, the story becomes much more involved when arbitrary deletions are allowed: the standard decomposition is far too rigid for the fully dynamic vertex-update model. 
We formally prove in \Cref{sec:original-decomp-lb} that strictly maintaining a valid standard $\eps$-sparse-dense decomposition (defined as in, e.g., \cite{Cohen-AddadLMNP21,Assadi022}) would necessarily require an expected amortized update time of $\Omega(\sqrt{n})$ per update. 

To see the difficulty, consider $O(\sqrt{n})$ disjoint cliques, each of size $O(\sqrt{n})$, so the graph is fully structured into dense components. 
Now insert $O(\sqrt{n})$ new vertices, each connecting to all existing vertices across all cliques. 
These new connections create many cross-cluster inconsistencies, causing every original vertex to become incident to a large number of ``sparse'' edges, and thus forcing essentially all vertices to be reclassified as sparse. 
This triggers a global recomputation of the decomposition and incurs $\Theta(n)$ changes. 
However, removing these $O(\sqrt{n})$ inserted vertices restores the original clique structure, so the algorithm will need to put the $\Theta(n)$ vertices back into the clusters, which induces a high time complexity!

\paragraph{From exact decomposition to partially correct tolerant decomposition.}
The above issue suggests a fundamental barrier for maintaining exact sparse-dense decompositions. 
However, the example also reveals that exact maintenance is unnecessary for achieving an $O(1)$-approximation in correlation clustering. 
In particular, even if we do not update cluster assignments when sparse vertices are inserted, the resulting inconsistencies can be charged to the cost of the sparse edges, preserving a constant-factor guarantee in the above example. 
As such, our first step is to relax the definition of dense structure. 
In what follows, we will see that the min-disagreement correlation clustering is surprisingly ``graceful'', in the sense that even some strong structural inconsistency in the decomposition can still lead to constant-factor approximation in the cost.
Conceptually, our new ``tolerant'' SDD framework provides the first \emph{separation} between the stronger decomposition problem and the correlation clustering objective.

To ensure efficiency under worst-case deletion, our framework tolerates three specific structural inconsistencies: 
\begin{enumerate}[label=(\arabic*).]
\item 
\textbf{Tolerant Clusters (Out-edge violations):} Clusters may have many outgoing positive edges, provided those external edges are sufficiently sparse and there are \emph{not} many internal non-edges inside. 
Intuitively, clusters remain internally cohesive but need not be externally isolated. 
This is useful because it makes the decomposition robust to vertex insertions that introduce many external connections, since we do not need to modify clusters in response. 
In fact, observe that such relaxation already overcomes the $\Omega(\sqrt{n})$ update time lower bound we discussed above: when high-degree sparse interlopers are inserted, we simply do not make any updates to the clusters. 
   
\item 
\textbf{Isolated Clusters (Non-edge violations):} Clusters may contain many non-edges, provide they have enough number of sparse edges \emph{outside the cluster}. Intuitively, clusters are relaxed from near-cliques to loosely coherent groups that may be internally incomplete. 
This avoids costly restructuring when internal density is violated due to outside clustering ``stealing'' vertices from this cluster (more details to be discussed later).
    
\begin{figure}[htbp]
  \centering
  \begin{subfigure}[b]{0.49\textwidth}
    \centering
    \begin{tikzpicture}[scale=0.8]
      \draw[thick, fill=blue!10] (0,0) circle (1.5);
      \node at (0,0.8) {\small Cluster $K$};
      
      \node[circle, fill=black, inner sep=1.5pt] (u1) at (0.4,0.2) {};
      \node[circle, fill=black, inner sep=1.5pt] (u2) at (-0.4,0.3) {};
      \node[circle, fill=black, inner sep=1.5pt] (u3) at (-0.2,-0.4) {};
      \node[circle, fill=black, inner sep=1.5pt] (u4) at (0.5,-0.3) {};
      \draw (u1) -- (u2) -- (u3) -- (u4) -- (u1) -- (u3);
      
      \node[circle, fill=black, inner sep=1.5pt] (v1) at (2.2,0.6) {};
      \node[circle, fill=black, inner sep=1.5pt] (v2) at (2.2,-0.6) {};
      \node[right] at (2.3,0) {\small singleton};
      \node[right] at (2.3,-0.5) {\small or disagree};
      
      \draw[blue, thick] (u1) -- (v1);
      \draw[blue, thick] (u4) -- (v1);
      \draw[blue, thick] (u4) -- (v2);
      \draw[blue, thick] (u3) -- (v2);
      
      \draw[dashed] (v1) -- (v2);
    \end{tikzpicture}
    \caption{Tolerant Clusters}
    \label{fig:tolerant_clusters}
  \end{subfigure}
  \hfill
  \begin{subfigure}[b]{0.49\textwidth}
    \centering
    \begin{tikzpicture}[scale=0.8]
      \draw[thick, fill=blue!10] (0,0) circle (1.5);
      \node at (0,0.8) {\small Cluster $K$};
      
      \node[circle, fill=black, inner sep=1.5pt] (u1) at (0.4,0.2) {};
      \node[circle, fill=black, inner sep=1.5pt] (u2) at (-0.4,0.3) {};
      \node[circle, fill=black, inner sep=1.5pt] (u3) at (-0.2,-0.4) {};
      \node[circle, fill=black, inner sep=1.5pt] (u4) at (0.5,-0.3) {};
      \draw (u2) -- (u3);
      \draw (u1) -- (u4);

      \draw[red, dotted, thick] (u1) -- (u2);
      \draw[red, dotted, thick] (u3) -- (u4);
      
      \node[circle, fill=black, inner sep=1.5pt] (s1) at (2.2,0.5) {};
      \node[circle, fill=black, inner sep=1.5pt] (s2) at (2.2,-0.5) {};
      \node[right] at (2.3,0) {\small many singleton};
      \node[right] at (2.3,-0.5) {\small or disagree};
      
      \draw (u1) -- (s1);
      \draw (u4) -- (s2);
    \end{tikzpicture}
    \caption{Isolated Clusters}
    \label{fig:isolated_clusters}
  \end{subfigure}
  \caption{Illustration of tolerant clusters and isolated clusters.}
  \label{fig:structural_inconsistencies}
\end{figure}
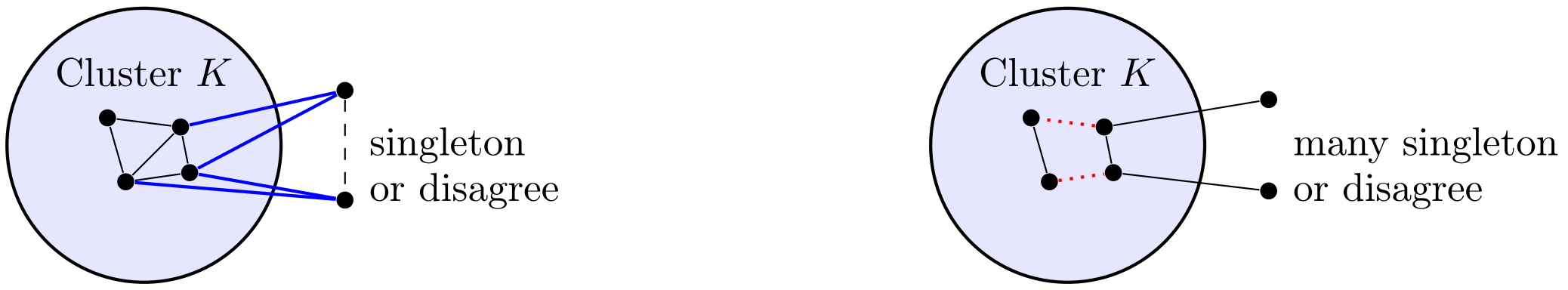
    \item \textbf{Partially Correct Singletons:} The pool of singletons (denoted as $\Vsingleton$) is permitted to include vertices that should belong to clusters, provided the total number of accumulated edges does not dominate the total number of edges that are \emph{actually} sparse. 
    Intuitively, singletons act as a buffer for uncertain assignments. This is useful because it allows us to defer reassignment decisions while still controlling the overall disagreement cost.
 
\end{enumerate}
At a high level, the first two types of tolerant structure can ensure that the algorithm does \emph{not} need to update the clusters too often, and the misclassification of the dense clusters can ease the required time complexity for cluster-forming subroutines. 
We call this new decomposition framework a \emph{partially correct tolerant decomposition}.

\paragraph{Why the relaxed structure still gives an $O(1)$-approximation.}
While our relaxed structure deviates considerably from the sparse-dense decomposition, we can nevertheless still guarantee a constant-factor approximation with the following analysis: 
\begin{itemize}
\item \textbf{Tolerant Clusters (Bounding out-edge cost):} 
As long as the number of non-edges inside a cluster is small, grouping the set of vertices would still induce a relatively small cost. 
The main risk of tolerating many outgoing edges is accidentally splitting a larger truly dense cluster. 
However, our rules ensure that these outgoing edges point primarily to either sparse vertices or to disagreement neighbors of the cluster. 
Hence, OPT must pay a cost comparable to these edges, allowing us to charge the cost to OPT. 

\item \textbf{Isolated Clusters (Bounding non-edge cost):}
When the number of non-edges inside a cluster is large, grouping $k$ vertices together joins many non-neighbors and can induce $\Theta(k^2)$ disagreement costs. 
However, our rules require that each of these vertices must have $\Omega(k)$ sparse incident edges outside of the cluster. 
Again, these edges induce $\Omega(k^2)$ cost for \emph{any} valid clustering (including OPT); as such, we can still keep the vertices together and charge the cost to OPT. 

\item \textbf{Partially Correct Singletons (Bounding misclassification cost):}
Finally, for $k$ dense vertices with degree around $2^i$, mistakenly putting them as singletons incurs roughly $k\cdot 2^i$ cost. 
We tolerate this \emph{only} when there are many sparse vertices in the graph, which induces cost much more than $k\cdot 2^i$ to OPT. 
Hence we can still keep the dense vertices as singletons and charge the cost to OPT. 
\end{itemize}

The formal proof is built on the charging argument similar to the techniques in \cite{Cohen-AddadLMNP21,Assadi022,AssadiSW23}, albeit it is considerably more involved. 
In a standard analysis, the clustering costs are charged to edges with unavoidable costs in the graph (and thus to OPT). 
One major complication in our setting is that since the pool of singleton vertices can still contain dense vertices, we cannot always find edge sets for the charging of edges going out of a cluster. 
Instead, our novel charging scheme charges these costs to the edges on singleton vertices, and then \emph{transfer} the charges of singleton vertices to OPT. 

\subsection{Tracking the set of singleton vertices and merging into clusters} 
The immediate advantage of allowing a tolerant structure in singleton vertices is that it becomes much easier to merge dense vertices into clusters. 
To understand why, we first give some intuition for what it means for a set of vertices to ``block'' clusters from forming.
In correlation clustering, a valid dense cluster must be both well-connected internally (forming a near-clique) and isolated externally (having few edges to outside vertices). 
Thus, even if a clique $V$ exists, a large external set $S$ with many connections to this clique can prevent $V$ from being recognized as a cluster because the clique no longer appears sufficiently isolated. 
In this sense, the external set $S$ acts as a ``blocking set'' that hides the clique and prevents it from being merged.

Consider a concrete example where a large blocking set $S$ of size $\Theta(n)$ hides two such groups of vertices. 
The first group is $n^{0.49}$ disjoint $O(1)$-sized cliques, each connecting to $O(1)$ vertices in $S$, and the second group is a single $\sqrt{n}$-sized clique connecting to $O(\sqrt{n})$ vertices in $S$. 
If we keep deleting the ``blocking vertices'' inside $S$ that correspond to the $O(1)$-sized cliques, those tiny cliques suddenly become perfectly isolated and should be clustered. 
However, if we insist that dense clusters must be formed immediately upon being unblocked, it is virtually impossible for us to efficiently locate these newly freed singleton vertices from the large pool of $\Vsingleton$ (here, and throughout, we use $\Vsingleton$ to denote the set of singleton vertices maintained by the algorithm). 
Note that notification-based approaches will \emph{not} work here since no random neighbors for a vertex $s\in S$ outside $S$ can be sampled unless we sample $\Omega(n)$ neighbors. 
On the other hand, if we only care about unclustered vertices that are ``dominating'' because their aggregate contribution to the approximation cost is large enough, then we do not need to search for or form any of the tiny $O(1)$-sized cliques until the $\sqrt{n}$-sized clique becomes unblocked. 
Therefore, since the large $\sqrt{n}$-sized clique dominates the singleton pool, it suffices to sample $O(2^i \log^2 n) = \widetilde{O}(\sqrt{n})$ vertices from the singleton vertices every $O(2^i)$ updates, i.e. with $\polylog{n}$ work each update, find the unblocked massive clique, form a cluster, and recurse.

\begin{figure}[htbp]
  \centering
  \begin{tikzpicture}[scale=0.9, every node/.style={scale=0.9}]
    
    \draw[thick, fill=gray!10] (0,0) ellipse (3.5cm and 1.2cm);
    \node at (0, -1.5) {\large $\Theta(n)$-size Blocking Set $S$};
    
    \node[circle, fill=black, inner sep=1.5pt] (s1) at (-2.5,0) {};
    \node[circle, fill=black, inner sep=1.5pt] (s2) at (-1.5,-0.5) {};
    \node[circle, fill=black, inner sep=1.5pt] (s3) at (-0.5,0.2) {};
    \node[circle, fill=black, inner sep=1.5pt] (s4) at (0.5,-0.4) {};
    \node[circle, fill=black, inner sep=1.5pt] (s5) at (1.5,0.3) {};
    \node[circle, fill=black, inner sep=1.5pt] (s6) at (2.5,-0.2) {};
    
    \node[red, font=\bfseries] at (-2.5,0) {$\times$};
    \node[red, font=\bfseries] at (-1.5,-0.5) {$\times$};
    \node[red, font=\bfseries] at (-0.5,0.2) {$\times$};

    \begin{scope}[shift={(-4, 3)}]
      \node[circle, fill=black, inner sep=1.5pt] (a1) at (0,0) {};
      \node[circle, fill=black, inner sep=1.5pt] (a2) at (0.5,0.5) {};
      \node[circle, fill=black, inner sep=1.5pt] (a3) at (0.5,-0.5) {};
      \draw (a1) -- (a2) -- (a3) -- (a1);
      
      \node[circle, fill=black, inner sep=1.5pt] (b1) at (1.5,0) {};
      \node[circle, fill=black, inner sep=1.5pt] (b2) at (2.0,0.5) {};
      \node[circle, fill=black, inner sep=1.5pt] (b3) at (2.0,-0.5) {};
      \draw (b1) -- (b2) -- (b3) -- (b1);
      
      \node at (2.7, 0) {\dots};
      
      \node[circle, fill=black, inner sep=1.5pt] (c1) at (3.5,0) {};
      \node[circle, fill=black, inner sep=1.5pt] (c2) at (4.0,0.5) {};
      \node[circle, fill=black, inner sep=1.5pt] (c3) at (4.0,-0.5) {};
      \draw (c1) -- (c2) -- (c3) -- (c1);
      
      \node[above] at (2, 0.7) {\small $n^{0.49}$ triangles};
    \end{scope}
    
    \draw[blue!70] (a3) -- (s1);
    \draw[blue!70] (b3) -- (s2);
    \draw[blue!70] (c3) -- (s3);

    \begin{scope}[shift={(3.5, 3)}]
      \draw[dotted, fill=blue!5] (0,0) circle (0.8cm);
      \node[circle, fill=black, inner sep=1.5pt] (x1) at (-0.3,0.3) {};
      \node[circle, fill=black, inner sep=1.5pt] (x2) at (0.3,0.3) {};
      \node[circle, fill=black, inner sep=1.5pt] (x3) at (0.4,-0.2) {};
      \node[circle, fill=black, inner sep=1.5pt] (x4) at (-0.1,-0.5) {};
      \node[circle, fill=black, inner sep=1.5pt] (x5) at (-0.5,-0.1) {};
      
      \draw (x1) -- (x2) -- (x3) -- (x4) -- (x5) -- (x1);
      \draw (x1) -- (x3); \draw (x1) -- (x4);
      \draw (x2) -- (x4); \draw (x2) -- (x5);
      \draw (x3) -- (x5);
      
      \node[above] at (0, 0.9) {\small $\sqrt{n}$-sized clique};
    \end{scope}
    
    \draw[blue!70] (x4) -- (s4);
    \draw[blue!70] (x4) -- (s5);
    \draw[blue!70] (x3) -- (s5);
    \draw[blue!70] (x5) -- (s4);
    \draw[blue!70] (x3) -- (s6);
    

  \end{tikzpicture}
  \caption{Illustration of a large blocking set $S$ that blocks potential clusters of different sizes. If we insist on forming clusters for all dense vertices, then deleting the vertices adjacent to the triangles becomes very hard to detect unless we go over the entire $\Vsingleton$ set. On the other hand, the tolerant structure in $\Vsingleton$ can effectively resolve the issue.}
  \label{fig:blocking_set_example}
\end{figure}
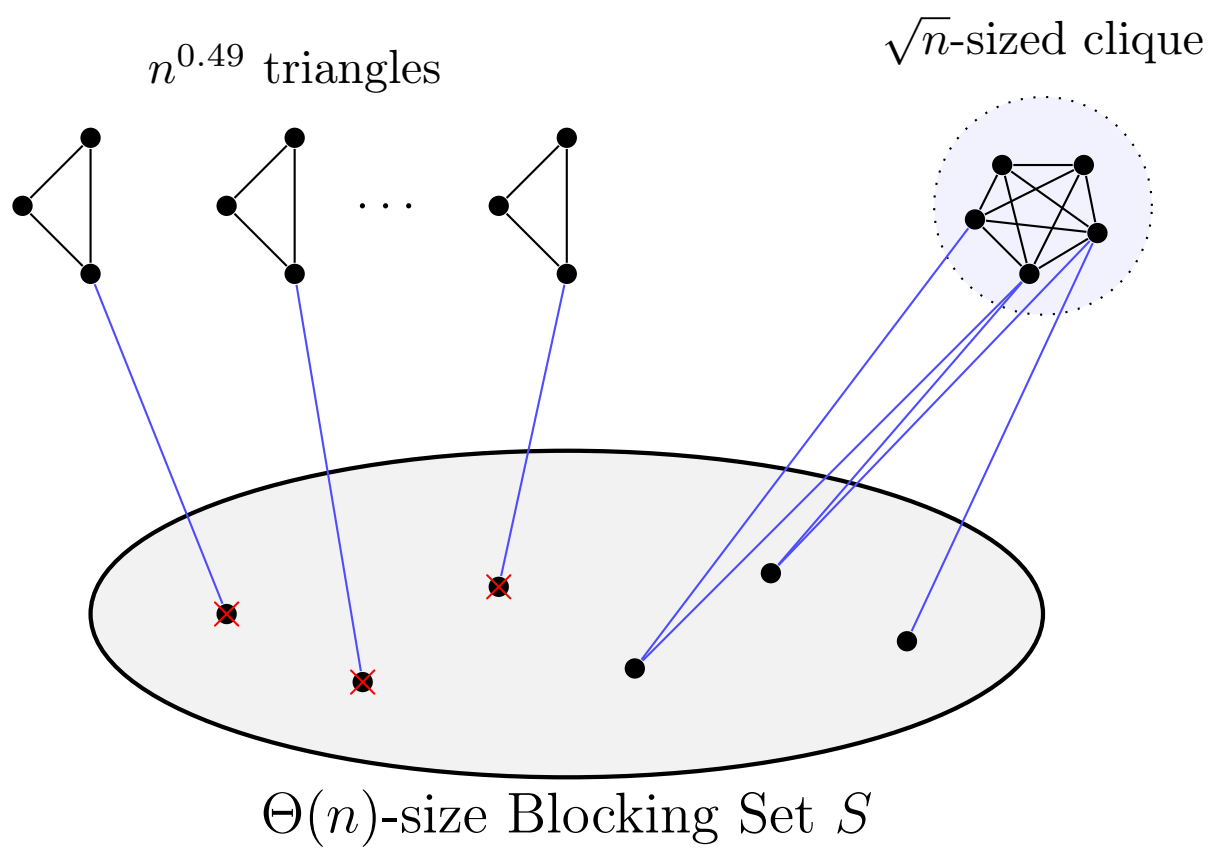

It turns out that we can formalize the notion of ``dominating'' for groups of vertices with degree around $2^i$.
In particular, we say a level $i$ is dominating if the $k$ dense vertices with degree around $2^i$ satisfy $k\cdot 2^i = \Omega(\card{\Vsingleton}/\log{n})$. 
The intuition behind this definition is that the quantity $k\cdot 2^i$ captures the total contribution of these vertices to the approximation cost; therefore, a dominating level contributes a significant fraction of the overall correlation clustering cost. 
A uniformly sampled vertex of $\Vsingleton$ is one of these $k$ dense vertices with probability $k / \card{\Vsingleton} \geq 1 / (2^i \log{n})$. Hence, $\Theta(2^i \log^2 n)$ samples detect a dominating level with high probability, which if we perform after every $O(2^i)$ updates costs $\polylog(n)$ per event.

\paragraph{Tracking a few vertices merging to a large cluster.}
Unfortunately, the reality is much more challenging than the ideal case for entire clusters waiting to be formed: it could happen that a few vertices for a dominating group need to be merged to an already-existing large cluster. 
Consider a scenario where a large cluster $K$ of size $n$ has only $O(1)$ vertices in $\Vsingleton$  waiting to merge into it. 
Furthermore, suppose there are also $n^{0.99}$ sparse vertices elsewhere. 
Since the $O(1)$ vertices have degree $2^i\approx n$, then level $i$ is dominating. 
However, uniform sampling over $\Vsingleton$ would hit these $O(1)$ dense vertices with probability $O(1/n^{0.99})$, making it impossible to sample these vertices in polylogarithmic time. 

\paragraph{Degree Bucketing Approach.} 
The above example shows that dominance alone is insufficient, as vertices at other degree scales can overwhelm uniform sampling. 
We therefore organize singleton vertices by degree and search for dominating structures across multiple scales. 
The algorithm partitions $\Vsingleton$ into $O(\log n)$ separate lists $\tilde{V}_{\mathrm{list},(i)}$, where each list contains vertices whose degrees are around $2^i$.
By separating vertices according to degree, the algorithm filters out the large number of low-degree sparse vertices and focuses sampling efforts on the degree scale where a dominating structure may exist.

To maintain these degree buckets efficiently across updates, the algorithm implements a top-down pruning and prioritization mechanism that separates vertices based on whether they have a ``low degree'' or ``high degree'' relative to the level $i$:

\begin{itemize}
\item 
Low-Degree Vertices (Safe Pruning): 
For any list $\tilde{V}_{\text{list},(i)}$, a vertex $u$ is considered to have a low degree for that level if its current degree drops below $2^i/10$. 
Such vertices are safely pruned (deleted) from $\tilde{V}_{\text{list},(i)}$. 
Structurally, a vertex with such a small degree can only become important for level $i$ again after many new vertex insertions, at which point the algorithm will track the new updates.

\item 
High-Degree Vertices (Priority-Based Deferral): 
Vertices in $\tilde{V}_{\text{list},(i)}$ with degrees substantially larger than $2^i$ (belonging to some level $j > i$) cannot simply be pruned because they might become part of a dominating cluster. 
To manage them efficiently, the algorithm adopts a top-down processing order, using the sparse populations in higher levels $j > i$ to decide whether to complete a scan of level $i$, because performing a full scan on level $i$ might visit too many sparse vertices of higher degree. 
Crucially, to determine whether to defer to a higher level, the algorithm can efficiently estimate if a higher level $j > i$ dominates level $i$ using $\widetilde O(2^i)$ samples. 
When a comparison prevents a full scan, the algorithm still attempts the scan with an allowance of $O(2^i\log^2 n)$ unsuccessful sparsity tests.
This ensures the top-down scheme remains both correct and polylogarithmically efficient under edge cases.
\end{itemize}

By utilizing the top-down prioritization and degree pruning, the algorithm isolates the appropriate dominating level. 
The key idea here is that instead of blindly sampling from the entire singleton pool, the algorithm can focus on the refined list $\tilde{V}_{\mathrm{list},(i)}$ corresponding to the dominating degree scale. 
This allows the algorithm to efficiently identify dense structures and merge their vertices into clusters while avoiding the global sampling bottleneck.

\subsubsection{Maintaining clusters}
We now discuss the subroutines that maintain clusters over time. 
The basic building blocks for our cluster-maintaining algorithm are \emph{cluster formation}, which groups vertices from $\Vsingleton$ into a new cluster, and \emph{cluster refreshment}, which periodically cleans up an existing cluster $K$ by removing sparse vertices and admitting new dense ones. 
Our refreshment algorithm tracks vertex insertions ($\inc(K)$) and deletions ($\del(K)$) made by the external sequence of updates. 
If a vertex is added to a cluster $K$, we will only track it if it is after the cluster formation or the last refreshment. 
Crucially, if a vertex is moved \emph{out of} $K$ by the algorithm, the $\del(K)$ does not track it, as it is only responsible for vertices deleted from the graph.

Because we only track vertex insertions and deletions, it might seem that no vertex could ever be moved out of $K$ between two refreshments. 
However, consider the scenario where a separate cluster $K'$ grows to be much larger than $K$.  
In this case, the vertices in $K$ might become densely connected to $K'$, meaning they should ideally merge into the larger cluster. 
To facilitate this without breaking our tracking rules, we implement a ``notification'' procedure that allows the larger $K'$ clusters to algorithmically reach in and extract vertices from $K$. 

\paragraph{Losing control of cluster sizes: tolerant structures to the rescue.} 
Unfortunately, this notification procedure comes with severe side effects. 
A cluster $K$ might not undergo any refreshment, yet its densely connected vertices can effectively abandon it to join another cluster $K'$.
We seemingly lose control of these clusters, leaving behind a group of clusters with many non-edges.
This is the ``vertices being stolen'' scenario we discussed in the intuitions for isolated clusters.

This is exactly where our Type (2) Tolerant Structure (Isolated Clusters) plays a crucial role. 
Though the vertices of $K$ were dense enough to join $K'$, the edges between the vertices in $K'$ and the remaining vertices of $K$ \emph{cannot be removed} unless the vertices are deleted. 
Therefore, the remaining vertices in $K$ now have a large number of edges with endpoints in $K'$. 
Since our relaxed decomposition permits a cluster to lack a number of internal edges as long as it has sufficiently dense external connections, this prevents the algorithm from being forced to immediately dismantle or repair the cluster, while still maintaining the desired approximation guarantees.

Finally, to handle the case where the outside cluster $K'$ is deleted (in which case the structural justification for $K$'s missing edges vanishes and $K$ must be dismantled), we use a ``lazy tracker'' to track all vertices that have \emph{ever} joined the cluster. 
The correctness of the algorithm then follows by maintaining the structural properties as prescribed by the partially correct tolerant decomposition.

\subsubsection{Amortization of the running time}
In our discussion about the tracking of the $\Vsingleton$ vertices, we established that maintaining the degree buckets is highly time-efficient. 
Therefore, what remains is to analyze two parts: 1) the overhead of updating statistics in the cluster-maintaining algorithm; and 2) the amortized time spent performing cluster formation and refreshment (i.e., distributing the cost of expensive operations across the sequence of updates that triggered them). 

For the first part, recall that we track $\inc(K)$ and $\del(K)$ to control the data stream updates.
The tracking of $\inc(K)$ can be accomplished in $O(\log{n})$ worst-case time per counter increment.
However, the analysis for $\del(K)$ is slightly harder, as we need to track the number of deleted vertices that \emph{ever joined} the cluster. 
When a vertex $v$ is deleted, it could change the counter for multiple clusters that $v$ joined before (efficiently stored in a doubly-linked list $\{K_1, \ldots, K_d\}$), and we need to bound the number of such clusters.
We show that the size of those clusters is geometrically increasing, assuming that they are not refreshed by the time when $v$ is deleted, and so there are at most $O(\log n)$ of them.

We next amortize the time complexity for cluster formation and refreshment to \emph{past updates in the graph}, including \emph{insertion} and \emph{deletion} of the vertices of the graph.
In particular, when a new cluster $K$ is formed, we aim to locate a sufficiently large group of data stream updates (e.g., $\Omega(|K|)$ updates), so that we can charge the cost of forming $K$ via an $O(1)$-to-$1$ mapping. 
For instance, if $K$ is formed or refreshed due to a sequence of $\eps|K|$ data stream updates, e.g., $\inc(K)$ or $\del(K)$ reaching $\eps|K|$ which triggers a refreshment, we can charge the efficiency cost of forming $K$ to those updates, losing only a $\frac{1}{\eps}$-multiplicative factor.

However, the main challenge is that vertices in $K$ may have been previously removed from an older cluster due to algorithmic identification, so we cannot locate new data stream updates with respect to vertices in $K$ that are available to charge. 
To solve this, one could attempt to trace these algorithmically removed vertices back to the insertions that are responsible for their removal, e.g., if vertices in $K$ were previously identified and moved to $\Vsparsealg$ by $K'$, then we charge the time of forming $K$ to that of forming $K'$. 
However, the charging mapping in this scenario might not be well-defined, e.g., it could happen that vertices in $K$ were previously moved to $\Vsparsealg$ by $K'$, while vertices in $K'$ were previously moved to $\Vsparsealg$ by an earlier cluster $K''$, resulting in an infinite chain, and we risk double-charging the same data stream updates. 

\paragraph{Forward-charging budgeting.}
Instead, we design a forward-charging budgeting scheme without tracking historical information, and thus avoiding the undesirable chain of chargings. 
We add a budget $\Psi$ (a constant to be determined) for each data stream update and transfer the budget to the clusters with which it is associated, e.g., we transfer a part of $\Psi$ to $K$ if a vertex is a data stream addition to $K$.
Then we show that every newly-formed cluster $K$ has a total budget of at least $\Omega(\eps|K|\Phi)$.
Therefore, the cluster uses the budget to pay for its own formation time, and then passes on the remaining budgets to all vertices it removes via identification. 
The key observation is that a cluster at most identifies $O(\eps|K|)$ vertices, as those vertices must be densely-connected to $K$, so the cost is safely charged if we have a larger total budget.
This process is captured by the following minted budget inequality:
\[\underbrace{C_1\eps|K|\Phi}_{\text{Unused Budgets}} \ge \underbrace{|K| \cdot \frac{\polylog(n)}{\poly(\eps)}}_{\text{Time for algorithmic operations}} + \underbrace{C_2\eps|K|\Phi}_{\text{Budgets being passed on}}.\]
Then the inequality holds if $C_1>C_2$ and $\Phi > \frac{\polylog(n)}{(C_1-C_2)\poly(\eps)}$.
With this inequality, we prove that the total amount of budgets $K$ has is larger than the amount of usage, including the time for algorithmic operations and the budgets that pass on to other vertices.
Therefore, the charging scheme works and our amortized update time is $\Psi$, i.e., the budget we assign to each data stream update.

\paragraph{Putting everything together.}
The final algorithm can therefore be viewed as dynamically maintaining the partially correct tolerant decomposition. 
The tolerant cluster definitions prevent the algorithm from overreacting to every insertion, and the partially correct singleton invariant ensures that temporary misclassifications do not accumulate enough error to break the overall approximation guarantee. 
If a vertex is deleted from a degree list $\Vsparsealgi$,it can be recovered later using the cluster notification procedures. 
On the other hand, all calls to the cluster formation and refreshment algorithms, including those induced by the singleton vertex-tracking algorithm, can be resolved by the charging scheme. 
This combination of flexible structural rules and efficient degree-bucketing enables constant-factor approximation for correlation clustering with vertex deletions in polylogarithmic amortized update time.

\paragraph{Robustness against a white-box adversary.}
We briefly introduce the idea to demonstrate the robustness of our algorithm and defer the formal proof to \Cref{sec:white-box}.
The key is that each randomized subroutine uses fresh samples to establish properties of the current graph, such as sparsity, the validity of a cluster, or the size of a degree level.
Although these outcomes determine the maintained clusters and lists, our structural lemmas, refreshment rules, and budgeting arguments apply deterministically to arbitrary update sequences whenever the local guarantees hold.
In particular, our proof only requires the randomness when a sample is freshly drawn, so no earlier sample must remain representative of a later graph.
Thus, we union bound across all calls of local randomized subroutines, and conditioned on their successes, we show the correctness of our algorithm against white-box adversary.


\section{Preliminaries}
\label{sec:prelim}
We introduce the basic notation and the notion of sparse-dense decomposition in this section.

\paragraph{Notation.} As standard, we denote a unweighted graph $G=(V,E)$ with $V$ as the set of the vertices and $E$ as the set of the edges. For any subset of vertices, $A \subseteq V$, we use $\bar{A}=V \setminus A$ to denote the complementary set of vertices in $G$. Let $\calA = \{A_{1},A_{2},\cdots\}$ be a family of collections of vertices. We say that $\calA$ is a \emph{partition} of $G$ if $i).$ $\cup_{i} A_i = V$, and $ii).$ for any pair $V_i, V_j \in \calA$, $V_i\cap V_j =\emptyset$.

For any vertex $v \in V$, we use $N(v)$ to denote the set of \emph{neighbors} of $v$, and $N[v]$ to denote the augmented neighborhood, i.e., $N[v]:=N(v)\cup \{v\}$. For a fixed set of vertices $U$, we use $N_U(v)$ to denote the set of neighbors of $v$ that are in $U$, i.e., $N_{U}(v):=N(v)\cap U$.

For correlation clustering, we have both the \emph{positive} and \emph{negative} subgraphs. Unless specified otherwise, we use $G^+=(V,E^+)$ to denote the positive subgraph and $G^-=(V,E^-)$ to denote the negative subgraph. We use $\OPT$ and $\cost(\ALG)$ to denote the optimal correlation clustering cost and the cost induced by a certain algorithm, respectively. Furthermore, for a special set of vertices $S \subseteq V$, when we write  $\OPT(S)$, we mean the clustering cost in the optimal solution induced by the set of edges incident on $S$, i.e., the set of edges in $\cup_{\substack{v\in S}\\ u\in V\setminus\{v\}} (u,v)$. 


\subsection{The Problem Description and the Model}
\label{subsec:problem-and-model}
We first introduce the problem of (min-disagreement) correlation clustering on signed complete graphs, which is the focus of this paper.

\begin{problem}[Min-disagreement correlation clustering]
Let $G=(V, E^+ \cup E^-)$ be a signed complete graph, where $V$ is the set of vertices, $E^+$ is the set of positive edges (similarity), and $E^-$ is the set of negative edges (dissimilarity). 

A clustering $\mathcal{C}$ of $G$ is a partition of the vertex set $V$ into a set of disjoint clusters, $\mathcal{C} = \{C_1, C_2, \ldots, C_k\}$, such that $V = \bigcup_{i=1}^k C_i$ and $C_i \cap C_j = \emptyset$ for $i \neq j$.

The set of disagreements with respect to a clustering $\mathcal{C}$ consists of two types of edges:
\begin{enumerate}
    \item Positive edges whose endpoints are in different clusters.
    \item Negative edges whose endpoints are in the same cluster.
\end{enumerate}

The cost for min-disagreement, denoted $\text{cost}(\mathcal{C})$, is the total number of such disagreeing edges. Formally, the cost is defined as:
$$
\text{cost}(\mathcal{C}) = \left| \{(u, v) \in E^+ \mid u \in C_i, v \in C_j, i \neq j \} \right| + \left| \{(u, v) \in E^- \mid u, v \in C_i \text{ for some } i \} \right|
$$

The goal of the \textbf{Min-Disagreement Correlation Clustering} problem is to find an (approximately) optimal clustering $\mathcal{C}^*$ that minimizes this cumulative cost.
\end{problem}

In its vanilla form, the graphs for correlation clustering are presented as an adjacency list of the \emph{entire} graph with edges and labels. However, as shown in \cite{Assadi022}, such a model does not support fast correlation clustering. 
As such, we instead consider the adjacency list model for the \emph{positive subgraph} $G^+=(V,E^+)$. This model was initiated by \cite{Assadi022}, and it has been used extensively in the dynamic setting \cite{Cohen-AddadLMP24,braverman2025fully,dalirrooyfard2025sparse}.

\begin{definition}[The adjacency list model of $G^+=(V,E^+)$, \cite{Assadi022,Cohen-AddadLMP24}]
Let $G=(V, E^+\cup E^-)$ be a signed complete graph, and let $G^+ = (V, E^+)$ be the subgraph of $G$ consisting of only the positive edges.  The \textbf{adjacency list model for the $G^+$ subgraph} is a query model where an algorithm can perform the following operations in $O(1)$ time:

\begin{enumerate}
    \item \textbf{Positive Degree Query:} For any given vertex $v \in V$, this query returns its degree in the $G^+$ subgraph, denoted as $\deg^+(v)$.

    \item \textbf{Positive Neighbor Query:} For any given vertex $v \in V$ and any integer $i$ such that $1 \le i \le \deg^+(v)$, this query returns the $i$-th vertex in the adjacency list of $v$ in $G^+$. This allows for iterating through all positive neighbors of a vertex.

    \item \textbf{Positive Edge Query:} For any two vertices $u,v$, this query returns whether $(u,v)\in E^+$.

    \item \textbf{Sampling from a Neighbor:} An algorithm can sample a random neighbor of a vertex $v$ from $N^+(v)$ in $O(1)$ time. In particular, we can first make a degree query to get $d = deg^+(v)$, then select a random integer $i$ from $[1, d]$, and finally make a neighbor query to retrieve the $i$-th neighbor.  
\end{enumerate}
\end{definition}

We then introduce the model for \emph{vertex-dynamic} updates on the adjacency list of $G^+$. The model was explored by \cite{Cohen-AddadLMP24,dalirrooyfard2025sparse}, although they did \emph{not} consider the case of \emph{worst-case deletion}. We introduce our model as follows.

\begin{definition}[Worst-case dynamic vertex insertion-deletion model, cf. \cite{Cohen-AddadLMP24,dalirrooyfard2025sparse}]
\label{def:vertex-update-CC-model}
 In our model, the $G^+$ subgraph started as $G^+ = (\emptyset, \emptyset)$. At each time $t$, a vertex can be
 \begin{itemize}
     \item inserted, which means $v$ is added to $G^+$ together with all the $\pl$ edges incident on $v$ whose other endpoints are among the vertices \emph{already exist} in $G$.
     \item deleted, which means $v$ is deleted together with all $\pl$ edges incident on $v$.
 \end{itemize}
 The adjacency list is updated after each vertex update, i.e., if $v$ already exists in the graph, and $u$ arrives with a $v$ as its neighbor, then $u$ is also added to the neighbor of $v$, denoting that $(u,v)$ is a $\pl$ edge. 
 Our amortized update time efficiency is compared to the number of \emph{vertex} insertions and deletions. The insertions and deletions are assumed to be arbitrary, i.e., they are fixed in advance in a worst-case. We also know an integer $n$ which is an upper bound of the number of vertices that ever exist in the graph.
\end{definition}

Compared to the existing work of \cite{Cohen-AddadLMP24,dalirrooyfard2025sparse}, our model notably allows arbitrary deletions to the vertices in the graph.

\paragraph{Number of vertices and updates.} We assume there are at least $10000$ vertices in the graph, since we can otherwise run $O(1)$ brute-force algorithms.
On the other hand, we focus on the case when the number of updates is bounded by $\poly(n)$, since this is the regime we can apply union bounds. 
Handling extremely long updates (e.g., exponentially many updates) remains an interesting open problem.

\subsection{Standard Technique Tools}
\label{subsec:standard-tech-tools}

We review some standard technical results that will be used repeatedly in our paper.

\paragraph{Concentration inequalities}
We use the following standard variant of the Chernoff-Hoeffding bound. 

\begin{proposition}[Chernoff-Hoeffding bound]\label{prop:chernoff}
	Let $X_1,\ldots,X_n$ be $n$ independent random variables with support in $[0,1]$. Define $X := \sum_{i=1}^{n} X_i$. Then, for every $\delta \in (0,1]$, there is 
	\begin{align*}
		\Pr\paren{\card{X - \expect{X}} > \delta\cdot \expect{X}} \leq 2 \cdot \exp\paren{-\frac{\delta^{2} \expect{X}}{3}}. 
	\end{align*}
 Furthermore, for every $\delta>0$, there is
     \begin{align*}
		\Pr\paren{\card{X - \expect{X}} > \delta\cdot \expect{X}} \leq 2 \cdot \exp\paren{-\frac{\delta^{2} \expect{X}}{2+\delta}}. 
	\end{align*}
\end{proposition}

\paragraph{Data structure that supports fast updates and search.} We use the following standard data structure that allows us to add, search, and delete an element from a (single) list of items in $O(\log{n})$ time.

\begin{proposition}
\label{prop:data-strcuture-stats}
There exists a deterministic algorithm and a corresponding data structure for maintaining a dynamic set $S$ of $n$ elements, supporting the following operations with the specified worst-case time complexities:
\begin{itemize}
    \item \textbf{Search:} Check for an element's existence in $O(\log n)$ time.
    \item \textbf{Insertion:} Add an element in $O(\log n)$ time.
    \item \textbf{Deletion:} Remove an element in $O(\log n)$ time.
    \item \textbf{Random Sampling:} Select an element uniformly at random in $O(\log n)$ time.
    \item \textbf{Return Size:} Return the size of the set in $O(\log n)$ time.
\end{itemize}
\end{proposition}

\subsection{The Standard Sparse-Dense Decomposition}
We now introduce the \emph{standard sparse-dense decomposition} that gives a \emph{partition} of the graph. The vertices are divided into \emph{almost-cliques}, which roughly means the vertices that can be ``bundled together'', and \emph{sparse vertices}, which roughly means the vertices that cannot join any almost-cliques. The technique has a rich history in theoretical computer science, and we use the version of \cite{Assadi022} for the purpose of correlation clustering.

\begin{definition}[Standard $\eps$-sparse-dense decomposition, \cite{Assadi022}]
\label{def:sdd}
Given a graph $G=(V,E)$, an $\eps$-sparse-dense decomposition $\{\Vsparse, K_1, \ldots, K_k\}$, denoted as $\SDD_{G,\eps}$, is a partition of $G$ consisting of:
\begin{itemize}
\item \textbf{Sparse vertices} $\Vsparse$: For a a fixed constant $\eta_0 \in (0,1]$, with $\epstilde=\eta_0\eps$, for any vertex $v\in \Vsparse$, either $N(v)=\emptyset$, or there are at least $\epstilde \cdot \deg{v}$ neighbors $u$ such that $\card{N[v] \sym N[u]} > \eta_0 \cdot \eps \cdot \max\set{|N[u]|,|N[v]|}$. 
\item \textbf{Dense vertices partitioned into \underline{$\eps$-almost-cliques} $K_1,\ldots,K_k$}: For every $i \in [k]$, each $K_i$ has the following properties. Let 
 $\Delta(K_{i})=\max_{v\in K_i}|N[v]|$ be the maximum closed degree of the vertices in $K_i$ (the degree is counted in $G$), then: 
\begin{enumerate}[leftmargin=20pt, label=$\roman*)$.]
\item\label{dec:non-neighbors}  Every vertex $v\in K_{i}$ has at most $\eps\cdot \Delta(K_{i})$ \emph{non-neighbors} inside ${K_{i}}$, excluding $v$ itself;
\item\label{dec:neighbors} Every vertex $v\in K_{i}$ has at most $\eps\cdot \Delta(K_{i})$ \emph{neighbors} outside ${K_{i}}$;
\item\label{dec:size} Size of each $K_i$ satisfies $(1-\eps)\cdot \Delta(K_{i}) \leq |K_{i}| \leq (1+\eps)\cdot \Delta(K_{i})$.
\end{enumerate}
\end{itemize}
\end{definition}
When we refer to a sparse-dense decomposition $\SDD_{G,\eps}$, we mean a partition of $G$ with the labels on the vertices, i.e., for each vertex $u\in \SDD_{G,\eps}$, there is a corresponding label indicating whether it is a sparse vertex, and if not, which almost-clique it belongs to.

We could similarly define whether a vertex is \emph{$\eps$-sparse} or $\eps$-dense. These notions can be formalized as follows.
\begin{definition}[cf. \cite{Assadi022,AssadiSW23}]
\label{def:sparse-dense-vertices}
Given a graph $G=(V,E)$, we define the $\eps$-sparse vertices and $\eps$-dense vertices as follows. 
\begin{itemize}
\item We say that a vertex $v$ is \emph{$\eps$-sparse} if either $N(v)=\emptyset$, or there exists at least $\eps \cdot \deg{v}$ neighbors $u$ such that $\card{N[v] \sym N[u]} > \eps \cdot \max\set{|N[u]|,|N[v]|}$. 
\item We say that a vertex $v$ is \emph{$\eps$-dense} if it is \emph{not} $\eps$-sparse.  
\end{itemize}
\end{definition}

When talking about sparsity with parameters, a vertex that is $\eps$-dense is also allowed to be $\eps'$-dense for $\eps'\geq \eps$.
Converse, a vertex that is $\eps$-sparse also implies that it can be $\eps''$ sparse for $\eps''\leq \eps$.

It is known in the literature that in a valid sparse-dense decomposition, no vertices of positive degree in an $\eps$-almost-clique can be more than $6\eps$-sparse, for $\eps\leq1/20$.
\begin{proposition}[\cite{Assadi022,AssadiSW23}]
\label{prop:cluster-vertex-are-dense}
Let $\SDD_{G,\eps}$ be a valid $\eps$-sparse-dense decomposition for graph $G$, where $\eps\leq1/20$. 
Then, for any $\eps$-almost clique $K$ and any vertex $v\in K$ of positive degree, $v$ is $6\eps$-dense.
\end{proposition}
\begin{proof}
Let $\Delta=\Delta(K)$.
For $u,v\in K$, the bounds on non-neighbors inside $K$ and neighbors outside $K$ give $|N[u]\sym N[v]|\leq4\eps\Delta$, while $|N[v]|\geq(1-2\eps)\Delta$.
Thus every positive neighbor of $v$ inside $K$ is in $6\eps$-agreement with $v$.
If $v$ has a neighbor outside $K$, then $\eps\Delta\geq1$, so $\deg(v)\geq(1-3\eps)\Delta$ and fewer than $6\eps\deg(v)$ neighbors lie outside $K$.
Otherwise every positive neighbor agrees.
Hence $v$ is $6\eps$-dense.
\end{proof}

As long as $\eps$ is a constant, the sparse-dense decomposition gives an $O(1)$ approximation. The formal statement is given as follows.
\begin{proposition}[\cite{Assadi022,AssadiSW23}]
\label{prop:sdd-approx}
Let $G=(V, E^+\cup E^-)$ be a correlation clustering instance, and let $\SDD'$ be a standard $\eps$-sparse-dense decomposition on $G^+=(V, E^+)$ such that every vertex $v\in \Vsparse$ is $\epstilde$-sparse for a fixed constant $\epstilde>0$ and every $K_i$ is an $\eps$-almost-clique. Then, for any sufficiently small \underline{constant} $\eps>0$, if we
\begin{itemize}
\item cluster each almost-clique $K_i$ in the same cluster;
\item cluster each sparse vertex as a singleton cluster
\end{itemize}
to form a correlation clustering, we can get an $O(1)$-multiplicative approximation to the optimal correlation clustering cost.
\end{proposition}

For a pair of vertices $(u,v)$, we can also define $\eps$-agreements between $u$ and $v$.
\begin{definition}
\label{lem:vertex-pair-agreement}
We say a pair of vertices $u$ and $v$ are in $\eps$-agreement if $\card{N[u]\sym N[v]}\leq \eps\cdot \max\set{|N[u]|,|N[v]|}$.
\end{definition}

\subsection{Existing algorithmic tools for sparse-dense decomposition}
In this section, we describe the existing technical tools for testing whether a vertex is sparse and for forming cliques. 

\begin{itemize}
    \item Testing whether a vertex is $\alpha\eps$-sparse/dense using $O(\frac{\log^{2}{n}}{\eps^2})$ time for any constant $\alpha$.
    \item Forming an $\eps$-almost-clique when a vertex is $\eps/20$-dense. 
\end{itemize}

%

\begin{lemma}
\label{lem:sparse-test}
There exists an algorithm such that given a vertex $v \in V$ and the adjacency list of the graph $G=(V,E)$, with high probability, in $O(\frac{\log^{2}{n}}{\alpha^2\eps^2})$ time answers ``whether $v$ is $\alpha\eps$-sparse'' for any \underline{constant} $\alpha$ with $0<\alpha\eps\leq1/100$, and $\eta=1/10$, such that
\begin{itemize}
\item If $v\in V$ is $\alpha \eps$-sparse, then the algorithm returns \emph{yes}.
\item If $v\in V$ is \emph{not} $(1-\eta)\alpha\eps$-sparse, then the algorithm returns \emph{no}.
\end{itemize}
\end{lemma}

\begin{proof}
Let $\varepsilon'=\alpha\varepsilon/42\le 1/4200$, so Lemma~3.2 of \cite{braverman2025fully} applies.
If $\deg(v)=0$, return \emph{yes}.

\emph{Low degree} ($\deg(v)<3000\log n/\varepsilon'$). We can simply decide \Cref{def:sparse-dense-vertices} exactly. For each
$u\in N(v)$, $|N[u]\triangle N[v]|=|N[u]|+|N[v]|-2|N[u]\cap N[v]|$, and the intersection
costs $O(\deg(v))$ pair queries. The total time is
$O(\deg(v)^2)=O(\log^2 n/(\alpha\varepsilon)^2)$.

\emph{High degree.} \cite{braverman2025fully} uses sparsity for open neighborhoods: $v$ is $\theta$-sparse
(for open neighborhoods) if at least $\theta\deg v$ neighbors $u$ satisfy
$|N(u)\triangle N(v)|\ge\theta\max\{\deg u,\deg v\}$. Run this test with parameter
$\varepsilon'$. With high probability it answers \emph{yes} if $v$ is
$42\varepsilon'$-sparse, and \emph{no} if $v$ is not
$38\varepsilon'$-sparse (for open neighborhoods). For adjacent $u,v$,
\begin{equation}\label{eq:oc}
  |N(u)\triangle N(v)|=|N[u]\triangle N[v]|+2,\qquad
  \max\{|N[u]|,|N[v]|\}=\max\{\deg u,\deg v\}+1.
\end{equation}
If $v$ is $\alpha\varepsilon$-sparse, it has at least
$\alpha\varepsilon\deg v=42\varepsilon'\deg v$ witnesses $u$. By \eqref{eq:oc}, each has
$|N(u)\triangle N(v)|>\alpha\varepsilon(\max\{\deg u,\deg v\}+1)+2
>42\varepsilon'\max\{\deg u,\deg v\}$. So $v$ is $42\varepsilon'$-sparse (for open
neighborhoods), and the test answers \emph{yes}.
If $v$ is $38\varepsilon'$-sparse (for open neighborhoods), it has at least
$\tfrac{38}{42}\alpha\varepsilon\deg v\ge 0.9\alpha\varepsilon\deg v$ witnesses $u$. By
\eqref{eq:oc}, each has
\[
  |N[u]\triangle N[v]|\ge\tfrac{38}{42}\alpha\varepsilon\max\{\deg u,\deg v\}-2
  >0.9\alpha\varepsilon(\max\{\deg u,\deg v\}+1),
\]
since $\tfrac{38}{42}-0.9=\tfrac{1}{210}$ and
$\max\{\deg u,\deg v\}\ge\deg v\ge 126000\log n/(\alpha\varepsilon)
>210(2+0.9\alpha\varepsilon)/(\alpha\varepsilon)$.
So $v$ is $0.9\alpha\varepsilon$-sparse. Contrapositively, a vertex that is not
$(1-\eta)\alpha\varepsilon$-sparse is not $38\varepsilon'$-sparse (for open neighborhoods),
and the test answers \emph{no}.
The test takes $O(\log^2 n/\varepsilon'^2)=O(\log^2 n/(\alpha\varepsilon)^2)$ time.
\end{proof}

\begin{lemma}
\label{lem:dense-vertex-clique-forming}
There is an algorithm that, given an $\varepsilon/20$-dense vertex $v$, for $\varepsilon\le 1/200$, that does not belong to any maintained cluster , with high probability 
does one of the following.
\begin{itemize}
\item \emph{Merge:} it adds $v$ to an existing cluster $C$ in $O(\log^2 n/\varepsilon^2)$
  time. Before and after admission, $C$ has a $4\varepsilon$-dense core of size at least
  $(1-8\varepsilon)|C|$ and the dense core after admission contains $v$. A merge happens whenever
  $\deg(v)\ge 10/\varepsilon$ and some cluster $C$ has
  $|N(v)\setminus C|<3\varepsilon\deg(v)$ and $|C|\le(1+4\varepsilon)\deg(v)$.
\item \emph{Form:} it forms an $(\varepsilon/2)$-almost-clique $C$ that includes $v$ such that
  $(1-\varepsilon/20)|N[v]|\le|C|\le(1+0.051\varepsilon)|N[v]|$ in
  $O(|C|\log^2 n/\varepsilon^2)$ time. $C$ contains every vertex that is in
  $(\varepsilon/5)$-agreement with $v$ and no $2\varepsilon$-sparse vertex. If
  $\deg(v)<10/\varepsilon$, then $C=N[v]$ is an isolated exact clique.
\end{itemize}
\end{lemma}

\begin{proof}
Let $K'$ consist of $v$ and its positive neighbors in $(\varepsilon/20)$-agreement with $v$.
Then $|K'|\ge(1-\varepsilon/20)|N[v]|$ and
$(1-\varepsilon/20)|N[v]|\le|N[u]|\le|N[v]|/(1-\varepsilon/20)$ for every $u\in K'$.
If $\deg(v)<10/\varepsilon$, fewer than $(\varepsilon/20)\deg(v)<1$ neighbors disagree with
$v$; every neighbor therefore satisfies
$|N[u]\triangle N[v]|\le(\varepsilon/20)|N[v]|/(1-\varepsilon/20)<1$. Thus $N[v]$ is an
isolated exact clique, which we return directly.

Otherwise, sample $O(\log n)$ neighbors of $v$ and collect their cluster labels, giving
$O(\log n)$ candidates. For each candidate $C$, estimate $|N(v)\setminus C|/\deg(v)$ to
additive error $\varepsilon/10$ using $O(\varepsilon^{-2}\log n)$ fresh neighbor samples,
accepting if the estimate is at most $3.1\varepsilon$ and $|C|\le(1+4\varepsilon)\deg(v)$.
A cluster satisfying the stated completeness condition is sampled and accepted with high
probability. For an accepted cluster, $T=\{v\}\cup(K'\cap C)$ satisfies
\[
  |T|\ge 1+(1-3.25\varepsilon)\deg(v).
\]
Every $u\in T$ has fewer than $3.302\varepsilon|N[v]|$ neighbors outside $T$ and fewer than
$0.051\varepsilon|N[v]|$ non-neighbors inside $T$. Removing $v$ adds at most
$1\le\varepsilon|N[v]|/10$ outside neighbors. The closed-degree bounds above,
$|T\setminus\{v\}|\ge(1-3.35\varepsilon)|N[v]|$, and
\[
  \frac{|T\setminus\{v\}|}{|C|}\ge\frac{1-3.25\varepsilon}{1+4\varepsilon}>1-8\varepsilon
\]
show that $T\setminus\{v\}$ and $T$ are the required $4\varepsilon$-almost-clique cores
before and after admission.

If no candidate is accepted, enumerate $N[v]$, sample $O(\log n)$ uniform members $w$ of
$N[v]$, and enumerate $N[w]$ whenever $|N[w]|\le 2|N[v]|$. Their union has
$O(|N[v]|\log n)$ candidate vertices and incidences. Every vertex in
$(\varepsilon/5)$-agreement with $v$ has more than $0.9|N[v]|$ common positive neighbors
with $v$ in $K'$, so all such vertices are enumerated with high probability. For each
candidate $u$, estimate $|N[u]\triangle N[v]|/\max\{|N[u]|,|N[v]|\}$ to additive error
$\varepsilon/100$ using $O(\varepsilon^{-2}\log n)$ samples from $N[u]$ and positive-edge
queries to $v$ (or equality with $v$). Accept estimates at most $0.3\varepsilon$.
Consequently, the output $C$ contains $K'$ and every $(\varepsilon/5)$-agreeing vertex,
while every accepted vertex is in $0.31\varepsilon$-agreement with $v$.

Each vertex in $C\setminus N[v]$ has at least
$(1-\varepsilon/20-0.31\varepsilon/(1-0.31\varepsilon))|N[v]|$ neighbors in $K'$, whereas
each member of $K'$ has at most $(\varepsilon/20)|N[v]|/(1-\varepsilon/20)$ neighbors outside
$N[v]$. Counting these edges gives
\[
  |C\setminus N[v]|\le\frac{(\varepsilon/20)|N[v]|}
  {(1-\varepsilon/20)\big(1-\varepsilon/20-0.31\varepsilon/(1-0.31\varepsilon)\big)}
  <0.051\varepsilon|N[v]|.
\]
Hence the stated size bounds hold. Agreement with $v$, together with $K'\subseteq C$ and
this bound on $|C\setminus N[v]|$, gives each member of $C$ fewer than
$0.37\varepsilon|N[v]|$ internal non-neighbors and fewer than $0.37\varepsilon|N[v]|$
external neighbors. Moreover, $|N[v]|\le\Delta(C)\le|N[v]|/(1-0.31\varepsilon)$ and
\[
  (1-\varepsilon/2)\Delta(C)\le(1-\varepsilon/20)|N[v]|\le|C|
  \le(1+0.051\varepsilon)|N[v]|\le(1+\varepsilon/2)\Delta(C),
\]
so $C$ is an $(\varepsilon/2)$-almost-clique. For $u,w\in C$, their normalized neighborhood
difference is at most $0.62\varepsilon/(1-0.31\varepsilon)^2<0.63\varepsilon$, while
$\deg(u)\ge(1-0.41\varepsilon)|N[v]|$. Thus fewer than $0.38\varepsilon$ of $u$'s neighbors
lie outside $C$, and all its neighbors inside $C$ are in $2\varepsilon$-agreement with it;
no member is $2\varepsilon$-sparse. All guarantees hold simultaneously by Chernoff bounds
and a union bound. Constant-time adjacency queries give $O(\log^2 n/\varepsilon^2)$ merge
time; enumerating and deduplicating candidates in a searchable dictionary, then testing
them, takes $O(|C|\log^2 n/\varepsilon^2)$ time.
\end{proof}

In fact, the algorithm for \Cref{lem:dense-vertex-clique-forming} does not have to follow the rigid definition of sparse-dense decomposition, but is rather capable of identifying vertices that are ``common neighbors'' of the almost-clique $C$. We will use the following lemma to handle certain subroutines in our algorithm.

\begin{lemma}
\label{lem:dense-vertex-clique-forming-generalize}
There exists an algorithm that, given parameters $0<\gamma\leq\delta/2<1$ and an $\eps$-almost-clique $C$ (\Cref{def:sdd}), where $\eps\leq1/2$, 
with high probability identifies a set of vertices $\bar{C}\subseteq V\setminus C$ such that
\begin{itemize}
    \item every vertex $u\notin C$ satisfying $\card{N(u)\cap C}\geq \delta\cdot \card{C}$ belongs to $\bar C$;
    \item no vertex in $\bar{C}$ satisfies $\card{N(u)\cap C}\leq \gamma\cdot \card{C}$.
\end{itemize}
The algorithm runs in $O(\card{C}\log{n}/\delta)$ time.
\end{lemma}
\begin{proof}
The algorithm is as follows.
\begin{tbox}
    \begin{itemize}
        \item Sample a multiset $S$ of $\lceil500 \log{n}/\delta\rceil$ vertices from $C$ independently uniformly at random.
        \item Enumerate the neighbors of the sampled vertices, counting multiplicities. Add all vertices $u\notin C$ such that $u$ has at least $400 \log{n}$ neighbors in $S$ to $\bar{C}$.
    \end{itemize}
\end{tbox}
Intersections with $S$ below count multiplicity.

For each $v$ with at least $\delta \card{C}$ neighbors in $C$, if we sample a vertex $u$ uniformly at random from $C$, the probability for $(u,v)$ to be an edge is at least $\delta$. Therefore, if we sample $\lceil 500 \log{n}/\delta\rceil$ vertices from $C$, the expected number of edges we identify between $v$ and $S$ is at least $500\log{n}$. By concentration inequality (\Cref{prop:chernoff}), we have
\begin{align*}
    \Pr\paren{\card{N(v)\cap S}\leq 400\log{n}}\leq \exp\paren{-\frac{(0.2)^2 500\log{n}}{3}}\leq 1/\poly(n).
\end{align*}
On the other hand, for any vertex $w$ with at most $\gamma \card{C}\leq \delta/2\cdot \card{C}$ neighbors in $C$, the expected number of neighbors is at most $251\log{n}$. Here rounding the sample size adds at most $\delta/2<1$ to $250\log n$, and $\log n\geq1$. Again, by concentration inequality, we have
\begin{align*}
    \Pr\paren{\card{N(w)\cap S}\geq 400\log{n}}\leq \exp\paren{-\frac{(0.59)^2 251\log{n}}{3}}\leq 1/\poly(n).
\end{align*}
A union bound gives both guarantees simultaneously for all vertices.
Each sampled vertex has degree at most $\Delta(C)\leq|C|/(1-\eps)\leq2|C|$.
Store $C$ in an array, mark its vertices, and count sampled-neighbor incidences in temporary per-vertex counters, resetting only the marks and counters touched by the call.
Enumeration and counting take $O(|C||S|)$ time, so
the total running time is $O(\card{C}\log{n}/\delta)$, as desired.
\end{proof}


For each of these subroutines, any prescribed failure probability $\xi>0$ is obtained by replacing each sample-size factor $\log n$ by a sufficiently large multiple of $\log(n/\xi)$, with the same replacement in its time bound.
All samples are fresh and independent after the graph and the input to the call have been fixed, and all inclusion and exclusion guarantees hold simultaneously.

\subsection{Notations}

\textcolor{black}{We keep a table of notions in \Cref{tab:algo_notation} to help the readers understand our algorithm and analysis.}

\begin{table}[h!]
\centering
\renewcommand{\arraystretch}{1.4}
\begin{tabular}{c|p{0.67\textwidth}}
\hline\hline
\textbf{Notation} & \textbf{Semantics} \\
\hline
\multicolumn{2}{c}{\textbf{Graph properties}} \\
\hline
$G^{(t)}=(V^{(t)},E^{(t)})$ & An input graph with vertex set $V^{(t)}$ and edge set $E^{(t)}$ for the time step $t$. \\
$n$ & The maximum number of vertices at any point, i.e., $\max_t \card{V_t}$. \\
$\degtu{t}{u}$ & The degree of a vertex $u$ at time step $t$. \\
$\Vsparse^{(t)}$ & The set of vertices that are $\eps$-sparse at time $t$. \\
\hline
\multicolumn{2}{c}{\textbf{Algorithmic partitioning}} \\
\hline
$\Vsparsealg^{(t)}$ & The set of active vertices treated as singleton by our algorithms at time $t$. \\
$\Vsparsealgi^{(t)}$ & The set of candidate vertices maintained at level $i$ by our algorithms at time $t$. \\
$\VsparseC^{(t)}$ & The set of positive-degree vertices in $\Vsparsealg^{(t)}$ that are $\zeta$-sparse at time $t$ \\
$\VsparseI^{(t)}$ & $\{v\in\Vsparsealg^{(t)}:\deg^{(t)}(v)>0\} \setminus \VsparseC^{(t)}$ \\
$\Vsparsei^{(t)}$ & The set of vertices $u \in \Vsparsealg^{(t)}$ with $\degtu{t}{u} \in [2^i, 2^{i+1})$. \\
$\VsparseCi^{(t)}$ & The set of vertices in $\Vsparsei^{(t)}$ that are $\zeta$-sparse at time $t$ \\
$\VsparseIi^{(t)}$ & $\Vsparsei^{(t)} \setminus \VsparseCi^{(t)}$ \\
$\Vsparsei^{(t)}(\beta\eps)$ & The set of vertices in $\Vsparsei^{(t)}$ that are $\beta\eps$-dense. \\
$\Vsparsealgi^{(t)}(\beta\eps)$ & The set of vertices in $\Vsparsealgi^{(t)}$ that are $\beta\eps$-dense. \\
\hline\hline
\end{tabular}
\caption{Notation and semantics for the variables and objects maintained by our algorithms.}
\label{tab:algo_notation}
\end{table}




\section{Correlation Clustering with a Partially Correct Tolerant Sparse-Dense Decomposition}
\label{sec:new-version-sdd}
We introduce a new version of sparse-dense decomposition in this section to obtain algorithms with $\polylog(n)$ update time. 
There are several nuanced issues if we want to simply use the original decomposition as in \Cref{def:sdd}. As it will be clear in later sections of the paper (and as in the discussion in \Cref{subsec:tech-overview}), we need to  ``tolerate'' the following aspects in our maintained clustering structure to ensure efficiency.
\begin{enumerate}
    \item \textbf{The ``wrong categorization'' of sparse vertices.} Ideally, we would want to keep a set of sparse vertices $\Vsparse$. However, some rather dense vertices might be in this set because the cost of the particular wrong cluster has \emph{not} dominated the overall cost. 
    \item \textbf{The outgoing edges on the vertices in clusters.} We would want the almost-cliques $K_1\cup \ldots \cup K_k$ to have few out-going edges. However, this might lead to large update time. 
    These vertices are sparse to all existing almost-cliques, which would cause a high update time complexity. 
    Therefore, we need to use a new notion that allows out-going edges.
    \item \textbf{The clusters without dense parts.} Finally, even for the non-neighbors inside the clusters, we will need to relax the requirements when the cluster size is proportional to the degree. This is necessary since in some cases, some of the dense vertices in a cluster might join other (much bigger) clusters, and we do \emph{not} have control over the those clusters.
\end{enumerate}
In light of the above discussion, we will argue that it suffices to maintain a data structure with the following properties:
\begin{enumerate}
    \item A \emph{sufficiently large fraction} of singleton vertices are actually sparse.
    \item Clusters will \emph{not} partition sufficiently dense collections of vertices (``\textbf{tolerate clusters}'').
    \item Clusters will \emph{not} contain sparse vertices with many non-edges between them \emph{unless} there are \emph{many edges going out of the cluster} (``\textbf{isolated clusters}''). 
\end{enumerate}
See \Cref{fig:structural_inconsistencies} for an illustration of the structural inconsistency we allow.

We will eventually argue that even with the above relaxations, the clustering can still maintain an $1/\poly(\eps)$-approximation, which translates into $O(1)$-approximation with constant $\eps$.

\subsection{A partially correct tolerant sparse-dense decomposition}
\label{subsec:properties-new-sdd}
We introduce the formal notion of \emph{partially correct tolerant sparse-dense decomposition} as follows.
\begin{definition}[Partially correct tolerant decomposition]
\label{def:partial-correct-new-sdd}    
Given a graph $G=(V,E)$, we say a partition $\{\Vsingleton, \cup_j \Ktolerant_j, \cup_j \Kisolate_j\}$ of $G$ is a partially correct tolerant $(\eps_s, \eps_\tau, \eps_l)$-sparse-dense decomposition if it satisfies the following properties , for fixed positive constants $C_t, C_l, \eta_d, \eta_l, C_0$. 
\begin{itemize}
\item \textbf{Partially correct singleton (``sparse'') vertices $\Vsingleton$}: 
\begin{enumerate}[label=$\roman*)$.]
    \item Let $\Vsingleton(i)$ be the set of vertices $v\in \Vsingleton$ such that $\deg(v)\in [2^i, 2^{i+1})$, where $i=0,\ldots,\lfloor\log_2 n\rfloor$ and indices outside this range denote empty buckets.
    \item For each such level $i$, we further partition the set into: 
    \begin{itemize}
        \item $\VsingletonC(i)$: the set of vertices $v\in \Vsingleton(i)$ that are $\eps_s$-sparse.
        \item $\VsingletonW(i):=\Vsingleton(i)\setminus \VsingletonC(i)$. 
    \end{itemize}
    \item For each such level $i$, 
    \begin{itemize}
        \item either $\sum_{j\in\{i-1,i,i+1\}}\card{\VsingletonC(j)}\geq \frac{1}{100}\cdot \card{\Vsingleton(i)}$.
        \item or $\sum_{v\in \VsingletonW(i)}\deg(v) \leq \frac{1}{C_0 \log{n}} \cdot \sum_{j} \sum_{v\in \VsingletonC(j)} \deg(v)$.
    \end{itemize}
    \item Let $\VsingletonC:=\cup_{i} \VsingletonC(i)$ and $\VsingletonW:=\cup_{i} \VsingletonW(i)$.
\end{enumerate}
\item \textbf{Tolerant clusters $\cup_j \Kdense_j$}: Each $\Kdense_j$ \emph{either} has $\Kdense_j = \{v\}$ and $N(v)=\emptyset$, \emph{or} each vertex $v \in \Kdense_{j}$ satisfies the following properties.
\begin{enumerate}[label=$\roman*)$.]
\item\label{line:new-non-neighbors-tolerant}  $v$ has at most $\eps_\tau\cdot |\Kdense_{j}|$ \emph{non-neighbors} inside ${\Kdense_{j}}$ not counting $v$ itself. 
\item\label{line:new-neighbors-tolerant} At least \textbf{one of the following} is true
\begin{itemize}
    \item \textbf{either} $v$ is $\eta_d \cdot \eps_\tau$-sparse.
    \item \textbf{or} let $K_v \subseteq N[v]$ be the set of vertices that are in $C_t\eps_\tau$-agreement with $v$, including $v$ itself, at least $\eps_\tau$-fraction of $K_v \setminus \Kdense_{j}$ is in $\Vsingleton$.
\end{itemize}
\end{enumerate}

\item \textbf{Isolated clusters $\cup_j \Kisolate_j$}: Each cluster $\Kisolate_j$ satisfies
\begin{enumerate}[label=$\roman*)$.]
    \item $\Kisolate_j$ is \textbf{not} a tolerant cluster. 
    \item For each $v\in \Kisolate_j$, at least \textbf{one of the following} is true
        \begin{itemize}
            \item \textbf{either} $v$ is $\eta_l \cdot \eps_l$-sparse
            \textbf{and} $\deg(v)\geq \card{\Kisolate_j}/2$; 
            \item \textbf{or} let $K_v \subseteq N[v]$ be the set of vertices that are in $C_l\eps_l$-agreement with $v$, including $v$ itself, we have at least $\eps_l$-fraction of $K_v \setminus \Kisolate_{j}$ is in $\Vsingleton$, \textbf{and} $\card{K_v\setminus \Kisolate_j}\geq \eps_l \cdot \card{\Kisolate_j}$.
                \end{itemize}
        \end{enumerate}
\end{itemize}
\end{definition}

Note that in the new definition, the vertices in a cluster might actually satisfy the properties for the \emph{sparse vertices}. 
As such, we forgo the \emph{separation} between sparse vertices and almost-cliques\footnote{We also get rid of the name ``almost-clique'' since these partitions are \emph{not} close to cliques.}, and we will simply check whether a given partition \emph{satisfies} the properties as prescribed by \Cref{def:partial-correct-new-sdd}. 
We further observe that a decomposition in the \emph{original definition} also satisfies the properties as prescribed by \Cref{def:partial-correct-new-sdd}.
\begin{lemma}
\label{lem:old-sdd-is-valid-for-new}
Let $\SDD_{G,\eps}$ be a standard $\eps$-sparse-dense decomposition (see \Cref{def:sdd}) with $\eps\leq 1/50$. Then, $\SDD_{G,\eps}$ is also a partially correct tolerant $(\eta_0 \cdot \eps, 2\eps, 0)$-sparse-dense decomposition (\Cref{def:partial-correct-new-sdd}), with $C_t\leq1$. 
\end{lemma}
\begin{proof}
    Isolated vertices are regarded as tolerant singleton clusters. All remaining singleton vertices in $\SDD_{G,\eps}$ are actually $\eta_0 \cdot \eps$-sparse. Therefore, there are no $\VsingletonW$, and all singleton vertices for all levels $i\in[O(\log{n})]$ are $\eta_0\eps$-sparse.
    For almost-cliques $K_j$ and any vertex $u\in K_j$, the number of internal non-neighbors is at most $\eps\Delta(K_j)\leq \eps|K_j|/(1-\eps)\leq2\eps|K_j|$.
For a positive neighbor $v$ outside $K_j$, either $v$ is a singleton or $v$ is in another almost-clique. In the latter case, let $\Delta$ be the larger maximum closed degree of the two almost-cliques. The size, internal non-neighbor, and exterior-degree bounds give
\[\card{N[u]\sym N[v]}\geq (1-3\eps)\Delta\geq \tfrac12\max\{|N[u]|,|N[v]|\}>2C_t\eps\max\{|N[u]|,|N[v]|\}.\]
Thus every $2C_t\eps$-agreeing positive neighbor outside $K_j$ is a singleton, which leads to the desired statement.
\end{proof}

Although \Cref{lem:old-sdd-is-valid-for-new} is a straightforward observation, it gives us a powerful tool: we could \emph{form} the decomposition according to \Cref{def:sdd}, and \emph{maintain} the decomposition according to \Cref{def:partial-correct-new-sdd}. 

\subsection{Approximation analysis of the partially correct tolerant sparse-dense decomposition}
\label{sebsec:approx-guarantee-partial-correct-sdd}

Throughout this section we suppress dependence on $C_t, C_l, \eta_d, \eta_l, C_0$ in $O(.)$.
We need to prove that if we could maintain a partially correct tolerant sparse-dense decomposition on $G^+ = (V, E^+)$, we can maintain an $O(1)$-approximation for correlation clustering.
A more formal statement is as follows.
\begin{theorem}
\label{thm:approximation-partial-correct-sdd}
Let $G=(V, E^+\cup E^-)$ be a correlation clustering instance, and let $\SDDpar$ be a partially correct tolerate $(\eps_s,  \eps_\tau, \eps_l)$-sparse-dense decomposition on $G^+ = (V, E^+)$ as defined by \Cref{def:partial-correct-new-sdd}. Then, for suitable choices of positive constant $\eps_s,  \eps_\tau,\eps_l \leq 1/500$, if we arrange the correlation clustering as follows:
\begin{itemize}
\item cluster the vertices in each cluster $\Ktolerant_j$ and $\Kisolate_j$ together;
\item cluster each vertex $v\in \Vsingleton$ as a singleton cluster.
\end{itemize}
Then, we can get an $O(1)$-multiplicative approximation to the optimal correlation clustering cost.
If there are no isolated clusters, we also allow $\eps_l=0$ and omit their cost analysis.
\end{theorem}


We prove \Cref{thm:approximation-partial-correct-sdd} for the rest of this section. The partially correct decomposition in \Cref{def:partial-correct-new-sdd} looks involved; nevertheless, compared to \Cref{prop:sdd-approx}, the changes in the costs can be categorized into the following aspects.
\begin{itemize}
    \item $\pl$ edges incident on the vertices in $\Vsingleton$, especially the costs induced by $\VsingletonW$ on all levels.
    \item $\pl$ edges going out of tolerant clusters $\Kdense_j$ and connecting to $\Vsingleton$ and other clusters.
    \item $\nl$ edges inside $\Kisolate_j$ clusters and the $\pl$ edges going out. 
\end{itemize}

At a high level, the proof of \Cref{thm:approximation-partial-correct-sdd} first bounds the costs induced by the vertices that are \emph{actually sparse} and the tolerant clusters. 
We then charge the costs for $\VsingletonW$ and the costs induced by $\Kisolate_j$ clusters to these costs.

\subsection*{A charging scheme} 
Before we formally prove the approximation guarantees, we introduce the charging framework of \cite{Assadi022,AssadiSW23}.
This is particularly important for the analysis of the cost induced by the $\eps_s$-\emph{sparse} vertices ($\VsingletonC$, not to be confused with $\Vsingleton$).
Recall that we deal with a signed complete graph $G=(V, E^+\cup E^-)$; we do \emph{not} care about the time index $t$ in this section since this argument is for the partition (which can be offline).
To be consistent with the notation of \cite{Assadi022,AssadiSW23}, we use $z \in V$ and edges $f$ incident on $z$ ($f \in E^+(z) \cup E^-(z)$) as the variable for the set of edges \emph{being charged}. 
The charging scheme is as follows.

\begin{tbox}
Charging scheme of~\cite{Assadi022,AssadiSW23}. 

\begin{enumerate}
	\item Initially, $\charge{z}{f} = 0$ for all $z \in V$ and $f \in E(z)$; 
	\item For any edge $e \in E$ with $\edgecost{\ALG}{e} = 1$, we will \underline{find a collection of vertex-edge pairs}, called the \textit{charge-set} of $e$: 
	\[
		\chargeset{e} \subseteq \set{(z,f) \mid z \in V, \text{~~} f \in E(z), \text{~and~} \edgecost{\OPT}{f} = 1}. 
	\]
	For simplicity of notation, we define $\chargeset{e} = \emptyset$ if $\edgecost{\ALG}{e} = 0$. 
	\item Increase $\charge{z}{f}$ for all $(z,f) \in \chargeset{e}$ by $\card{\chargeset{e}}^{-1}$. 
\end{enumerate}
\end{tbox}

The charging scheme works because every charged vertex-edge pair includes an edge that $\OPT$ must also pay for. Consequently, limiting the \emph{maximum} charge on any single pair to a constant value is sufficient to demonstrate that $\ALG$ provides an $O(1)$-approximation. This observation is formally stated in the lemma below, drawn from \cite{Assadi022,AssadiSW23}.

\begin{lemma}[\cite{Assadi022,AssadiSW23}]
\label{lem:charging}
	Suppose there is a choice of $\chargeset{e}$ for edges $e \in E$ in the charging scheme such that for all $z \in V$ and $f \in E(z)$, we have $\charge{z}{f} \leq \alpha$ for some $\alpha \geq 1$.  Then, the algorithm $\ALG$ is a $2\alpha$-approximation to the optimal correlation clustering cost, i.e., $\cost(\ALG) \leq 2\alpha \cdot \OPT$. 
\end{lemma}
By~\Cref{lem:charging}, we just need to define charges for the edges in a way that we only write $O(1)$ charges to any vertex-edge pair. Following the approach in \cite{Assadi022,AssadiSW23}, this analysis is broken into two parts: handling sparse vertices and handling almost-cliques.

\subsection*{Costs induced by \texorpdfstring{$\pl$}{pl} edges on $\Vsingleton$} 
We observe that since we put vertices in $\Vsingleton$ as \emph{singleton vertices}, only the $\pl$ edges will induce costs. To bound the cost induced by the vertices in $\Vsingleton$, we take a two-step approach:
\begin{itemize}
    \item We first analyze the costs induced by the vertices that are \emph{actually} $\eps_s$-sparse, i.e., the vertices in $\VsingletonC$.
    \item We then charge the costs induced by $\VsingletonW$ to the costs induced by $\VsingletonC$.
\end{itemize}

\paragraph{Cost induced by $\pl$ edges on $\VsingletonC$.} For the vertices that are $\eps_s$-sparse, we can directly invoke the lemma in \cite{Assadi022,AssadiSW23} since the properties for sparse vertices in \Cref{def:partial-correct-new-sdd} has not changed from the same notion in \Cref{def:sdd}\footnote{Except that we classify vertices with $N^{+}(u)=\emptyset$ as clusters; however, this does not change the cost since these vertices are put into singleton clusters with $0$ cost anyway.}.
\begin{lemma}[\cite{Assadi022,AssadiSW23}]
\label{lem:charging-sparse}
	There exist sets $\chargeset{e}$ such that for every $\eps_s$-sparse vertex $v$ and $e \in E(v)$, there is 
	\[
	\text{for all vertex-edge pairs $(z,f)$:} \quad \spcharge{z}{f} = O(\eps_s^{-2}). 
	\] 
\end{lemma}
The same charging bound applies to the number of positive edges incident on any set of $\eps_s$-sparse vertices: apply the argument to a clustering that makes those vertices singletons. With closed neighborhoods, disagreement witnesses exclude the two endpoints, and the charging argument is unchanged.

As such, the total costs induced by the sparse vertices can be bounded by $O(\eps_s^{-2})\cdot\OPT$ as follows.
\begin{lemma}
    \label{lem:correct-sparse-vertices}
    Given the positive subgraph $G^+=(V,E^+)$ of a correlation clustering instance, and 
    let $\SDDpar_{G,\eps}=\{\Vsingleton, \cup_j \Kdense_j, \cup_j \Kisolate_j\}$ be a partially correct tolerate $(\eps_s,  \eps_\tau, \eps_l)$-sparse-dense decomposition  on the positive subgraph $G^+$. Then, let $\cost(\ALG, \VsingletonC)$ be the cost induced by the $\pl$ edges on $\VsingletonC$. We have that
    \begin{align*}
        \cost(\ALG, \VsingletonC) \leq O(\frac{1}{\eps_s^2})\cdot \OPT.
    \end{align*}
\end{lemma}

\paragraph{Cost induced by $\pl$ edges on $\VsingletonW$.} In general, the vertices in $\VsingletonW$ might induce high cost since each $v\in \VsingletonW$ can pay $\deg^+(v)$ cost. 
However, the catch here is that the \emph{total cost} induced by the vertices in $\VsingletonW$ cannot dominate the cost in $\Vsingleton$.
In what follows, we will prove the approximation guarantees induced by the $\pl$ edges on the vertices of $\VsingletonW$ by charging to the costs induced by the $\pl$ edges on $\VsingletonC$.

\begin{lemma}
    \label{lem:wrong-sparse-vertices}
    Given the positive subgraph $G^+=(V,E^+)$ of a correlation clustering instance, and 
    let $\SDDpar_{G,\eps}=\{\Vsingleton, \cup_j \Kdense_j, \cup_j \Kisolate_j\}$ be a partially correct tolerate $(\eps_s,  \eps_\tau, \eps_l)$-sparse-dense decomposition  on the positive subgraph $G^+$. Then, let $\cost(\ALG, \VsingletonW)$ be the cost induced by the $\pl$ edges on $\VsingletonW$. We have that
    \begin{align*}
        \cost(\ALG, \VsingletonW) \leq O(1)\cdot \cost(\ALG, \VsingletonC) .
    \end{align*}
\end{lemma}
\begin{proof}
    We analyze the two cases in $\VsingletonW$ for each level $i \in [O(\log{n})]$:
    \begin{itemize}
        \item Case $A).$ $\sum_{j\in\{i-1,i,i+1\}}\card{\VsingletonC(j)}\geq \frac{1}{100}\cdot \card{\Vsingleton(i)}$. In this case, note that since a vertex $v\in \VsingletonC(i)$, we have $\deg^+(v)\geq 2^i$. Therefore, counting an incident edge at most twice, we have that
        \begin{align*}
            \cost(\ALG, \VsingletonC(i))\geq \tfrac12\,2^i\cdot \card{\VsingletonC(i)}.
        \end{align*}
        On the other hand, the cost induced by $\VsingletonW(i)$ is upper-bounded by
        \begin{align*}
          \cost(\ALG, \VsingletonW(i)) &\leq 
            \cost(\ALG, \Vsingleton(i)) \\
            &\leq 2^{i+1}\cdot \card{\Vsingleton(i)}\\
            & \leq 200\cdot 2^i \cdot \sum_{j\in\{i-1,i,i+1\}}\card{\VsingletonC(j)} \tag{by the condition of Case $A).$}\\
            &\leq 800 \cdot \sum_{j\in\{i-1,i,i+1\}}\cost (\ALG, \VsingletonC(j)).
        \end{align*}
        \item Case $B).$ $\sum_{v\in \VsingletonW(i)}\deg^{+}(v) \leq \frac{1}{C_0 \log{n}} \cdot \sum_{i} \sum_{v\in \VsingletonC(i)} \deg^{+}(v)$. In this case, we know that 
        \begin{align*}
        \cost(\ALG, \VsingletonW(i)) &\leq \sum_{v\in \VsingletonW(i)}\deg^+(v) \\
        & \leq \frac{1}{C_0 \log{n}} \cdot \sum_{i} \sum_{v\in \VsingletonC(i)} \deg^{+}(v)\\
        & \leq O(\frac{1}{\log{n}}) \cdot \cost(\ALG, \VsingletonC). 
        \end{align*}
    \end{itemize}
    Therefore, in either case, we have that 
    \begin{align*}
    & \cost(\ALG, \VsingletonW(i)) \\
    & \leq 800 \cdot \sum_{j\in\{i-1,i,i+1\}}\cost (\ALG, \VsingletonC(j)) + O(\frac{1}{\log{n}}) \cdot \cost(\ALG, \VsingletonC),
    \end{align*}
    which implies 
    \begin{align*}
        & \sum_{i=0}^{O(\log{n})} \cost(\ALG, \VsingletonW(i)) \\
        & \leq \sum_{i=0}^{O(\log{n})} \paren{800 \cdot \sum_{j\in\{i-1,i,i+1\}}\cost (\ALG, \VsingletonC(j)) + O(\frac{1}{\log{n}}) \cdot \cost(\ALG, \VsingletonC)}\\
        &\leq  O(1)\cdot \sum_{i=0}^{O(\log{n})}\sum_{j\in\{i-1,i,i+1\}}\cost(\ALG, \VsingletonC(j)) \\ &\quad{}+ O(\log{n})\cdot \frac{1}{C_0 \log{n}} \cdot \cost(\ALG, \VsingletonC) \tag{there are at most $O(\log{n})$ levels}\\
        &\leq O(1)\cdot \cost (\ALG, \VsingletonC),
    \end{align*}
    where the last line follows since $\sum_{j=0}^{O(\log{n})}\cost(\ALG, \VsingletonC(j)) \leq 2\cost (\ALG, \VsingletonC)$ and each level is accounted for at most $3$ times.
    As such, we conclude that $\cost(\ALG, \VsingletonW)\leq\allowbreak \sum_i \cost(\ALG, \VsingletonW(i)) \leq\allowbreak O(1)\cdot \cost(\ALG, \VsingletonC)$, as desired by the lemma statement.
\end{proof}

Combining \Cref{lem:correct-sparse-vertices,lem:wrong-sparse-vertices} leads to the following approximation guarantee for the $\pl$ edges on $\Vsingleton$.
\begin{lemma}
    \label{lem:partial-correct-sparse-vertices}
    Given the positive subgraph $G^+=(V,E^+)$ of a correlation clustering instance, and 
    let $\SDDpar_{G,\eps}=\{\Vsingleton, \cup_j \Kdense_j, \cup_j \Kisolate_j\}$ be a partially correct tolerate $(\eps_s,  \eps_\tau, \eps_l)$-sparse-dense decomposition  on the positive subgraph $G^+$. Then, let $\cost(\ALG, \Vsingleton)$ be the cost induced by the $\pl$ edges on $\Vsingleton$. We have that
    \begin{align*}
        \cost(\ALG, \Vsingleton) \leq O(\frac{1}{\eps_s^2})\cdot \OPT.
    \end{align*}
\end{lemma}

\subsection*{The cost induced by tolerant clusters}
We now turn to the tolerant clusters with potentially many neighbors outside. We will eventually show that this will \emph{not} blow up the approximation by too much. To the above end, we analyze the two sets of edges that induce costs: the costs for the $\pl$ edges going \emph{out of} the cluster and the costs for the $\nl$ edges \emph{within} the same cluster. In the remainder of the writing for the cost induced by tolerant clusters, we use $\Ktolerate$ (without the index $j$ for $\Ktolerate_j$) to denote the tolerant cluster when the context is clear.



\paragraph{The $\pl$ edges going \emph{out} of the clusters.} We first handle the case for the $(+)$ edges going \emph{out of} almost-cliques. 
For each $\Ktolerate_j$, we let $\Ktolerate_j(\textnormal{sparse})$ be the set of vertices that are $\eta_d \eps_\tau$-sparse, and we let $\Ktolerate_j(\textnormal{singleton})$ 
be the set of vertices $v$ such that $K_v\setminus\Ktolerate_j$ has at least $\eps_\tau$-fraction in $\Vsingleton$. Note that $\Ktolerate_j = \Ktolerate_j(\textnormal{sparse})\cup \Ktolerate_j(\textnormal{singleton})$. We first observe that by \Cref{lem:charging-sparse}, the $\pl$ edges incident on $\cup_j \Ktolerate_j(\textnormal{sparse})$ is at most $O(1/\eps^2_\tau)\cdot \OPT$.
\begin{lemma}
\label{lem:low-out-degree-edge-cost-sparse}
The total cost induced by $\pl$ edges going out of $\Ktolerate_j$ for any $j$ and incident on $\cup_j \Ktolerate_j(\textnormal{sparse})$ is at most $O(1/\eps^2_\tau)\cdot \OPT$.
\end{lemma}

We now need to handle the edges incident on $\cup_j \Ktolerate_j(\textnormal{singleton})$. We further divide these edges into the following categories:
\begin{itemize}
    \item $E^{+}(\Ktolerate(\textnormal{singleton}), K_v)$: the set of edges going out of $\Ktolerate_j$ for any $j$, and the other endpoint is in $K_v$, where $v\in\Ktolerate_j(\textnormal{singleton})$ is the first endpoint.
    \item $E^{+}(\Ktolerate(\textnormal{singleton}), \textnormal{disagree})$: the set of edges going out of $\Ktolerate_j$ for any $j$, and the other endpoint is \emph{not} in $K_v$, where $v\in\Ktolerate_j(\textnormal{singleton})$ is the first endpoint.
\end{itemize}
We use $\cost(\ALG, E^{+}(\Ktolerate(\textnormal{singleton}), K_v))$ and $\cost(\ALG, E^{+}(\Ktolerate(\textnormal{singleton}), \textnormal{disagree}))$ to denote the cost induced by the two parts, respectively. The first cost is straightforwardly charged to the cost of $\Vsingleton$ as follows.

\begin{lemma}
\label{lem:low-out-degree-edge-cost-agree}
\[\cost(\ALG, E^{+}(\Ktolerate(\textnormal{singleton}), K_v)) \leq O(\frac{1}{\eps^2_s}\cdot \frac{1}{\eps_\tau}\cdot \OPT).\]
\end{lemma}
\begin{proof}
    We first note
    \begin{align*}
    \cost(\ALG, E^{+}(\Ktolerate(\textnormal{singleton}), K_v))& \leq \card{E^{+}(\Ktolerate(\textnormal{singleton}), K_v)}\\
    & \leq \sum_j\sum_{v\in\Ktolerate_j(\textnormal{singleton})}|K_v\setminus\Ktolerate_j|\\
    &\leq \frac{1}{\eps_\tau}\cdot \card{\cup_j \cup_{v\in \Ktolerate_j(\textnormal{singleton})}E^{+}(v, \Vsingleton\cap K_v)}\\
    &\leq \frac{1}{\eps_\tau}\cdot \card{E^+(\Vsingleton)}.
    \end{align*}
    The second-last inequality holds since there are at least $\eps_\tau$ fraction in $\Vsingleton$, and the charged edges are disjoint: each joins a unique vertex $v$ in a tolerant cluster to a singleton. Therefore, using \Cref{lem:partial-correct-sparse-vertices}, we have
    \begin{align*}
        \cost(\ALG, E^{+}(\Ktolerate(\textnormal{singleton}), K_v)) \leq \frac{1}{\eps_\tau}\cdot \cost(\ALG, \Vsingleton) \leq O(\frac{1}{\eps^2_s}\cdot \frac{1}{\eps_\tau}\cdot \OPT),
    \end{align*}
    which is as desired.
\end{proof}

We emphasize in that in the proof of \Cref{lem:low-out-degree-edge-cost-agree}, we charge the $\pl$ edge costs to the $\pl$ edges incident on $\Vsingleton$, which is set of the singleton vertices kept by the algorithm. We will later use the cost upper bound for the $\pl$ edges on $\Vsingleton$ (\Cref{lem:partial-correct-sparse-vertices}) to transfer the charges. 
We never rely on any bound for the number of $\pl$ edges for $\Vsparse$ (the truly sparse edges) in this proof.

Next, we prove that the cost induced by the edges in $E^{+}(\Ktolerate(\textnormal{singleton}), \textnormal{disagree})$ can be charged into $O(1/\eps^2_\tau)\cdot \OPT$ of cost. 
\begin{lemma}
\label{lem:low-out-degree-edge-cost-disagree}
There exists $\chargeset{e}$ for every $\pl$ edges in $E^{+}(\Ktolerate(\textnormal{singleton}), \textnormal{disagree})$ such that 
\begin{align*}
	\textnormal{for all vertex-edge pairs $(z,f)$:} \quad \occharge{z}{f} \leq O(\frac{1}{\eps_\tau}). 
\end{align*}
Therefore, the cost induced by $\pl$ edges in $E^{+}(\Ktolerate(\textnormal{singleton}), \textnormal{disagree})$ is at most $O(\frac{1}{\eps_\tau})\cdot \OPT$, i.e.,
\begin{align*}
    \cost(\ALG, E^{+}(\Ktolerate(\textnormal{singleton}), \textnormal{disagree})) \leq O(\frac{1}{\eps_\tau})\cdot \OPT.
\end{align*}
\end{lemma}
\begin{proof}
We are talking about edges whose two endpoints are not in $C_t\eps_\tau$-agreement.
Therefore, the proof is very similar to the original charging argument used for $\pl$ edges for sparse vertices in \cite{Assadi022,AssadiSW23}. 
We argue that for any $\pl$ edges whose endpoints have a high level of disagreement, clustering the endpoints together would cause a sufficiently large amount of costs. 
To this end, we introduce the following technical claim from \cite{Assadi022,AssadiSW23}. 
\begin{claim}[\cite{Assadi022,AssadiSW23}]
\claimlab{helper-charge}
Let $\theta \in (0,1)$ be a parameter and $\Etilde$ be any collection of edges in $E^+$ in the signed complete graph $G=(V, E^+ \cup E^-)$ such that for all $\xi = (\alpha,\beta) \in \Etilde$, 
\begin{align}
	\card{N^{+}[\alpha] \sym N^{+}[\beta]} \geq \theta \cdot \max\set{\degp{\alpha}+1,\degp{\beta}+1}. 
\end{align}
Then, there is a choice of $\chargeset{\xi}$ for all $\xi \in \Etilde$ such that $\charge{z}{f} = O(\theta^{-1})$ for all vertex-edge pairs $(z,f)$ in $G$. 
\end{claim}
For completeness, if $\OPT$ separates the endpoints of $\xi$, charge $\xi$ itself at one endpoint. Otherwise, for each vertex in the closed-neighborhood symmetric difference, choose an incident edge on one of the two endpoints that $\OPT$ misclassifies, and use the resulting vertex-edge pairs as the charge set. The endpoints of $\xi$ are not in this symmetric difference, so these are valid third-vertex witnesses. A fixed pair at $z$ receives at most $1/(\theta\deg^+(z))$ from each incident positive edge, hence at most $O(\theta^{-1})$ in total.

For any fixed $\Ktolerant_j$, let $(u,v)$ be a $\pl$ edge such that $v\in \Ktolerant_j$ and \linebreak[4]$u\in N^+(v)\setminus(K_v\cup\Ktolerant_j)$ such that $u$ and $v$ are not in $C_t\eps_\tau$-agreement. 
By definition, edge $(u,v)$ satisfies
\begin{align*}
    \card{N^{+}[u]\sym N^{+}[v]} > C_t\eps_\tau \cdot \max\{\degp{u}+1, \degp{v}+1\}.
\end{align*}
Hence, we could apply \claimref{helper-charge} to show that
\begin{align*}
    \textnormal{for all vertex-edge pairs $(z,f)$:} \quad \discharge{z}{f} \leq O(\eps_\tau^{-1}). 
\end{align*}
The ``therefore'' part then follows from \Cref{lem:charging}.
\end{proof}

\paragraph{The $\nl$ edges inside the clusters.} We now deal with the $\nl$ edges inside the same clusters. 
We note that the argument is essentially unchanged from the proof for the same lemma in \cite{Assadi022,AssadiSW23}: at a high level, for this part of the argument, we only use the fact that for each pair of vertices $u,v \in \Ktolerate$, we have $\card{N[u]\cap N[v]}\geq (1-O(\eps))\cdot \card{\Ktolerate}$. This property still holds as long as each vertex $v$ has at most $\eps\cdot \card{\Ktolerate}$ non-neighbors inside $\Ktolerate$. We provide the lemma and the proof in what follows for completeness.
\begin{lemma}\label{lem:charging-in-cliques}
There exists $\chargeset{e}$ for every $(-)$-edge inside any single tolerant cluster such that 
	\[
	 \textnormal{for all vertex-edge pairs $(z,f)$:} \quad \iccharge{z}{f} \leq 1. 
	\] 
\end{lemma}
\begin{proof}
In a nutshell, the lemma states the simple fact that if a collection of vertices $\Ktolerant$ only has a few $\nl$ edges in between, then putting them together will \emph{not} induce too much extra cost. The lemma is a formalization of the intuition. In this proof, we use $\eps$ as a short-hand notation as $\eps_\tau$ since the context is clear.

Let $C^*_1, C^*_2, \cdots, C^*_\ell$ be the $\ell$ clusters in $\OPT$. For $\nl$ edges that are split by two clusters of the optimal clustering, we analyze the cases as follows.

\begin{enumerate}[label=(\roman*).]
\item \textbf{If for all $j\in [\ell]$, $\card{C^*_j \cap \Ktolerate}\leq (1-4\eps)\cdot \card{\Ktolerate}$.} In this case, by the requirement of property \ref{line:new-non-neighbors-tolerant} in \Cref{def:partial-correct-new-sdd}, every $v\in \Ktolerate$ has at most $\eps\card{\Ktolerate}$ non-neighbors from the $G^+$ subgraph in $\Ktolerate$, i.e., at most $\eps\card{\Ktolerate}$ neighbors in $\Ktolerate$ that connects to $v$ by a $\nl$ edge. As such, the total number of $(-)$ edges inside $\Ktolerate$ is at most 
\begin{align*}
\frac{1}{2}\cdot \sum_{v\in \Ktolerate}\paren{\text{number of $\nl$ edges from $v$ to $\Ktolerate$}}\leq \frac{\eps}{2}\cdot \card{\Ktolerate}^2. 
\end{align*}

On the other hand, we define
\[\chargeset{e} = \{(x,\{x,y\})\mid \{x,y\}\in E^+,\ x\in C_j^*\cap\Ktolerate,\ y\in C_i^*\cap\Ktolerate,\ i\ne j\},\]
and we show that $\card{\chargeset{e}}\geq \frac{\eps}{2}\cdot \card{\Ktolerate}^2$. Note that by the constraints on the number of $\nl$ edges, for each vertex $v\in C_j^*\cap\Ktolerate$, the number of edges that cross the clusters is at least
\begin{align*}
\card{\Ktolerate}-\card{C_j^*\cap\Ktolerate}-\eps\card{\Ktolerate} &\geq \card{\Ktolerate}-(1-4\eps)\card{\Ktolerate}-\eps\card{\Ktolerate}\\
&=3\eps\card{\Ktolerate}.
\end{align*}
There are $\card{\Ktolerate}$ vertices in $\Ktolerate$; as such, we conclude that in this case, there is 
\begin{align*}
\card{\chargeset{e}} \geq 3\eps \cdot \card{\Ktolerate}^2.
\end{align*}
Therefore, by our charging scheme, the number of charges we can write on any vertex-edge pair is at most $1$, as desired.
\item \textbf{If there exists a $j\in[\ell]$ such that $\card{C^*_j \cap \Ktolerate}> (1-4\eps)\cdot \card{\Ktolerate}$.} In this case, we analyze the $\nl$ edges on each vertex $v$ that are split by the optimal clustering. Assign each such edge to one endpoint $v$ outside $C^*_j$. Note that by the guarantee in \Cref{def:partial-correct-new-sdd}, there are at most $\eps \card{\Ktolerate}$ such $\nl$ edges for each vertex $v$.
We define the $\chargeset{e}$ as
\begin{align*}
\chargeset{e} = \{(v,\{v,x\})\mid \{v,x\}\in E^+,\ x \in C^*_j\cap \Ktolerate\} \\ \text{for all separated $(-)$ edges $e$ assigned to $v$}.
\end{align*}
We claim that the size of $\chargeset{e}$ is at least $\eps \card{\Ktolerate}$. To see this, note that the number of $\pl$ edges that connect $v$ and $C^*_j\cap \Ktolerate$ is at least 
\begin{align*}
& \paren{\text{total number of edges}} - \paren{\text{number of $\nl$ edge in $\Ktolerate$}} \\
& \geq (1-4\eps)\card{\Ktolerate}-\eps\card{\Ktolerate} \tag{by the properties prescribed in \Cref{def:partial-correct-new-sdd}}\\
& = (1-5\eps)\card{\Ktolerate}\\
&\geq \eps \card{\Ktolerate},
\end{align*}
where the last inequality holds true for any $\eps\leq 1/50$. Therefore, the charge written on any pair in $\chargeset{e}$ is at most $1$. Furthermore, note that the definition of $\chargeset{e}$ is \emph{disjoint} for different assigned vertices $v\in \Ktolerate$, which concludes the proof of the case. 
\end{enumerate}
Finally, note that if the optimal clustering does \emph{not} split the $\nl$ edge, we can simply write a charge of $1$ on the pair consisting of the edge itself and one endpoint, which also gives us the desired statement as in \Cref{lem:charging-in-cliques}. These negative-edge pairs are disjoint from the positive-edge pairs above, and different tolerant clusters use disjoint internal edges.
\end{proof}

Combining \Cref{lem:low-out-degree-edge-cost-sparse,lem:low-out-degree-edge-cost-agree,lem:low-out-degree-edge-cost-disagree} and \Cref{lem:charging-in-cliques} leads to the following lemma for the approximation guarantees for the tolerant clusters.
\begin{lemma}
    \label{lem:cost-tolerant-cluster}
    Given the positive subgraph $G^+=(V,E^+)$ of a correlation clustering instance, and 
    let $\SDDpar_{G,\eps}=\{\Vsingleton, \cup_j \Kdense_j, \cup_j \Kisolate_j\}$ be a partially correct tolerate $(\eps_s,  \eps_\tau, \eps_l)$-sparse-dense decomposition on the positive subgraph $G^+$. Let $\cost(\ALG, \cup_j \Ktolerant_j)$ be the cost induced by the $\pl$ edges on going out of the vertices for all $\cup_j \Ktolerant_j$ and the $\nl$ edges inside the $\cup_j \Ktolerant_j$ clusters. We have
    \begin{align*}
        \cost(\ALG, \cup_j \Ktolerant_j) \leq O(\frac{1}{\eps_\tau^2}+\frac{1}{\eps_s^2\eps_\tau})\cdot \OPT.
    \end{align*}
\end{lemma}

\subsection*{The cost induced by isolated clusters} 
We now provide an analysis of the cost of the isolated clusters as follows.
\begin{lemma}
    \label{lem:cost-bad-cluster}
    Given the positive subgraph $G^+=(V,E^+)$ of a correlation clustering instance, and 
    let $\SDDpar_{G,\eps}=\{\Vsingleton, \cup_j \Kdense_j, \cup_j \Kisolate_j\}$ be a partially correct tolerate $(\eps_s,  \eps_\tau, \eps_l)$-sparse-dense decomposition on the positive subgraph $G^+$. Let $\cost(\ALG, \cup_j \Kisolate_j)$ be the cost induced by the $\pl$ edges on going out of the vertices for all $\cup_j \Kisolate_j$ and the $\nl$ edges inside the $\cup_j \Kisolate_j$ clusters. We have
    \begin{align*}
        \cost(\ALG, \cup_j \Kisolate_j) \leq O(\frac{1}{\eps_l^3}+\frac{1}{\eps^2_l\cdot \eps^2_s})\cdot \OPT.
    \end{align*}
\end{lemma}
\begin{proof}
Fix an isolated cluster $\Kisolate_j$, by definition, 
for each $v\in \Kisolate_j$, recall that at least one of the following cases will happen:
\begin{itemize}
    \item Either Case A). the vertex is $\eta_l\eps_l$ sparse and $\deg(v)\geq \card{\Kisolate_j}/2$;
    \item Or Case B). at least $\eps_l$ of $\card{K_v \setminus \Kisolate_j}$ is in $\Vsingleton$, and $\card{K_v \setminus \Kisolate_j}\geq \eps_l \cdot \card{\Kisolate_j}$.
\end{itemize}
Define $\Kisolate_j(\textnormal{sparse})$ as the vertices satisfying Case A), and $\Kisolate_j(\textnormal{disagree})$ as the remaining vertices in $\Kisolate_j$, which satisfy Case B).

\paragraph{Costs induced by the $\pl$ edges incident on $\cup_j \Kisolate_j$.}
Let $E^{+}(\Kisolate_j(\textnormal{sparse}))$ be the set of positive edges with at least one endpoint in $\Kisolate_j(\textnormal{sparse})$, and let $\cost\paren{E^{+}(\Kisolate_j(\textnormal{sparse}))}$ be the induced cost using the clustering in \Cref{thm:approximation-partial-correct-sdd}. 
By \Cref{lem:charging-sparse}, we know that the cost induced by the total number of $\pl$ edges in $\cup_{j} E^{+}(\Kisolate_j(\textnormal{sparse}))$ induces at most $O(1/\eps_l^2)\cdot \OPT$, i.e.,
\begin{align*}
    \cost\paren{\cup_j E^{+}(\Kisolate_j(\textnormal{sparse}))} \leq \card{\cup_{j} E^{+}(\Kisolate_j(\textnormal{sparse}))} \leq O(\frac{1}{\eps_l^2})\cdot \OPT.
\end{align*}
This is an upper bound for the cost of the $\pl$ edges going out of $E^{+}(\Kisolate_j(\textnormal{sparse}))$. 

For the edges going out of $\Kisolate_j(\textnormal{disagree}))$, we can similarly define $E^{+}(\Kisolate_j(\textnormal{disagree}))$ as the set of $\pl$ edges with one endpoint in $\Kisolate_j(\textnormal{disagree})$ and the other outside $\Kisolate_j$ and the induced cost $\cost\paren{E^{+}(\Kisolate_j(\textnormal{disagree}))}$.
We note that case B) has the same exterior-fraction requirement as the tolerant vertices.
Therefore, by \Cref{lem:low-out-degree-edge-cost-agree} and \Cref{lem:low-out-degree-edge-cost-disagree}, the $\pl$ edges going out of $\Kisolate_j(\textnormal{disagree})$ account for at most $O(1/\eps_l+1/(\eps_s^2\eps_l))\cdot \OPT$ cost, i.e., 
\begin{align*}
    \cost\paren{\cup_j E^{+}(\Kisolate_j(\textnormal{disagree}))} \leq \card{\cup_{j} E^{+}(\Kisolate_j(\textnormal{disagree}))} \leq O(\frac{1}{\eps_l}+\frac{1}{\eps_s^2}\cdot\frac{1}{\eps_l}) \cdot\OPT.
\end{align*}
A more concrete analysis can be done by writing the charges as we did in \Cref{lem:low-out-degree-edge-cost-agree} and \Cref{lem:low-out-degree-edge-cost-disagree} using parameter $\eps_l$ (as opposed to $\eps_\tau$), and we omit the calculations to avoid repetition.

Therefore, the cost induced by $\pl$ edges going out of clusters can be bounded by
\begin{align*}
    \cost\paren{\cup_j E^+(\Kisolate_j)} & \leq \cost\paren{\cup_j E^{+}(\Kisolate_j(\textnormal{sparse}))} + \cost\paren{\cup_j E^{+}(\Kisolate_j(\textnormal{disagree}))} \\
    & = O(\frac{1}{\eps_l^2}+ \frac{1}{\eps_s^2}\cdot\frac{1}{\eps_l})\cdot \OPT.
\end{align*}

\paragraph{Costs induced by the $\nl$ edges incident on $\cup_j \Kisolate_j$.}
We define a set of ``charging edge'' $\Echarge_{\Kisolate_j, v}$ for each $\Kisolate_j(\textnormal{sparse})$ and $\Kisolate_j(\textnormal{disagree})$ as follows.
\begin{itemize}
    \item For each $v\in \Kisolate_j(\textnormal{sparse})$, $\Echarge_{\Kisolate_j, v}$ is the edges $(u,v)$ for $u\in N^{+}(v)$;
    \item For each $v\in \Kisolate_j(\textnormal{disagree})$, $\Echarge_{\Kisolate_j, v}$ is the edges $(u,v)$ for $u\in N^{+}(v)\setminus \Kisolate_j$.
\end{itemize}
We use $\Echarge_{j, v}$ as a short-handed notation. 
By our calculations for the cases for the $\pl$ edge going out of the clusters, we have 
\begin{align*}
    \card{\cup_{j,v} \Echarge_{j, v}} \leq O(\frac{1}{\eps_l^2}+ \frac{1}{\eps_s^2}\cdot\frac{1}{\eps_l})\cdot \OPT.
\end{align*}
For the costs induced by the $\nl$ edges, let $E^-(\cup_j \Kisolate_j)$ be set of $\nl$ edges with both endpoints in \emph{the same} $\Kisolate_j$, we have
\begin{align*}
    \card{E^-(\cup_j \Kisolate_j)} &\leq \sum_{j} \card{\Kisolate_j}^2. 
\end{align*}
We prove that $\sum_j\card{\Kisolate_j}^2 \leq O(1/\eps_l)\cdot \card{\cup_{j,v} \Echarge_{j, v}}$. To see this, note that for any vertex $v\in \Kisolate_j(\textnormal{sparse})$, since we have $\card{\Echarge_{j, v}} = \deg^{+}(v) \geq \card{\Kisolate_j}/2$. On the other hand, for $v\in \Kisolate_j(\textnormal{disagree})$, we have $\card{\Echarge_{j, v}} \geq \card{K_v \setminus \Kisolate_j}\geq \eps_l \cdot \card{\Kisolate_j}$. Therefore, we have
\begin{align*}
    2\card{\cup_{j,v} \Echarge_{j, v}} &\geq \sum_{j}\sum_{v\in \Kisolate_j}\card{\Echarge_{j, v}} \tag{an edge has at most two endpoints}\\
    &\geq \eps_l \cdot \sum_{j}\sum_{v\in \Kisolate_j} \card{\Kisolate_j} \\
    & = \eps_l \cdot \sum_j\card{\Kisolate_j}^2.
\end{align*}
Therefore, we have 
\begin{align*}
    \card{E^-(\cup_j \Kisolate_j)} \leq O(1/\eps_l)\cdot \card{\cup_{j,v} \Echarge_{j, v}} \leq O(\frac{1}{\eps_l^3}+ \frac{1}{\eps_s^2}\cdot\frac{1}{\eps_l^2})\cdot \OPT,
\end{align*}
as desired.
\end{proof}

\paragraph{Finalizing the proof of \Cref{thm:approximation-partial-correct-sdd}.} Combining \Cref{lem:partial-correct-sparse-vertices,lem:cost-tolerant-cluster,lem:cost-bad-cluster} leads to the approximation guarantees of $O(1/\poly(\eps_s, \eps_\tau, \eps_l))$-approximation using the clustering of \Cref{thm:approximation-partial-correct-sdd}. Picking $\eps_s, \eps_\tau, \eps_l$ as sufficiently small constants leads to the desired $O(1)$-approximation.



\newcommand{\ctilde}{\ensuremath{\widetilde{c}}\xspace}

\section{The Algorithm}
\label{sec:dynamic-alg-vertex-update}

We present our main algorithm for dynamic correlation clustering in this section. 
The overall goal is to maintain a partially-correct tolerant SDD as defined in \Cref{def:partial-correct-new-sdd}.
At a very high level, our idea is a generalization of the decomposition-based algorithm in \cite{Cohen-AddadLMP24,braverman2025fully} that maintains (various versions of) agreement decomposition. 
The key differences are: $i).$ using our relaxed definition of decomposition in \Cref{sec:new-version-sdd}, we are able to make clique refreshment \emph{less frequently}, which results in better efficiency; 
and $ii).$ we handle the formation of almost-cliques induced by vertex deletions using a set of entirely novel ideas.

\subsection{The main algorithm}
We present our main algorithm as in \Cref{alg:vertex-dynamic-alg} to \Cref{alg:check:n:singleton}. 

\paragraph{Maintaining clusters.}
To begin with, \Cref{alg:vertex-dynamic-alg} handles vertex updates and maintains the clusters formed by the algorithm.
In summary, for each \textbf{insertion} of vertex $v$, it conducts a test to decide whether $v$ should join an existing cluster, form a new cluster, or join in $\Vsparsealg$; and for each \textbf{deletion} of vertex $v$, it removes $v$ from the current cluster or $\Vsparsealg$ list and notifies all cluster that $v$ has \emph{ever} joined.
In particular, we apply \Cref{lem:sparse-test} to test whether $v$ is $\eps/20$-sparse, and if it returns yes, we directly add it to the $\Vsparsealg$ data structure (in the following, \emph{$\eps$-sparsity test returns yes} means \Cref{lem:sparse-test} reports that the vertex is $\eps$-sparse).
Otherwise, we first check if there is a cluster with a $\beta\eps$-dense core($\beta=4$) that $v$ can merge to, as defined below:
\begin{definition}[Dense core]
\label{def:dense:core}
For a graph $G=(V,E)$, let $S\subseteq V$ be a set of vertices.
We say that $S$ is an $\eps$-dense core if it is an $\eps$-almost-cliques (see \Cref{def:sdd}).
A cluster $K$ contains an $\eps$-dense core if there exists $K'\subseteq K$ such that $K'$ is an $\eps$-dense core with $\card{K'}\geq (1-O(\eps))\cdot \card{K}$.
\end{definition}
If the test returns no but $v$ cannot join any existing cluster, we call \Cref{alg:new-ac} to form a new $\eps$-almost-clique according to \Cref{def:sdd} that includes all vertices that are in $\eps/5$-agreement with $v$.
Here, this new cluster may take vertices from other existing clusters.
We always form a cluster according to the classical SDD definition; however, later on we maintain them according to our partially correct tolerant definition.
Note that the constant threshold ($\beta \eps$) for a vertex to join an existing cluster is slightly looser than that ($\eps$) for forming a new cluster, i.e., $\beta > 1$. 
This is to prevent edge cases where we are forced to form a new $\eps$-almost-clique upon each new insertion, which is prohibitive in time.

To avoid correctness issues due to vertex deletion, we perform a notification procedure (called ``identification'') mentioned in \Cref{subsec:tech-overview} for each newly formed cluster.
The cluster discovers all vertices that connect to at least $\alpha$-fraction of its vertices (later we set $\alpha=1/2$), and adds them to $\Vsparsealg$.
This step is necessary in the following scenario: 
\begin{enumerate}
    \item Cluster $K$ is formed and then blocked by some blocking set.
    \item Cluster $K'$ is formed and is densely connected to $K$. However, $K$ cannot join $K'$ since the vertices in $K$ are still sparse due to the blocking set.
    \item The blocking set is deleted, making $K$ dense with $K'$.
\end{enumerate}
Since the size of $K$ might be negligible compared to the scale of the blocking set, there is no efficient way of notification from the perspective of the blocking set, as mentioned in \Cref{subsec:tech-overview}.
Therefore, we need to inform $K$ when $K'$ is formed by our identification step, so $K$ is moved to $\Vsparsealg$ and captured by other algorithmic steps.
Note that we only need to remove a vertex if $\card{K'}\ge\frac{1}{2}\card{K}$, i.e., we do not want to remove a vertex from a massive cluster if it is identified by a $O(1)$-size cluster, as doing this would create massive false-negative edges whose costs cannot be charged.

Finally, for each cluster $K$ formed by the algorithm, we keep two counters $\inc(K)$ and $\del(K)$ to track the additions and removals in a cluster due to vertex insertions and deletions.
Whenever these counters reach $\eps\cdot c(K)$, where $c(K)$ is the size of $K$ when it is formed, we call \Cref{alg:refresh-ac} to refresh the cluster, keeping dense vertices, removing sparse vertices, and detecting vertices that is dense with other group of vertices.
To clarify, we use the words \textbf{insertion} and \textbf{deletion} to denote a data stream operation on the graph; and we use the words \textbf{addition} and \textbf{removal} to denote an algorithmic operation with respect to a cluster.
Here, $\inc(K)$ tracks the number of distinct vertices added to $K$ that are inserted to the graph \emph{after} $K$ is formed.
This allows us to charge the time easily to those data stream operations while still giving a correct cluster, since we can upper bound the number of added vertices that are inserted to the graph \emph{before} $K$ is formed by $O(\eps) \cdot c(K)$.
In addition, $\del(K)$ tracks the number of vertices deleted from the graph that has \emph{ever} joined $K$.
This step is essential for the correctness of our isolated clusters.
For these clusters, a large portion of its (previously) dense vertices are removed but we still keep the cluster since the cost is comparable to that of keeping all the remaining vertices as singleton points.
However, when the dense vertices outside $K$ are deleted, there are less edges between inside and outside of $K$, so at some point we cannot charge the cost and must dismantle $K$.
We cannot track these updates if we only track the deletion of vertices inside $K$.
To this end, \Cref{alg:vertex-dynamic-alg} ensures that the clusters maintained by our algorithm satisfy \Cref{def:partial-correct-new-sdd}.
We still need a sub-routine to handle the correctness of $\Vsparsealg$, which is deferred to \Cref{alg:deletion-clique-forming}.

\begin{Algorithm}
\label{alg:vertex-dynamic-alg}
\textbf{An algorithm that maintains an $O(1)$-approximation for correlation clustering anytime.}\\
\textbf{Input:} A graph $G=(V,E)$ with one vertex update per time. 
\smallskip

\begin{enumerate}
\item \textbf{Initialization} for each cluster $K$, we maintain the following counters: 
\begin{enumerate}
\item $ c(K)$: number of vertices inside $K$ when $K$ is formed
\item $\inc(K)$: number of distinct vertices added to $K$ that are inserted to the graph \emph{after} $K$ is formed
\item $\del(K)$: number of vertices deleted from the graph that \emph{ever} join $K$
\end{enumerate}

\item \textbf{Updates} for each $t\in[T]$:
\begin{enumerate}
    \item\label{line:insertion-formation} If $v$ is an \textbf{insertion}:
        \begin{enumerate}
    \item\label{line:v-sparsity-test-insert} Do sparsity test on $v$ (see \Cref{lem:sparse-test}) and check whether $v$ is $\eps/20$-sparse.
    \item If the test returns yes, add it to $\Vsparsealg$. Otherwise, do the following:
    \begin{enumerate}
        \item\label{line:merge:vertex} Run the algorithm in \Cref{lem:dense-vertex-clique-forming} and check whether $v$ could be merged to an existing cluster $K'$ with a $\beta\eps$-dense core. If there is such $K'$, add $v$ to $K'$ and increase $\inc(K')$ by $1$. 
        \item\label{line:call-clique-forming-alg} Otherwise, form a new $\eps$-almost-clique $K$ by \Cref{alg:new-ac}. 
    \end{enumerate}
\end{enumerate}
\item\label{line:deletion-dismantle} If $v$ is a \textbf{deletion}: 
\begin{enumerate}
\item If $v\in \Vsparsealg^{(t)}$, delete $v$ from $\Vsparsealg$ and $\Vsparsealgi^{(t)}$ for every $i$.
\item\label{line:deletion-dismantle-execution} If $v$ is in an cluster $K$, remove $v$ from $K$
\item Let $K_1, \ldots, K_L$ be the clusters maintained at $t$ that $v$ was in and increase $\del(K)$ by $1$ (see the detailed implementation in \Cref{lem:track-del-counter}).

\end{enumerate}


\item\label{line:vertex-dynamic-refresh-ac} 
\textbf{Maintaining clusters:}
\begin{enumerate}
    \item \label{line:refresh} \textbf{Addition-triggered refreshment:} For every cluster $K$, if at any point we have $\inc(K) \ge \eps \cdot c(K)$, 
     run \Cref{alg:refresh-ac} to refresh the cluster.
\item \label{line:dismantle} \textbf{Removal-triggered refreshment:} For every cluster $K$, if at any point we have $\del(K) \ge \eps \cdot c(K)$, run \Cref{alg:refresh-ac} to refresh the cluster.
\end{enumerate}

\item\label{line:run-the-new-algorithm} \textbf{Handling deletion-induced clique forming}: Run \Cref{alg:deletion-clique-forming} with $\zeta = \frac{\eps}{500}$.
\end{enumerate}
\end{enumerate}
\end{Algorithm}

\begin{Algorithm}
\label{alg:new-ac}
\textbf{An algorithm that forms a new cluster.}\\
\textbf{Input: graph $G=(V,E)$ updated by arbitrary insertions and deletions, a vertex $v$.}
\begin{enumerate}
    \item Run the algorithm in \Cref{lem:dense-vertex-clique-forming} to form an $\eps/2$-almost-clique $K$ according to \Cref{def:sdd}. 
    \item \label{line:identify} \textbf{Identification:} Run the algorithm of \Cref{lem:dense-vertex-clique-forming-generalize} with $\delta=\alpha$ and $\gamma=\alpha/2$ to identify a set $S \subset N(K)\setminus K$ of vertices, which includes all vertices $u$ satisfying $\card{N(u)\cap K}\geq \alpha\cdot \card{K}$. For each $u \in S$:
    \begin{itemize}
    \item If $u$ is in no cluster and not in $\Vsparsealg$ (e.g., it was removed by \Cref{alg:check:n:singleton}), add $u$ to $\Vsparsealg$ and $\Vsparsealgi$ for every $i$. If $u$ is in $\Vsparsealg$, add $u$ to $\Vsparsealgi$ for every $i$. 
    \item If $u$ is in some cluster $K'$ and $\card{K}\ge \frac{1}{2} c(K')$, i.e., $c(K')$ is the size of $K'$ when it is formed, remove $u$ from $K'$ and add $u$ to $\Vsparsealg$ and $\Vsparsealgi$ for every $i$.
    \end{itemize}
    \item Initiate the counters for $K$: $c(K) = \card{K}$, $\inc(K) = \del(K) = 0$.
\end{enumerate}
\end{Algorithm}

\begin{Algorithm}
\label{alg:refresh-ac}
\textbf{An algorithm that refreshes an cluster.}\\
\textbf{Input: graph $G=(V,E)$ updated by arbitrary insertions and deletions, an cluster $K$.}
\begin{enumerate}
\item Remove the old cluster before processing its remaining vertices.
\item\label{line:sparsity-test-dismantle} Do sparsity test on every $v \in K$ (see \Cref{lem:sparse-test}) and check whether $v$ is $\eps/20$-sparse. Repeat the following on a vertex whose test returns no until $K$ is empty or there is no such vertex. 
    \begin{enumerate}
        \item Run the algorithm in \Cref{lem:dense-vertex-clique-forming} and check whether $v$ could be merged to an existed cluster $K'$ with a $\beta\eps$-dense core. If there is such $K'$, add $v$ to $K'$. 
        \begin{itemize}
            \item If $v$ is inserted by the data stream after $K'$ is formed \textbf{and} $v$ did \emph{not} join $K'$ before, increase $\inc(K')$ by $1$. 
        \end{itemize}
        \item Otherwise, form a new $\eps$-almost-clique $\bar{K}$ by \Cref{alg:new-ac}. 
    \end{enumerate}
    \item For the remaining vertices in $K$, add them to $\Vsparsealg$ and $\Vsparsealgi$ for all $i\in [L]$.
\end{enumerate}
\end{Algorithm}

\begin{remark}
\label{rem:constant}
In the above algorithms, we use the following constant choices: $\beta = 4$, $\alpha = \frac{1}{2}$, and $\eps$ is a universal constant satisfying $\eps \le \frac{1}{19000}$.
\end{remark}

\begin{remark}
For cluster $K$ with initial size $c(K)$ smaller than $\frac{1}{\eps}$, 
due to the definition of SDD in \Cref{def:sdd}, $K$ is an exact clique, and moreover, we refresh it whenever there is an insertion or a deletion to $K$ (i.e. $\inc(K)$ or $\del(K)$ changes by 1) by \Cref{alg:refresh-ac}. Algorithmic removals do not trigger refreshment.
\end{remark}

\paragraph{Maintaining $\Vsparsealg$.}

For those sparse vertices added to $\Vsparsealg$ by \Cref{alg:vertex-dynamic-alg}, they could become dense later on, so we need \Cref{alg:deletion-clique-forming} to track these changes. 
A natural start point is to sample from $\Vsparsealg$ to estimate the number of dense vertices in $\Vsparsealg$ that belong to each degree level $[2^i, 2^{i+1})$ and form clusters on that degree level whenever it becomes ``dangerous'', i.e., dense vertices occupy a large portion of that degree level.
However, as we mentioned in \Cref{subsec:tech-overview}, there could be a vertex of degree $\Theta(n)$ that should join a cluster and $n^{0.9}$ vertices of degree $\Theta(1)$.
Although the cost is essentially higher for the high degree vertex so we should prioritize it, uniformly sampling from $\Vsparsealg$ cannot detect the vertex with high degree.
Therefore, we need to maintain a list for each degree level which we can directly scan through.
Unfortunately, there is \emph{no} known efficient method to track the degree bucketing \emph{exactly}, e.g., a trivial method requires to query the degree of all vertices in level $i$ for every $2^i$ updates, giving an $\frac{n}{2^{i}}$ amortized time in the worst case.

Instead of exactly tracking the degree, we keep an approximate degree bucketing list $\Vsparsealgi$ for each $i \in [L]$, where $L=1+\lceil\log_2n\rceil=O(\log n)$, storing the sparse vertices in the $i$-th degree level.
We abuse the notation $[L]$ to denote $\{0,\ldots, L-1\}$ in the following.
The main challenge is that the degree of a vertex can change drastically after it joins $\Vsparsealg$, so if we only add a sparse vertex to one list based on its degree when it enters, we may lose track of it later.
Thus, we add the vertex to all degree lists, but then the number of low-degree vertices could be significantly larger than that of the vertices with the correct degree in each list, causing efficiency issues.
Therefore, we design the sparse-list data structure to prioritize the high-degree vertices and prune the lists in a top-down manner.
We implement it as follows:
\begin{itemize}
    \item \textbf{Low degree}: We add all vertices to $\Vsparsealgi$ for each $i$ when they join $\Vsparsealg$.
    When we prune $\Vsparsealgi$, we simply remove all vertices with a low degree without further operation.
    Note that adding and removing low degree vertices in each level only incurs a $\log n$ blow-up to the amortized update time as there are $\log n$ levels.
    Moreover, this would not incur correctness issues due to degree increments. 
    For instance, a vertex $v$ in level $j$ may become dense and shift to level $i$ for $i \gg j$, but we deleted $v$ from $\Vsparsealgi$ earlier as it was a low-degree vertex in level $i$.
    In this case, we show that a large portion of the newly-added dense neighbors of $v$, i.e., those that cause the degree of $v$ to increase, is added to $\Vsparsealgi$, so we identify $v$ via its neighbors, and there is no need to add $v$ back to $\Vsparsealgi$.
    \item \textbf{High degree}:
    When our algorithm determines a level $i$ as dangerous, before pruning $\Vsparsealgi$, we first check the size of all higher degree-levels to ensure that there is no level $j>i$ with more sparse vertices, so that we afford the time to check all high degree vertices in $\Vsparsealgi$.
    Then, the high degree vertices are remained in $\Vsparsealgi$ after pruning as long as they cannot form or join any cluster.
    This top-down construction prioritizes high degree vertices and permits a full scan of $\Vsparsealgi$ when no higher degree-level blocks it.
\end{itemize}

We keep a counter in \Cref{alg:deletion-clique-forming} for each level that counts both graph updates by data stream and changes in membership of $\Vsparsealg$.
With the sparse-list data structure, we call \Cref{alg:estimate:size} every $\Theta(\eps 2^i)$ counts to estimate the size of $\VsparseIi$ and fully scan $\Vsparsealgi$ if level $i$ satisfies none of the following properties, for which we can tolerate the cost of level $i$.
\begin{itemize}
    \item \textbf{Non-dominant}: the number of dense vertices in level $i$ is significantly smaller than the number of sparse vertices in $\Vsparsealg$ (see Line~\ref{line:dom}).
    \item \textbf{Non-emergent}: the number of dense vertices in level $i$ is smaller than the number of sparse vertices in level $i$ up to a constant factor (see Line~\ref{line:emerg}).
    \item \textbf{Negligible:} the number of dense vertices in level $i$ is smaller than the number of sparse vertices in some higher degree level $j$ (see Line~\ref{line:negli}).
\end{itemize}
We emphasize that we estimate the size of the actual degree bucket $\Vsparsei$, i.e., vertices in $\Vsparsealg$ with degree $[2^i, 2^{i+1})$, to detect the dangerous level, since we can afford this time, but we cannot access the vertices in $\Vsparsei$, so we scan $\Vsparsealgi$ instead.
Once level $i$ passes all thresholds, we conduct the ``check and form cluster'' operation with $\Vsparsealgi$ by \Cref{alg:check-and-try-cluster}, which eliminates all sufficiently dense vertices from level $i$.
If a comparison fails, we still attempt the same scan, but stop at $2^iL^2$ visits of higher-degree vertices who does not depart $\Vsparsealg$ because it is sparse.
Thus we can still charge the time to the counts we wait until checking this level even when it does not pass the tests.
This step is essential since an ``unimportant'' level that does not pass the tests may become dominant, dangerous, and non-negligible before the next check is conducted.

We remark that the estimate of a level $i$ might not be accurate if the vertices inside keep bouncing between two levels, e.g., at some times their degrees are $2^i+1$ and at other times their degrees are $2^i-1$.
To handle these edge cases, we instead use the size of three degree levels as a characterization of level $i$.
For an integer $i$ and a positive odd number $a$, we define $S_i(a)=\{i-k, i-(k-1), \ldots, i, \ldots, i+(k-1), i+k, ~\mathrm{where}~ a=2k+1\}$.
We mainly use this notation in our algorithm to denote the $3$-level $S_i(3)=\{i-1,i,i+1\}$, i.e., we estimate the number of dense vertices in degree level $[2^{i-1},2^{i+2})$.

\begin{Algorithm}
\label{alg:deletion-clique-forming}
\textbf{An algorithm that triggers the construction of clusters.}\\
\textbf{Input: graph $G=(V,E)$ updated by arbitrary insertions and deletions and a parameter $\zeta =\eps/500$ (we use a different notation here for analysis purpose).}
\smallskip
\begin{enumerate}
    \item \textbf{Pre-processsing:} if there are less than $300\log^{4}{n}$ vertices in the graph
    \begin{enumerate}
        \item Run offline algorithm to obtain an $O(1)$-approximation at every time.
        \item Query all the degrees in the graph. 
    \end{enumerate}
    \item \textbf{Initialization} for each level $i\in[L]$, we maintain the following:
    \begin{enumerate}
        \item $\Vsparsealgi$: the list of candidate singleton vertices for level $i$.
        \item\label{line:level-counter-definition} $c_i$: number of graph insertions and deletions, plus actual entries to and departures from $\Vsparsealg$, since the most recent level-$i$ check started.
    \end{enumerate}
    \item\label{line:level-counter-increments} \textbf{Counter updates:} Increase every $c_i$ by one for each graph insertion or deletion and for each entry to or departure from $\Vsparsealg$, including changes during any subroutine.

    \item \textbf{For each time $t \in [T]$:}
    \begin{enumerate}
    \item If a vertex $u$ is inserted to $\Vsparsealg^{(t)}$, add $u$ to $\Vsparsealgi^{(t)}$ for each $i \in [L]$; and if $u$ is included in an cluster, remove it from $\Vsparsealg^{(t)}$ and $\Vsparsealgi^{(t)}$ for each $i \in [L]$.
    \item\label{line:check-and-run-almost-clique-alg} While $c_i\ge h_i:=\max\{1,\lfloor\zeta2^i/10\rfloor\}$ for some level $i$, choose such a level and \textbf{check} it as follows:
    \begin{enumerate}
    \item\label{line:level-counter-reset} Set $c_i\gets0$.
    Changes during this check count toward the next check.
    \item Run \Cref{alg:estimate:size} with $\gamma = O(2^i \log^2 n)$ samples and count samples in $\cup_{j\in S_i(3)}\Vsparsej^{(t)}$ for which the $5\zeta$-sparsity test returns no. Let $\bar{u}_i$ be the estimate.
    \item Initiate a dummy variable $\mathrm{FLAG}_i\gets 0$. 
    \item \label{line:dom} \textbf{Ignore non-dominant level:} If $\bar{u}_i< \frac{1}{100\cdot 2^iL}\cdot \card{\Vsparsealg^{(t)}}$, set $\mathrm{FLAG}_i\gets 1$.
    \item \label{line:emerg} \textbf{Ignore non-emergent level:}
    Run \Cref{alg:estimate:size} with $\gamma = O(2^i \log^2 n)$ samples and count samples in $\Vsparsei^{(t)}$ for which the $5\zeta$-sparsity test returns yes.
    Let $\bar{u}'_i$ be the estimate.
    If $\bar{u}_i < \frac{1}{5} \bar{u}'_i$, set $\mathrm{FLAG}_i\gets 1$. 
    \item \label{line:negli} \textbf{Ignore negligible level:} For each $j$ in ordered list $\{ L-1, L-2, \ldots, \max\{0, i-4\}\} \backslash \{i \}$
    \begin{itemize}
    \item Run \Cref{alg:estimate:size} with $\gamma = O(2^i \log^2 n)$ samples and count samples in $\Vsparsej^{(t)}$ for which the $5\zeta$-sparsity test returns yes.
    Let $\bar{u}_j$ be the estimate.
    If $\bar{u}_j > 100L \cdot \bar{u}_i$, terminate the loop and set $\mathrm{FLAG}_i\gets 1$.
    \end{itemize}
    \item \label{line:check:and:try:almost-clique} \textbf{Check and form cluster:} Run \Cref{alg:check-and-try-cluster} with $\Vsparsealgi$ and $\mathrm{FLAG}_i$.
    \end{enumerate}
    \item When every $c_i<h_i$, run \Cref{alg:check:n:singleton} before returning the clustering.
    \end{enumerate}
\end{enumerate}
\end{Algorithm}

\begingroup\tcbset{lines before break=8}
\begin{Algorithm}
\label{alg:check-and-try-cluster}
\textbf{Check and try cluster.}\\
\textbf{Input:} the list $\Vsparsealgi$ and a flag
$\mathrm{FLAG}_i\in\{0,1\}$, where $0$ specifies a full scan
and $1$ specifies a partial scan.
\smallskip
\begin{enumerate}
    \item Set $\mathrm{FAIL}_i\gets0$.

    \item Scan $\Vsparsealgi$ in list order, including entries
    appended during the scan.
    Each entry is visited at most once unless it is removed and created again.
    For each visited vertex $u$, do the following:
    \begin{enumerate}
        \item\label{line:deg:del}
        If $\deg(u)<2^i/10$, delete $u$ from $\Vsparsealgi$
        and continue to the next entry.

        \item Run the $12\zeta$ sparsity test of
        \Cref{lem:sparse-test} on $u$.
        \begin{enumerate}
            \item If the test returns yes, increase
            $\mathrm{FAIL}_i$ by one and leave $u$ in the list.

            \item If the test returns no:
            \begin{enumerate}
                \item Run the algorithm in
                \Cref{lem:dense-vertex-clique-forming} to check
                whether $u$ can be merged to an existing cluster
                $K'$ with a $\beta\eps$-dense core.
                If there is such $K'$, add $u$ to $K'$.
                In this case, if $u$ is inserted by the data stream
                after $K'$ is formed and $u$ did not join $K'$ before,
                increase $\inc(K')$ by $1$.

                \item\label{line:run-new-ac-alg-deletion-induced-forming}
                If no such $K'$ exists, form a new
                $\eps$-almost-clique $K$ by \Cref{alg:new-ac}.
            \end{enumerate}
        \end{enumerate}

        \item If $\mathrm{FLAG}_i=1$ and
        $\mathrm{FAIL}_i=2^iL^2$, stop and return.
    \end{enumerate}

    \item Return when no unvisited entry remains.
\end{enumerate}
\end{Algorithm}
\endgroup

\begin{Algorithm}
\label{alg:estimate:size}
\textbf{An algorithm that estimates the size of a subset $S$ of $\Vsparsealg^{(t)}$.}\\
\textbf{Input: graph $G=(V,E)$ updated by arbitrary insertions and deletions, number of samples $\gamma$, and a set $S$ in level $i$ to test size (given implicitly).}
\begin{enumerate}
    \item Sample $\gamma$ vertices independently and uniformly with replacement from $\Vsparsealg^{(t)}$ with a sufficiently large constant. 
    \item Let $\widehat{\gamma}$ be the number of samples belonging to $S$, and return $\widehat{s}^{(t)} = \frac{\card{\Vsparsealg^{(t)}}}{\gamma} \cdot \widehat{\gamma}$.
\end{enumerate}
\end{Algorithm}

\begingroup\tcbset{lines before break=8}
\begin{Algorithm}
\label{alg:check:n:singleton}
\textbf{An algorithm that removes degree-zero vertices from $\Vsparsealg^{(t)}$.}\\
\textbf{Input: graph $G=(V,E)$ updated by arbitrary insertions and deletions, and $\Vsparsealg^{(t)}$.}
\begin{enumerate}
    \item Run \Cref{alg:estimate:size} with $\gamma = O(\log^2 n)$ samples to estimate the fraction of degree-zero vertices in $\Vsparsealg$ with additive error $1/100$.
    \item If the estimated fraction is less than $1/20$, finish the cleanup.
    \item Otherwise:
    \begin{enumerate}
        \item Traverse $\Vsparsealg$ and remove its degree-zero vertices from the active set and all lists, retaining them as singleton clusters.
        \item Run the checks in Line~\ref{line:check-and-run-almost-clique-alg} of \Cref{alg:deletion-clique-forming} until every $c_i<h_i$.
    \end{enumerate}
\end{enumerate}
\end{Algorithm}
\endgroup

\begin{remark}
\label{rem:complete-cluster-operations}
We specify the following implementation orders:
\begin{itemize}
    \item Throughout the algorithms, each merge or formation, including identification and all resulting refreshments, is completed before the calling algorithm continues.
In particular, all cluster operations and required refreshments are completed before continuing or finishing a scan, testing the while condition in \Cref{alg:deletion-clique-forming}, or returning the clustering.
\item Each level check is completed before starting another, and all operations for the current graph insertion or deletion are completed before processing the next one.
\item Before any merge or formation, remove all clusters whose insertion or deletion counter has reached its threshold, and set aside their remaining vertices for \Cref{alg:refresh-ac}.
Whenever a vertex is added to a cluster, remove it from any set awaiting refreshment.
\end{itemize}

\end{remark}



\section{Correctness of Each Cluster}
\label{sec:cor:cluster}

In this section, we show that each cluster maintained by our algorithm satisfies \Cref{def:partial-correct-new-sdd}, as stated in the following theorem statement.

\begin{theorem}
\label{thm:cor:cluster}
At any time, let $K_1, \ldots, K_L$ be the clusters maintained by the algorithm, then for each $i \in [L]$, $K_i$ is either an $\eps_\tau$-tolerant cluster or an $\eps_l$-isolated cluster for universal constants $\eps_\tau$ and $\eps_l$.
\end{theorem}

\subsection{Proof Outline}

In this section, we outline the proof for the correctness of each cluster in \Cref{sec:cor:cluster} and the correctness of $\Vsparsealg$ in \Cref{sec:cor:vs}.
Essentially, we show that the clusters maintained by \Cref{alg:vertex-dynamic-alg} satisfies \Cref{def:partial-correct-new-sdd}, so our algorithm gives an $O(1)$-approximation for correlation clustering by \Cref{thm:approximation-partial-correct-sdd}.

We begin by introducing a group of auxiliary lemmas.
Our key technical lemma is \Cref{lem:core}, stating that for an $\eps$-dense core $K$ of size $|K|$, there could be at most $\frac{\eps |K|}{\alpha(1-\eps)}$ vertices that connect to at least $\alpha$-fraction of vertices in $K$.
This lemma is heavily used in the following analysis, e.g., we use it to bound the number of additions to a cluster as well as the number of identified vertices when a cluster is formed.
Next, consider a cluster $K$, which is formed at time $t_0$ and refreshed at time $t^*$.
Building on \Cref{lem:core}, we show the following lemmas that characterize the dynamic of $K$ after its formation:
\begin{itemize}
    \item \Cref{lem:num:alg:add}: there are at most $O(\eps)|K|$ additions to $K$ during $[t_0,t^*]$.
    \item \Cref{lem:dense:change}: if $K$ is still ``active'', i.e., vertices could be merged to its dense core, then its dense core has at most $O(\eps)|K|$ changes during $[t_0,t^*]$.
    \item \Cref{lem:not:refreshed}: the degree of a vertex added to $K$ can decrease by at most $O(\eps)\card{K^{(t_0)}}$ before $K$ is refreshed.
    \item \Cref{lem:two:clique}: if a vertex $v$ is added to $K_1$ and later to $K_2$, and both clusters are still maintained, then $\card{K_2^{(t_2)}}>\frac32\card{K_1^{(t_1)}}$, where $t_1,t_2$ are their formation times.
\end{itemize}

With these auxiliary lemmas, we are able to show the core \Cref{thm:Vsingleton}.
Informally, it ensures the validity of $\Vsparsealg$, stating that for any group of $\eps$-dense vertices $K$ with size roughly $2^i$, at least one of the following happens:
\begin{enumerate}
    \item There is a cluster $K'$ such that at least $(1-O(\eps))$-fraction of vertices in $K$ are in $K'$ and all vertices in $K \setminus K'$ are in both $\Vsparsealg$ and $\Vsparsealgi$.
    \item At least $O(\eps)$-fraction of $K$ is in both $\Vsparsealg$ and $\Vsparsealgi$.
\end{enumerate}
\Cref{thm:Vsingleton} ensures that when our algorithm determines to scan $\Vsparsealgi$ to form clusters, for each $\eps$-dense core $K$ with size roughly $2^i$, either it is mostly contained in an existing cluster $K'$ and the rest of $K$ are in $\Vsparsealgi$, so we merge all vertices to $K'$, or at least $O(\eps)$ fraction of $K$ is inside $\Vsparsealg$, so we form a new cluster by \Cref{alg:new-ac}.

We discuss the rough idea to prove \Cref{thm:Vsingleton}.
In the first case, since $K'$ contains a vertex in dense core $K$, \Cref{lem:dense:change,lem:not:refreshed} bound the initial size of $K'$ by $O(\card{K})$.
Then if a vertex outside $K'$ were in another cluster $K^*$, either identification when $K'$ is formed would remove it from $K^*$, or the two initial clusters $K'$ and $K^*$ would contain almost all of $K_v$, so they overlap a lot.
The latter contradicts \Cref{lem:two:clique}.
Thus all vertices of $K$ outside $K$ are in $\Vsparsealg$ and $\Vsparsealgi$.
In the second case, either the last $O(\eps)$-fraction of vertices of $K$ inserted into the graph are in $\Vsparsealg$ and $\Vsparsealgi$, or we consider the last addition of a vertex of $K$ to its current cluster $K^+$.
At that time, almost all of $K$ was already present, so $K^+$ contained almost all of $K$.
Since case (i) does not happen, $K^+$ contains less than $(1-O(\eps))$-fraction of $K$ now, so $O(\eps)|K|$ such vertices have been removed from $K^+$.
By the choice of the last addition, these vertices are singletons at the current time.
In both cases, their positive degrees remain large enough to keep them in $\Vsparsealgi$.

\paragraph{Correctness of each cluster.}

We show that after a cluster $K$ is formed, if it is not refreshed by our algorithm, then it is either a tolerant cluster or an isolated cluster.
By \Cref{lem:num:alg:add}, there are at most $O(\eps)|K|$ additions to $K$ during $[t_0,t^*]$.
Thus, the fraction of non-neighbors inside $K$ mainly depends on the number of vertices that were once in $K$ during $[t_0,t^*]$, outside $K$ now, and still present in the graph.
They are removed from $K$ mainly due to algorithmic operations.
If this quantity is small, then the cluster remains a tolerant cluster because there can be at most $O(\eps) \card{K}$ non-neighbors for each vertex inside.
On the contrary, if there are more than $\Theta(\card{K})$ such vertices, the size of $K$ decreases by a lot, so the ratio of non-neighbors inside could become larger for each vertex, making $K$ no longer a tolerant cluster.
In this case, many vertices are removed from $K$ by our algorithm, but most of them cannot be deleted by the data stream, otherwise it will trigger the refreshment of $K$.
Thus, there are many edges between the inside and the outside of $K$, making $K$ satisfies the requirement of an isolated cluster.
Recall that in \Cref{def:partial-correct-new-sdd}, we also have some requirements for $v \in K$ if it is instead dense with some other vertices.
In particular, if a vertex $v$ in $K$ is dense with a group of vertices $K_v$ outside $K$, then at least $\eps$-fraction of these vertices should be in $\Vsparsealg$ so that we can charge the cost of keeping $v$ in $K$.
Fortunately, this is guaranteed by the validity of $\Vsparsealg$ in \Cref{thm:Vsingleton}.

\paragraph{Correctness of $\Vsparsealg$.}
For the correctness of $\Vsparsealg$, whenever a level becomes ``dangerous'' in the sense that we cannot charge the cost of not forming clusters on that level, either the size comparisons in \Cref{alg:deletion-clique-forming} triggered a full scan at its last check, or the partial scan at that check must have eliminated dense vertices. 
Indeed, in the latter case, the sparse vertices present at the check, together with subsequent entries into $\Vsparsealg$, account for too few unsuccessful sparsity tests to exhaust the scan's budget $2^i \polylog(n)$, so we traverse through all dense vertices.
Therefore, in either case we scan and prune the list $\Vsparsealgi$ and, again by \Cref{thm:Vsingleton}, we eliminate all dense vertices from $\Vsparsealg$ in level $i$, so that it is no longer dangerous.

\begin{figure}[htbp]
\centering
\begin{tikzpicture}[
    box/.style={draw=black!80, rectangle, rounded corners, align=center, minimum width=2.8cm, minimum height=0.8cm, fill=blue!5, thick},
    groupbox/.style={draw=black!40, dashed, rounded corners, fill=gray!5, inner sep=12pt},
    arrow/.style={-{Stealth[scale=1.2]}, thick, draw=black!70}
]

\node[box] (L65) at (0, 0) {\Cref{lem:num:alg:add}};
\node[box] (L66) at (0, -1.8) {\Cref{lem:dense:change}};
\node[box] (L67) at (0, -3.6) {\Cref{lem:not:refreshed}};
\node[box] (L68) at (0, -5.4) {\Cref{lem:two:clique}};

\begin{scope}[on background layer]
    \node[groupbox, fit=(L65) (L68), label=below:{\small Cluster Dynamic}] (group) {};
\end{scope}

\node[box] (L61) at (0, 1.8) {\Cref{lem:core} \\ Core Lemma};

\node[box] (CorrCluster) at (4.8, -0.9) {Correctness of \\ each cluster};
\node[box] (Thm4) at (4.8, -4.2) {\Cref{thm:Vsingleton} \\ Validity of $\Vsparsealg$};

\node[box] (Tolerant) at (11, 0.6) {$O(\eps)$-Tolerant};
\node[box] (Isolate) at (11, -2.4) {$O(\eps)$-Isolated};
\node[box] (CorrectVs) at (11, -4.2) {Correctness \\ of $\Vsparsealg$};


\draw[arrow] (L61.south) -- (group.north);

\draw[arrow] (L65.east) -- (CorrCluster.165);
\draw[arrow] (L66.east) -- (CorrCluster.195);

\draw[arrow] (L66.east) -- (Thm4.150);
\draw[arrow] (L67.east) -- (Thm4.180);
\draw[arrow] (L68.east) -- (Thm4.210);

\draw[arrow] (Thm4.north) -- (CorrCluster.south);

\draw[arrow] (Thm4.east) -- (CorrectVs.west);

\draw[arrow] (CorrCluster.east) -- node[above, sloped, align=center, font=\footnotesize, yshift=2pt] {$\le O(\epsilon)$ former \\ members outside $K$} (Tolerant.west);
\draw[arrow] (CorrCluster.east) -- node[below, sloped, align=center, font=\footnotesize, yshift=-2pt] {$> O(\epsilon)$ former \\ members outside $K$} (Isolate.west);

\end{tikzpicture}
\caption{Control flow for the proof outline.}
\label{fig:control_flow}
\end{figure}
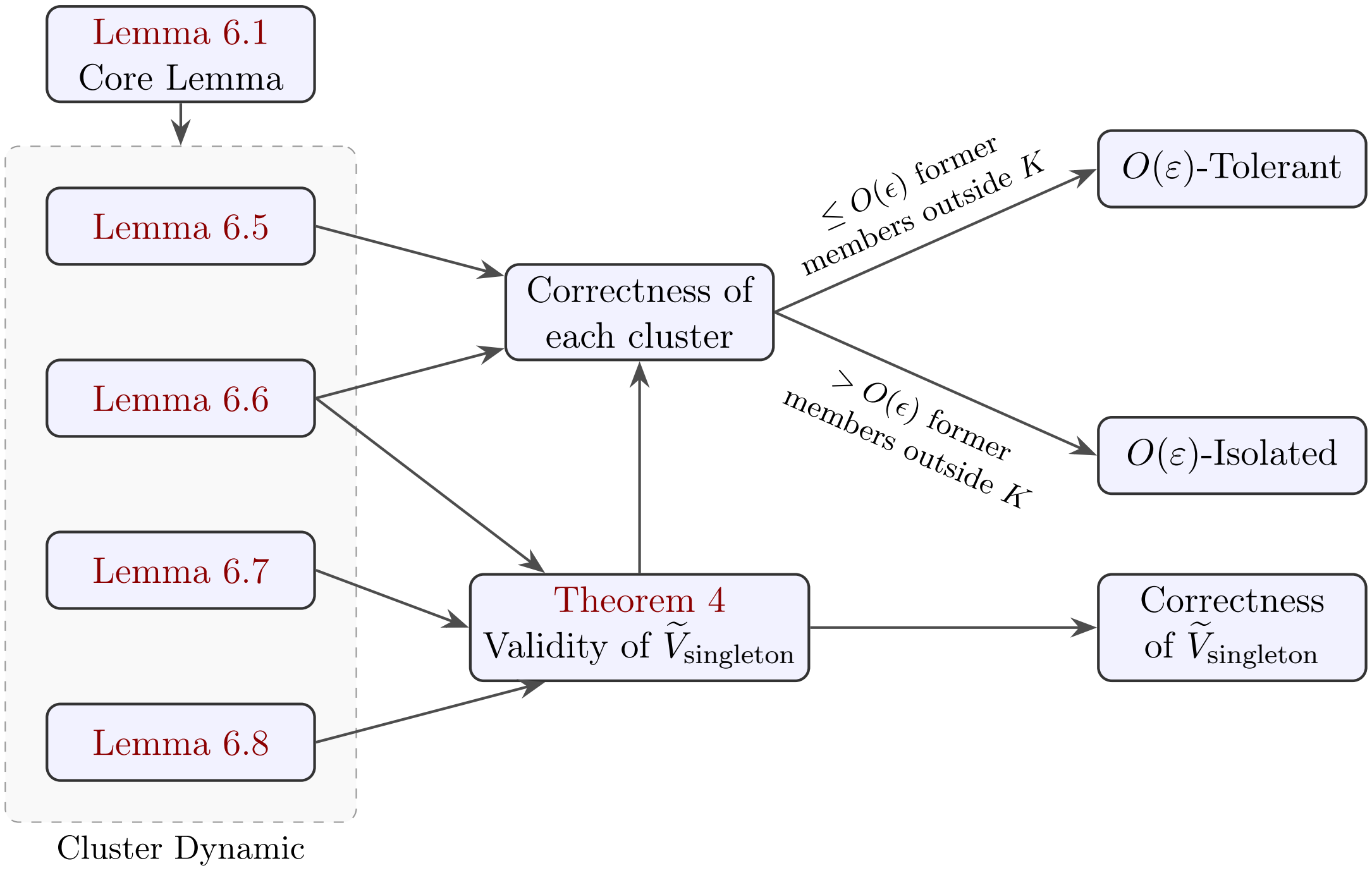

We provide the control flow of the above outline in \Cref{fig:control_flow}.

\subsection{Preliminaries}
In this section, we introduce definitions and terminologies used in the proof.
Recall that we define an $\eps$-dense core to be an $\eps$-almost-clique contained by a cluster $K$, we use $C^{(t)}$ to denote the dense core of $K$ at time $t$ if it exists, and we use $N^{(t)}(v)$ to denote the neighborhood of vertex $v$ at time $t$.
We write $N^{(t)}(u)$ for the positive neighborhood of $u$ at time $t$, and $N^{(t)}[u]:=N^{(t)}(u)\cup\{u\}$ for its closed positive neighborhood.
We often use $t_{v}^+$ to denote the time when a vertex $v$ is added to a specific cluster $K$ given by the context.
We remark that any cluster formed during refreshment is regarded as a new cluster, and an empty cluster is discarded.
We next state a core technical lemma which states that the number of densely-connected neighbors of an $\eps$-dense core must be bounded, which is used later to bound the number of additions to a cluster.

\begin{lemma}
\label{lem:core}
We use $E(v,S) = \{(v,s)\in E^+: s \in S\}$ to denote the set of edges between a vertex $v$ and a set of vertices $S$. 
Let $S \subset V$ be an $\eps$-dense core and let $H_\alpha =\{v \in V \setminus S: \card{E(v,S)} \ge \alpha \card{S} \}$, then we have $\card{H_\alpha} \le \frac{\eps}{\alpha(1-\eps)} \cdot \card{S}$.
\end{lemma}
\begin{proof}
Recall that in \Cref{def:sdd}, each vertex in the dense core $S$ has at most $\eps \card{S}/(1-\eps)$ neighbors outside $S$, so $S$ has at most $\eps \card{S}^2/(1-\eps)$ out-edges.
On the other hand, for each vertex in $H_\alpha$, it is outside of $S$ and connects to at least $\alpha \card{S}$ vertices in $S$.
Thus, there are at least $\card{H_\alpha} \cdot \alpha \card{S}$ edges between $H_\alpha$ and $S$.
Then we have $\card{H_\alpha} \cdot \alpha \card{S} \le \eps \card{S}^2/(1-\eps)$, giving us $\card{H_\alpha} \le \frac{\eps}{\alpha(1-\eps)} \cdot \card{S}$.
\end{proof}

To begin with, we have the following observations of our algorithm.
\begin{observation}
\obslab{obs:cor:refresh}
When a cluster $K$ is formed by \Cref{alg:new-ac}, $K$ is an $\eps$-dense core as defined in \Cref{def:dense:core}.
\end{observation}

\begin{observation}
\obslab{obs:v:insert}
Whenever a vertex $v$ is added to a cluster $K$, $K$ has an $\beta\eps$-dense core, and moreover, $v$ belongs to the dense core.
\end{observation}

Next, we fix a time $t$ and consider an arbitrary cluster $K$ maintained by our algorithm at $t$.
We aim to show that $K$ satisfies the definition of a tolerant cluster or an isolated cluster in \Cref{def:partial-correct-new-sdd} at $t$.
Let $t_0$ be the time that $K$ is formed by \Cref{alg:new-ac} and let $K^{(t_0)}$ be the set of vertices that is in $K$ at $t_0$, then by \obsref{obs:cor:refresh}, we have that $K^{(t_0)}$ is an $\eps$-dense core, i.e., $C^{(t_0)} = K^{(t_0)}$.
The goal is then to show that $K$ satisfies \Cref{def:partial-correct-new-sdd} in the following sequence of dynamic inputs.
We now define four types of operations on $K$ after it is formed.
\begin{definition}
We define the following operations on a cluster: 
\begin{itemize}
    \item We use insertion/deletion to denote a data stream update on the graph; we use addition/removal to denote an algorithmic operation on a vertex with respect to a cluster.
    \item \textbf{Data stream removal:} Vertex $v$ is deleted by the data stream, where $v$ has been added to $K$ at some point at or after $t_0$ ($v$ may or may not be in $K$ when it is deleted from the graph).
    Data stream removals are counted by $\del(K)$ in \Cref{alg:vertex-dynamic-alg}.
    \item \textbf{Algorithmic removal:} Vertex $v$ is removed from cluster $K$ by our algorithm, but $v$ still exists in the graph. 
    \item \textbf{Data stream addition:} Vertex $v$ is added to cluster $K$, where $v$ is inserted by the data stream after $t_0$.
    Data stream additions are counted by $\inc(K)$ in \Cref{alg:vertex-dynamic-alg}.
    \item \textbf{Algorithmic addition:} Vertex $v$ is added to cluster $K$, where $v$ was in the graph at $t_0$.
\end{itemize}
\end{definition}

We first state the following observation, saying that there are at most $\eps \card{K^{(t_0)}}$ data stream operations before $K$ is refreshed.
The proof of the observation is straightforward as we keep two counters $\inc(K)$ and $\del(K)$ in \Cref{alg:vertex-dynamic-alg} for these two kinds of operation, respectively; and when $\inc(K) \ge \eps \card{K^{(t_0)}}$ or $\del(K) \ge \eps \card{K^{(t_0)}}$, we refresh $K$ (see Line~\ref{line:refresh} and Line~\ref{line:dismantle}).
\begin{observation}
\obslab{obs:num:operations}
\label{obs:num:adv}
There are at most $\eps \card{K^{(t_0)}}$ data stream removals and data stream additions to $K$ before $K$ is refreshed.
\end{observation}



\subsection{Core Technical Lemmas}
We write $N^{(t)}(u)$ for the positive neighborhood of $u$ at time $t$, and $N^{(t)}[u]:=N^{(t)}(u)\cup\{u\}$ for its closed positive neighborhood.
Throughout, $\beta=4$, $\alpha=1/2$, and $\eps\le1/19000$, as in \Cref{rem:constant}.

We first bound the number of vertices added to a cluster before it is refreshed.
Intuitively, every vertex added algorithmically already exists when the cluster is formed and must have many positive neighbors in the initial cluster. \Cref{lem:core} then bounds the number of such vertices.

\begin{lemma}
\label{lem:num:alg:add}
Suppose that $K$ is formed at $t_0$ and is not refreshed before $t$.
At most $2\eps\card{K^{(t_0)}}$ distinct vertices outside $K^{(t_0)}$ are added to $K$ by algorithmic additions, and at most $3\eps\card{K^{(t_0)}}$ such vertices are added in total.
\end{lemma}
\begin{proof}
By \obsref{obs:cor:refresh}, $K^{(t_0)}$ is an $\eps$-dense core.
By \Cref{lem:core}, fewer than $2\eps\card{K^{(t_0)}}$ vertices outside $K^{(t_0)}$ have at least $2\card{K^{(t_0)}}/3$ positive neighbors in it at $t_0$.
Thus it suffices to show that every algorithmic addition is one of these vertices: \obsref{obs:num:operations} allows at most $\eps\card{K^{(t_0)}}$ further vertices from data stream additions.

First suppose that $\card{K^{(t_0)}}\ge(1-\eps)/\eps$.
We prove the stronger auxiliary fact that every surviving vertex that has belonged to $K$ retains at least $(1-12\eps)\card{K^{(t_0)}}$ positive neighbors in $K^{(t_0)}$ until refreshment.
For vertices of $K^{(t_0)}$, this follows directly from \Cref{def:sdd} and \obsref{obs:num:operations} (the missing self-neighbor is absorbed using $1\le\eps\card{K^{(t_0)}}/(1-\eps)$).

We consider the distinct vertices outside $K^{(t_0)}$ in the order in which each is first added to $K$, including both algorithmic and data stream additions.
A vertex removed from $K$ and later added back is not counted again.
We prove the above claim by induction in this order.
Suppose the bound holds for all vertices preceding the next vertex $v$.
Then every earlier vertex added algorithmically belongs to the set counted above, so there are at most $2\eps\card{K^{(t_0)}}$ such vertices.
Together with the at most $\eps\card{K^{(t_0)}}$ vertices added by data stream additions, this gives at most $3\eps\card{K^{(t_0)}}$ distinct outside vertices added before $v$.
Immediately after adding $v$, their number is at most $3\eps\card{K^{(t_0)}}+1\le5\eps\card{K^{(t_0)}}$.

Let $C$ be the dense core after this addition, we show $C\cap K^{(t_0)}\neq \emptyset$.
By \obsref{obs:v:insert} and \Cref{lem:dense-vertex-clique-forming}, $C$ contains $v$ and more than half of the resulting cluster.
The cluster was nonempty before the addition, so $C$ contains another vertex already in $K$.
Moreover, $\card C\le(1+5\eps)\card{K^{(t_0)}}$.
By \Cref{def:dense:core}, each vertex of $C$ has at most
\[
 \frac{\beta\eps}{1-\beta\eps}\card C=\frac{4\eps}{1-4\eps}\card C
 \le\frac92\eps\card{K^{(t_0)}}
\]
positive neighbors outside $C$, and at most the same number of non-neighbors inside $C$.
If $C$ contained no vertex of $K^{(t_0)}$, that other vertex would have at least $(1-12\eps)\card{K^{(t_0)}}$ positive neighbors outside $C$ by induction, which contradicts with it is in $C$.
Then choose an arbitrary $w\in C\cap K^{(t_0)}$.

Consider any later time $\tau$ when $v$ survives and $K$ has not been refreshed.
A vertex of $K^{(t_0)}$ missing from $N^{(\tau)}(v)$ belongs to at least one of four sets:
\begin{enumerate}[label=(\roman*)]
    \item the initial non-neighbors of $w$, excluding $w$ itself: at most $2\eps\card{K^{(t_0)}}$ by \Cref{def:sdd};
    \item the positive neighbors of $w$ outside $C$ when $v$ was added: at most $\frac92\eps\card{K^{(t_0)}}$;
    \item the non-neighbors of $v$ inside $C$ when $v$ was added: at most $\frac92\eps\card{K^{(t_0)}}$;
    \item the initial vertices deleted by time $\tau$: at most $\eps\card{K^{(t_0)}}$ by \obsref{obs:num:operations}.
\end{enumerate}
Consequently,
\[
 \card{N^{(\tau)}(v)\cap K^{(t_0)}}
 \ge\left(1-\left(2+\frac92+\frac92+1\right)\eps\right)\card{K^{(t_0)}}
 =(1-12\eps)\card{K^{(t_0)}}.
\]
For an algorithmic addition, $v$ already existed at $t_0$, so these surviving positive edges also existed then.
Thus $v$ had more than $2\card{K^{(t_0)}}/3$ initial positive neighbors, proving the induction step and all relevant size bounds.
Repeated additions do not change the number of distinct vertices counted.

Finally, suppose that $\card{K^{(t_0)}}<(1-\eps)/\eps$.
By \Cref{def:sdd} and integrality, $K^{(t_0)}$ is a clique with no positive edges leaving it at $t_0$.
Since $\eps\card{K^{(t_0)}}<1$, \obsref{obs:num:operations} permits no data stream addition or removal before refreshment.
A first algorithmic addition from outside $K^{(t_0)}$ would have no positive edge to any other vertex of $K$.
Its dense core would contain it and another vertex, giving it at least half the core as non-neighbors, contrary to \Cref{def:dense:core}.
Hence no outside vertex is added and no initial vertex is deleted before refreshment.
In this case, every vertex that has belonged to $K$ has a closed neighborhood containing $K^{(t_0)}$.
\end{proof}

We next show that a dense core must retain almost all of $K^{(t_0)}$.
This is because a vertex shared by the dense core and $K^{(t_0)}$ still has many initial positive neighbors by \Cref{lem:num:alg:add}, and only a small number of them can lie outside the core.
This also means that if the cluster is ``active'' in the sense that it admits new additions, its dense core cannot change too much.

\begin{lemma}
\label{lem:dense:change}
Suppose that $K$ is formed at $t_0$, is not refreshed before $t$, and has a $\beta\eps$-dense core at $t$.
For every $\tau\in[t_0,t]$,
\[
 \card{K^{(t_0)}\setminus K^{(\tau)}}\le7\eps\card{K^{(t_0)}},
 \qquad
 \card{K^{(\tau)}\setminus K^{(t_0)}}\le3\eps\card{K^{(t_0)}}.
\]
Whenever a $\beta\eps$-dense core $C^{(\tau)}$ exists, these bounds also hold with $C^{(\tau)}$ in place of $K^{(\tau)}$, and $\card{K^{(\tau)}\setminus C^{(\tau)}}\le10\eps\card{K^{(t_0)}}$.
\end{lemma}
\begin{proof}
By \Cref{lem:num:alg:add}, at most $3\eps\card{K^{(t_0)}}$ vertices outside $K^{(t_0)}$ have been added to $K$.
This proves the second bound and gives $\card{K^{(\tau)}}\le(1+3\eps)\card{K^{(t_0)}}$ at every time $\tau$.

First consider a time $\tau$ when a dense core $C^{(\tau)}$ exists.
This core contains an initial vertex: otherwise, it has at most $3\eps\card{K^{(t_0)}}$ vertices, whereas the auxiliary bound in the proof of \Cref{lem:num:alg:add} gives every vertex of the core at least $(1-12\eps)\card{K^{(t_0)}}$ positive neighbors outside it, contradicting \Cref{def:dense:core}.
Choose $u\in C^{(\tau)}\cap K^{(t_0)}$.
Every vertex of $K^{(t_0)}\setminus C^{(\tau)}$  is an initial
non-neighbor of $u$, a deleted vertex, or a positive neighbor of $u$ outside $C^{(\tau)}$.
These three sets have size at most $\frac32\eps\card{K^{(t_0)}}$, $\eps\card{K^{(t_0)}}$, and $\frac92\eps\card{K^{(t_0)}}$, respectively, by \Cref{def:sdd}, \obsref{obs:num:operations}, and \Cref{def:dense:core}.
Here we use $\eps/(1-\eps)\le3\eps/2$ and $4\eps(1+3\eps)/(1-4\eps)\le9\eps/2$.
Consequently,
\[
 \card{K^{(t_0)}\setminus C^{(\tau)}}
 \le\left(\frac32+1+\frac92\right)\eps\card{K^{(t_0)}}
 =7\eps\card{K^{(t_0)}}.
\]
Together with the addition bound, this proves all assertions about $C^{(\tau)}$, including $\card{K^{(\tau)}\setminus C^{(\tau)}}\le10\eps\card{K^{(t_0)}}$.

Finally, suppose that more than $7\eps\card{K^{(t_0)}}$ initial vertices are outside $K$ at some earlier time.
If there is no later addition, they remain outside through $t$, and there could not be a dense core at $t$, which is a contradiction.
Otherwise, immediately before the next addition, the same vertices are still outside $K$.
The merge guarantee in \Cref{lem:dense-vertex-clique-forming} requires a dense core before this addition, contradicting the bound just proved.
Thus the first bound also holds at times when no dense core exists.
\end{proof}

The next lemma states that the degree of a dense vertex cannot decrease too much before its cluster is refreshed.
A vertex in a dense core has few positive neighbors outside it.
Thus, a large decrease in its degree must delete many vertices that have belonged to the cluster, triggering refreshment even if they were removed from the cluster earlier.

\begin{lemma}
\label{lem:not:refreshed}
Suppose that $v$ belongs to a $\lambda\eps$-dense core of $K$ at time $t_1\ge t_0$, where $\lambda\in\{1,\beta\}$.
If $v$ has not been deleted and $K$ is not refreshed before $t\ge t_1$, then
\[
 \degtu{t}{v}\ge\degtu{t_1}{v}-(\lambda+2)\eps\card{K^{(t_0)}}.
\]
\end{lemma}
\begin{proof}
Let $C^{(t_1)}$ be the dense core containing $v$ at time $t_1$.
By \Cref{def:dense:core,lem:num:alg:add}, the number of positive neighbors of $v$ outside $C^{(t_1)}$ at that time is at most
\[
 \frac{\lambda\eps}{1-\lambda\eps}\card{C^{(t_1)}}
 \le\frac{\lambda\eps(1+3\eps)}{1-\lambda\eps}\card{K^{(t_0)}}
 \le(\lambda+1)\eps\card{K^{(t_0)}}.
\]
Suppose the claimed degree bound fails.
Then more than $(\lambda+2)\eps\card{K^{(t_0)}}$ positive neighbors of $v$ at time $t_1$ have been deleted by time $t$.
At most $(\lambda+1)\eps\card{K^{(t_0)}}$ of them lie outside $C^{(t_1)}$, so more than $\eps\card{K^{(t_0)}}$ lie inside it.
All these vertices have belonged to $K$, and their deletions are counted by $\del(K)$ even if they were removed from $K$ earlier.
This contradicts \obsref{obs:num:operations}, since $K$ has not been refreshed.
\end{proof}

Finally, we show that if a vertex is added to two clusters consecutively, then the second one must have a substantially larger size.
Since if not, a vertex is added to two clusters of comparable initial size, then the two initial clusters must overlap substantially: the vertex has many neighbors in the first and few neighbors outside the dense core of the second.
Identification when the more recently formed cluster is formed would then remove every remaining vertex from the earlier cluster, so both clusters cannot remain maintained.

\begin{lemma}
\label{lem:two:clique}
Suppose that $v$ is added to $K_1$ and later to a distinct cluster $K_2$, formed at times $t_1,t_2$, respectively.
If both clusters are still maintained after all cluster operations and refreshments are complete, then
\[
 \card{K_2^{(t_2)}}>\frac32\card{K_1^{(t_1)}}.
\]
\end{lemma}
\begin{proof}
Suppose for contradiction that $\card{K_2^{(t_2)}}\le3\card{K_1^{(t_1)}}/2$.

First we show that the initial clusters overlap substantially.
Let $\tau$ be the time when $v$ is added to $K_2$, and let $C$ be its dense core immediately afterward.
The neighborhood bound in the proof of \Cref{lem:num:alg:add} gives
$\card{N^{(\tau)}[v]\cap K_1^{(t_1)}}\ge(1-12\eps)\card{K_1^{(t_1)}}$.
Of these vertices, those outside $K_2^{(t_2)}$ either lie outside $C$ or belong to $C\setminus K_2^{(t_2)}$.
By \Cref{def:dense:core,lem:num:alg:add}, the two sets have size at most $4\eps(1+3\eps)\card{K_2^{(t_2)}}/(1-4\eps)$ and $3\eps\card{K_2^{(t_2)}}$, respectively.
Thus
\begin{align*}
 \card{K_1^{(t_1)}\cap K_2^{(t_2)}}
 &\ge(1-12\eps)\card{K_1^{(t_1)}}
 -\left(\frac{4\eps(1+3\eps)}{1-4\eps}+3\eps\right)\card{K_2^{(t_2)}}\\
 &>(1-23\eps)\card{K_1^{(t_1)}}.
\end{align*}

We next show that identification removes the earlier cluster.
Let $K,K'$ be the earlier and later clusters, formed at $t_0,t'_0$, respectively.
Every vertex in the displayed intersection is present when $K'$ is formed.
Consider any vertex $u$ remaining in $K$ after the vertices of $K'$ have been added.
Since $u\notin(K')^{(t'_0)}$, the neighborhood bound in the proof of \Cref{lem:num:alg:add}, applied at this time, gives
\begin{align*}
 \card{N(u)\cap(K')^{(t'_0)}}
 &\ge\card{K_1^{(t_1)}\cap K_2^{(t_2)}}-12\eps\card{K^{(t_0)}}\\
 &>(1-41\eps)\card{K_1^{(t_1)}}
 >\frac12\card{(K')^{(t'_0)}}.
\end{align*}
Here both initial cluster sizes are at most $3\card{K_1^{(t_1)}}/2$.
Moreover,
\[
 \card{(K')^{(t'_0)}}>(1-23\eps)\card{K_1^{(t_1)}}
 >\frac12\card{K^{(t_0)}}.
\]
Thus $u$ is identified and passes the size comparison in Line~\ref{line:identify} of \Cref{alg:new-ac}.
Every remaining vertex of $K$ is consequently removed.
The empty cluster is discarded, contradicting that both clusters are still maintained.
\end{proof}

\subsection{Validity of \texorpdfstring{$\Vsparsealg$}{Vs} and \texorpdfstring{$\Vsparsealgi$}{Vsi}}
We now present our core theorem on the validity of $\Vsparsealg$ and $\Vsparsealgi$ in \Cref{thm:Vsingleton}.

\begin{theorem}
\label{thm:Vsingleton}
Fix a time $t$ after all cluster operations and refreshments are complete.
For any $\eps$-dense vertex $v$, let $K_v$ be the set of vertices in $N^{(t)}[v]$ that are in $\eps$-agreement with $v$, and suppose that $\card{K_v}\in[2^i,2^{i+1})$.
At least one of the following happens:
\begin{enumerate}
    \item There exists a cluster $K$ at $t$ such that at least a $(1-91\eps)$-fraction of $K_v$ is in $K^{(t)}$, and all vertices in $K_v\setminus K^{(t)}$ are in both $\Vsparsealg^{(t)}$ and $\Vsparsealgj^{(t)}$ for every $j\in S_i(5)$.
    \item At least a $40\eps$-fraction of $K_v$ is in both $\Vsparsealg^{(t)}$ and $\Vsparsealgj^{(t)}$ for every $j\in S_i(5)$.
\end{enumerate}
\end{theorem}
\begin{proof}
    Note that $\card{K_v}\ge2$.
All numerical inequalities below use $\beta=4$ and $\eps\le1/19000$, as in \Cref{rem:constant}.
Density and agreement (\Cref{def:sparse-dense-vertices,lem:vertex-pair-agreement}) give, for every $u\in K_v$,
\[
\begin{aligned}
 \deg^{(t)}(u)+1
 &\le\frac{\card{K_v}}{(1-\eps)^2}
 \le1.02\card{K_v},\\
 \card{K_v\setminus N^{(t)}[u]}
 &\le\frac{\eps}{(1-\eps)^2}\card{K_v}
 \le1.02\eps\card{K_v}.
\end{aligned}
\]
Consider any maintained cluster $K'$ formed at $t'_0$ and a vertex $w\in(K')^{(t)}\cap K_v$.
By the neighborhood bound in the proof of \Cref{lem:num:alg:add},
\[
 (1-12\eps)\card{(K')^{(t'_0)}}
 \le\deg^{(t)}(w)+1
 \le1.02\card{K_v},
\]
and hence
\[
 \card{(K')^{(t'_0)}}
 \le\frac{1.02}{1-12\eps}\card{K_v}
 \le1.11\card{K_v}.
\]
Whenever $u\in K_v$ is added to such a cluster, its dense core contains $u$ and has at most $(1+3\eps)\card{(K')^{(t'_0)}}$ vertices, by \obsref{obs:v:insert} and \Cref{lem:num:alg:add}.
Therefore, by \Cref{def:dense:core}, $u$ has at most
\[
 \frac{4\eps}{1-4\eps}(1+3\eps)\card{(K')^{(t'_0)}}
 \le\frac{4\cdot1.11(1+3\eps)}{1-4\eps}\eps\card{K_v}
 \le4.60\eps\card{K_v}
\]
positive neighbors outside that core.
If a subset $A\subseteq K_v$ is already present at this addition time $s\le t$, the preceding non-neighbor bound and \Cref{lem:num:alg:add} imply
\[
\begin{aligned}
 \card{(K')^{(s)}\cap K_v}
 &\ge\card A-(1.02+4.60)\eps\card{K_v}\\
 &=\card A-5.62\eps\card{K_v},\\
 \card{(K')^{(t'_0)}\cap K_v}
 &\ge\card A-(5.62+3\cdot1.11)\eps\card{K_v}\\
 &=\card A-8.95\eps\card{K_v}.
\end{aligned}
\]

\textbf{Proof of statement (1).}
Suppose that $\card{K^{(t)}\cap K_v}\ge(1-91\eps)\card{K_v}$, and let $A=K^{(t_0)}\cap K_v$.
By \Cref{lem:num:alg:add} and the size bound above,
\[
\begin{aligned}
 \card A
 &\ge\bigl(1-(91+3\cdot1.11)\eps\bigr)\card{K_v}\\
 &=(1-94.33\eps)\card{K_v}
 >3\card{K_v}/4.
\end{aligned}
\]
Suppose that $u\in K_v\setminus K^{(t)}$ is in another cluster $K'$ at $t$, and let $s$ be the time of its last addition to $K'$.
If $s<t_0$, then $u$ remains in $K'$ from $s$ through $t$, and hence $u\notin K^{(t_0)}$.
It has at least
\[
\begin{aligned}
 \card A-1.02\eps\card{K_v}
 &\ge\bigl(1-(94.33+1.02)\eps\bigr)\card{K_v}\\
 &=(1-95.35\eps)\card{K_v}\\
 &>\frac{1.11}{2}\card{K_v}
 \ge\card{K^{(t_0)}}/2
\end{aligned}
\]
positive neighbors in $K^{(t_0)}$, and
\[
 \card{K^{(t_0)}}
 >\frac34\card{K_v}
 >\frac{1.11}{2}\card{K_v}
 \ge\card{(K')^{(t'_0)}}/2.
\]
Identification in Line~\ref{line:identify} of \Cref{alg:new-ac} therefore removes $u$, contradicting the choice of $s$.
If $s>t_0$, all vertices of $A$ are present at $s$, so
\[
\begin{aligned}
 \card{(K')^{(t'_0)}\cap K_v}
 &\ge\card A-8.95\eps\card{K_v}\\
 &\ge\bigl(1-(94.33+8.95)\eps\bigr)\card{K_v}\\
 &=(1-103.28\eps)\card{K_v}
 >3\card{K_v}/4.
\end{aligned}
\]
The two initial clusters share a vertex of $K_v$ and both have sizes in $(3\card{K_v}/4,1.11\card{K_v}]$, contradicting \Cref{lem:two:clique}, since $1.11<(3/2)(3/4)=1.125$.
These are the only possibilities, since additions to distinct clusters occur at different steps.
Thus every vertex of $K_v\setminus K^{(t)}$ is a singleton.

\textbf{Proof of statement (2).}
Suppose that no cluster contains $(1-91\eps)\card{K_v}$ vertices of $K_v$ at $t$.
Let $K_{v,40\eps}$ contain the last $\lceil40\eps\card{K_v}\rceil$ vertices of $K_v$ inserted into the graph.
If all are singletons, they suffice.
Otherwise, among all vertices of $K_v$ currently in clusters, choose a vertex $w$ whose last addition to its current cluster is latest.
Let $s$ be that time and $K$ that cluster.
Note that more than $(1-40\eps)\card{K_v}$ current vertices of $K_v$ (at least all vertices in $K_v \setminus K_{v,40\eps}$) are already present at $s$.
The bound above gives
\[
 \card{K^{(s)}\cap K_v}
 >\bigl(1-(40+5.62)\eps\bigr)\card{K_v}
 =(1-45.62\eps)\card{K_v}.
\]
Every vertex of $(K^{(s)}\cap K_v)\setminus K^{(t)}$ is now a singleton: it was in $K$ at $s$, so if it were now in another cluster, its last addition would be after $s$, contradicting the choice of $w$.
Recall that we assume no cluster contains $(1-91\eps)\card{K_v}$ vertices of $K_v$ at $t$, so there are more than
\[
 (91-45.62)\eps\card{K_v}
 =45.38\eps\card{K_v}
 >40\eps\card{K_v}
\]
such vertices.

\textbf{Validity of $\Vsparsealgi.$}
Finally, consider any singleton $u$ found above.
We prove that it is never removed from the list $\Vsparsealgi$ due to low degree.
In statement (1), use its last removal from a cluster at or after $t_0$, if there is one; otherwise, use identification at $t_0$ or its later insertion.
In statement (2), use its insertion if it was never in a cluster, and its last removal otherwise; for vertices removed from $K^{(s)}$, this removal occurs after $s$.
After the corresponding operations are complete, $u$ is added to $\Vsparsealg$ and all degree lists and remains a singleton through $t$.
At least $(1-94.33\eps)\card{K_v}$ current vertices of $K_v$ are already present then: use $A$ in statement (1), and the more than $(1-40\eps)\card{K_v}$ vertices present at insertion or at $s$ in statement (2).
These vertices survive through $t$, so throughout this interval,
\[
\begin{aligned}
 \deg(u)
 &\ge\bigl(1-(94.33+1.02)\eps\bigr)\card{K_v}-1\\
 &=(1-95.35\eps)\card{K_v}-1
 >\card{K_v}/2-1.
\end{aligned}
\]
Since degrees are integers,
$\deg(u)\ge\lfloor\card{K_v}/2\rfloor\ge2^{i-1}$.
For every $j\in S_i(5)$, we have $j\le i+2$, so
\[
 \frac{2^j}{10}
 \le\frac{2^{i+2}}{10}
 =\frac45\,2^{i-1}
 <2^{i-1}.
\]
Thus the degree check in \Cref{alg:check-and-try-cluster} and \Cref{alg:check:n:singleton} cannot remove $u$ from the required lists.
\end{proof}

\subsection{Correctness of High Degree Clusters}

Note that by \obsref{num:adv} and \Cref{lem:num:alg:add}, we bound the number of data stream removals, data stream additions, and algorithmic additions.
Then intuitively, the cluster still satisfies the requirements of an $O(\eps)$-tolerant cluster for a sufficiently large constant if we only have these three types of operation.
However, vertices in a cluster could be largely removed due to algorithmic removals, which does not recognize by our counters in \Cref{alg:vertex-dynamic-alg}.
Thus we count the distinct vertices that have belonged to $K$ during $[t_0,t]$, remain in the graph at $t$, and lie outside $K^{(t)}$, each counted once.
Now, we categorize into two cases based on \textbf{this number}: I. there are at most $(2\beta+13)\eps \card{K^{(t_0)}}$ such vertices; and II. there are more than $(2\beta+13)\eps \card{K^{(t_0)}}$ such vertices.
Throughout this subsection, $\card{K^{(t_0)}}\ge(1-\eps)/\eps$.

\subsection*{Case I: at most $(2\beta+13)\eps \card{K^{(t_0)}}$ such vertices}

We first show that if this number is at most $(2\beta+13)\eps \card{K^{(t_0)}}=21\eps \card{K^{(t_0)}}$, then the number of non-neighbors inside $K$ is also bounded, satisfying the requirement of a tolerant cluster.
\begin{lemma}
\label{lem:tol:non-neighbor}
Suppose that there are at most $21\eps\card{K^{(t_0)}}$ surviving vertices that have belonged to $K$ during $[t_0,t]$ and lie outside $K^{(t)}$.
Then each vertex $v \in K$ has at most $(4\beta+22)\eps \card{K^{(t)}}$ non-neighbors inside $K$, i.e., $K$ satisfies the first property of a tolerant cluster (see \Cref{def:partial-correct-new-sdd}).
\end{lemma}
\begin{proof}
By \Cref{lem:num:alg:add}, each $v\in K^{(t)}$ has at most $12\eps\card{K^{(t_0)}}$ non-neighbors in $K^{(t_0)}$, and at most $3\eps\card{K^{(t_0)}}$ vertices of $K^{(t)}$ lie outside $K^{(t_0)}$.
Thus $v$ has at most $15\eps\card{K^{(t_0)}}$ non-neighbors in $K^{(t)}$.
The assumption in \textbf{Case I} and \obsref{obs:num:operations} give $\card{K^{(t)}}\ge(1-22\eps)\card{K^{(t_0)}}$, then we have $15\eps\card{K^{(t_0)}}\le(4\beta+22)\eps\card{K^{(t)}}$.
This proves the statement.
\end{proof}

The bound of $15\eps\card{K^{(t_0)}}$ non-neighbors does not use the assumption of Case I and also holds for smaller clusters, by the closed-neighborhood bound in \Cref{lem:num:alg:add}.
We remark that the above analysis also shows that \Cref{lem:tol:non-neighbor} is satisfied at any time during $[t_0,t]$ at which the hypothesis of Case I holds.
Next, we show that for dense vertices with a large number of neighbors outside, at least a portion of these neighbors are inside $\Vsparsealg$, such that the second requirement of a tolerant cluster is also satisfied and we can charge the cost.

\begin{lemma}
\label{lem:tol:vs}
Suppose that $v$ belongs to $K$ and is $\eps$-dense at time $t$, let $K_v$ be the set of vertices in $N[v]$ that are in $\eps$-agreement with $v$, then at least $40\eps$-fraction of $K_v \setminus K$ is in $\Vsparsealg$, i.e., $K$ satisfies the second property of a tolerant cluster (see \Cref{def:partial-correct-new-sdd}).
\end{lemma}
\begin{proof}
The proof directly follows from \Cref{thm:Vsingleton}.
Its first statement cannot hold for a cluster $K'\ne K$, since then $v\in K_v\setminus K'$ would belong to $\Vsparsealg$, contradicting $v\in K$.
If it holds for $K$, all of $K_v\setminus K$ belongs to $\Vsparsealg$.
Otherwise, the second statement places at least $40\eps\card{K_v}\ge40\eps\card{K_v\setminus K}$ vertices in $\Vsparsealg$, and all are outside $K$.
\end{proof}

Set $\eps_\tau=(4\beta+22)\eps=38\eps$, $\eps_l=\eps$, $\eta_d=1/38$, and $\eta_l=1$.
The agreement constants in \Cref{def:partial-correct-new-sdd} are $C_t=1/38$ and $C_l=1$, respectively, so both use agreement parameter $\eps$.
Combining \Cref{lem:tol:non-neighbor} and \Cref{lem:tol:vs}, we have the following main result for \textbf{Case I}.
\begin{lemma}[Case I: tolerant clusters]
\label{lem:tol}
$K$ is a $(4\beta+22)\eps$-tolerant cluster at time $t$.
\end{lemma}

\subsection*{Case II: more than $(2\beta+13)\eps \card{K^{(t_0)}}$ such vertices}

If this number is large, the size of $K$ decreases a lot, so the analysis in \Cref{lem:tol:non-neighbor} would not go through in general.
We remark that this does not necessarily mean that $K$ is an isolated cluster, e.g., an exact clique will not become an isolated cluster due to algorithmic removals since there is no non-neighbor inside.
For this case, \Cref{lem:tol:non-neighbor} is automatically true and \Cref{lem:tol:vs} is true given \Cref{thm:Vsingleton}, therefore $K$ is still a tolerant cluster.
In the following, we suppose that $K$ is not a $(4\beta+22)\eps$-tolerant cluster, which requires more than $(2\beta+13)\eps \card{K^{(t_0)}}$ such vertices by \Cref{lem:tol}.
We next show that, under this assumption, $K$ satisfies the property of an isolated cluster.

\begin{lemma}
For any vertex $v \in K$ and time $t$, suppose that there are more than $21\eps\card{K^{(t_0)}}$ surviving vertices that have belonged to $K$ during $[t_0,t]$ and lie outside $K^{(t)}$, then at least one of the following holds
\begin{itemize}
    \item \textbf{either} $v$ is $\eps$-sparse \textbf{and} $\degtu{t}{v}\geq \card{K^{(t)}}/2$;
    \item \textbf{or} $v$ is $\eps$-dense and let $K_v$ be the set of vertices in $N[v]$ that are in $\eps$-agreement with $v$ at $t$, $\card{K_v\setminus K^{(t)}}\geq \eps \card{K^{(t)}}$.
\end{itemize}
\end{lemma}
\begin{proof}
By neighborhood bound in the proof of \Cref{lem:num:alg:add}, every current vertex has at least $(1-12\eps)\card{K^{(t_0)}}$ positive neighbors in $K^{(t_0)}$ that survive at $t$.
Since $|K^{(t)}|\le(1+3\eps)\card{K^{(t_0)}}$, we have $\deg^{(t)}(v)\ge|K^{(t)}|/2$.
Therefore, it suffices to assume that $v$ is $\eps$-dense and prove the second statement, i.e., $\card{K_v\setminus K^{(t)}}\geq\eps\cdot\card{K^{(t)}}$.
We next divide into two cases: \textbf{Case (a):} $\degtu{t}{v}\ge2\card{K^{(t_0)}}$; and \textbf{Case (b):} $\degtu{t}{v}<2\card{K^{(t_0)}}$.

For \textbf{Case (a)}, by the $\eps$-density of $v$, by \Cref{lem:num:alg:add} we have
\[
 \card{K_v}\ge(1-\eps)\cdot\degtu{t}{v}
 \ge2(1-\eps)\cdot\card{K^{(t_0)}}
 \ge2(1-4\eps)\cdot\card{K^{(t)}}.
\]
Therefore, $K_v$ must have at least $\frac12\card{K^{(t)}}$ vertices outside $K^{(t)}$, which completes the proof.

For \textbf{Case (b)}, more than $21\eps\card{K^{(t_0)}}$ surviving vertices that have belonged to $K$ lie outside $K$, and by \Cref{lem:num:alg:add} at most $3\eps\card{K^{(t_0)}}$ of them are outside $K^{(t_0)}$.
So more than $18\eps\card{K^{(t_0)}}$ surviving vertices of $K^{(t_0)}$ lie outside $K$.
At most $12\eps\card{K^{(t_0)}}$ fail to be positive neighbors of $v$, and fewer than $\eps\deg^{(t)}(v)<2\eps\card{K^{(t_0)}}$ neighbors fail agreement with $v$.
Hence
\[
 \card{K_v\setminus K^{(t)}}>4\eps\card{K^{(t_0)}}>\eps\card{K^{(t)}}.
\]
\end{proof}

We remark that \Cref{lem:tol:vs} still holds due to the validity of $\Vsparsealg$ shown by \Cref{thm:Vsingleton}.
Therefore, $K$ satisfies all the requirements for an isolated cluster in \Cref{def:partial-correct-new-sdd}.
We have the following main result for \textbf{Case II}.
\begin{lemma}
\label{lem:iso}
$K$ is an $\eps$-isolated cluster at time $t$.
\end{lemma}

\subsection{Correctness of Low Degree Clusters}
In the previous subsections, we establish correctness for clusters that accumulate up to $\eps|K^{(t_0)}|$ data stream updates before being refreshed.
Those estimates assumed $\card{K^{(t_0)}}\ge(1-\eps)/\eps$.
For $\card{K^{(t_0)}}<(1-\eps)/\eps$, \Cref{lem:num:alg:add} shows that every vertex that has belonged to $K$ lies in $K^{(t_0)}$ and none has been deleted.
Algorithmic removals and re-entries therefore preserve an internally exact clique.
We now prove that such clusters trivially satisfy the properties of the tolerant clusters.
The crucial observation here is that if $\card{K^{t_0}}<\frac{1-\eps}{\eps}$, then $K^{(t_0)}$ can only be an \emph{exact} clique due to \Cref{def:sdd}, since we allow at most $\eps\card{K^{(t_0)}}/(1-\eps)<1$ out-neighbors.

\begin{lemma}[Correctness of Low Degree Clusters]
\label{lem:correct_low_degree}
Let $K$ be a cluster maintained by the algorithm at time $t$ such that $\card{K^{t_0}}<\frac{1-\eps}{\eps}$.
Then $K$ satisfies the properties of an $\eps_\tau$-tolerant cluster at time $t$.
\end{lemma}
\begin{proof}
By \Cref{lem:num:alg:add}, $K^{(t)}\subseteq K^{(t_0)}$ and $K^{(t_0)}$ is a clique, so there are no internal non-neighbors.
For each vertex, the second tolerant-cluster condition follows either from its $\eps$-sparsity or from \Cref{lem:tol:vs}.
\end{proof}

An output singleton omitted from $\Vsparsealg$ by \Cref{alg:check:n:singleton} also has no internal non-neighbors.
If its vertex is $\eps$-sparse, it satisfies the tolerant-cluster condition directly.
Otherwise, the first statement of \Cref{thm:Vsingleton} is impossible because the vertex belongs to neither $\Vsparsealg$ nor a maintained cluster.
The second statement supplies the required fraction of $K_v$ in $\Vsparsealg$.

Combining \Cref{lem:tol}, \Cref{lem:iso}, and \Cref{lem:correct_low_degree}, we complete the proof of \Cref{thm:cor:cluster} for the correctness of each cluster.


\section{Correctness of \texorpdfstring{$\Vsparsealg$}{Vsparse}}
\label{sec:cor:vs}
In this section, we show that $\Vsparsealg$ satisfies the requirements in \Cref{def:partial-correct-new-sdd}, which completes the proof of the correctness.
We note that we use a different accuracy parameter $\zeta$ for the sparsity to distinguish from previous sections, but it is within a constant-fraction of $\eps$, e.g., $\zeta=\frac{\eps}{500}$ as stated in \Cref{alg:vertex-dynamic-alg}.
To begin with, we state the definition of important levels as mentioned in \Cref{sec:dynamic-alg-vertex-update}.
\begin{definition}[Important levels]
Let $\Vsparsei(2\zeta)$ be the set of vertices in $\Vsparsei$ that are  $2\zeta$-dense, and let $\VsparseCi$ be the set of vertices in $\Vsparsei$ that are $\zeta$-sparse.
For each level $i$, we have the following definitions
\begin{itemize}
    \item Level $i$ is \textbf{dominant} if $\card{\Vsparsei(2\zeta)} > \frac{1}{10 \cdot 2^i \log n}\card{\Vsparsealg}$;
    \item Level $i$ is \textbf{emergent} if $\card{\Vsparsei(2\zeta)} > 10 \card{\bigcup_{j\in S_i(3)}\VsparseCj}$;
    \item Level $i$ is \textbf{non-negligible} if $\card{\Vsparsei(2\zeta)} > \frac{1}{10 \log n}\card{\bigcup_{j'\in S_{j}(3)}\VsparseCjp}$ for all $j \in [L]$ with $j \ge i-4$.
\end{itemize}
\end{definition}
The central goal is to demonstrate \cref{lem:main:alg}, which states that each level $i$ is either \textbf{non-dominant}, or \textbf{non-emergent}, or \textbf{negligible}, otherwise it will be captured by our size check and form cluster operation in \Cref{alg:deletion-clique-forming}.
Therefore, the number of ``incorrect vertices'', i.e., the vertices being $\frac{\eps}{C}$-dense for some universal constant $C$, in each level is strictly upper bounded so that it satisfies \Cref{def:partial-correct-new-sdd}.

We begin our analysis by showing the correctness of the size estimation subroutine \Cref{alg:estimate:size}.
This lemma is crucially used afterwards to show that it passes the threshold to trigger cluster formation when a level $i$ is dominant, emergent, and non-negligible. 
\begin{lemma}
\label{lem:est-size}
With high probability, the estimate of the size of a fixed subset $S$ given by \Cref{alg:estimate:size} satisfies: (1). if $\card{S}\ge \frac{\card{\Vsparsealg}}{10 \cdot 2^i \log n}$, it gives a $1.1$-approximation to $\card{S}$; and (2). if $\card{S}< \frac{\card{\Vsparsealg}}{10 \cdot 2^i \log n}$ it has an additive error of at most $\frac{\card{\Vsparsealg^{(t)}}}{100 \cdot 2^i \log n}$.
\end{lemma}
\begin{proof}
Recall that in \Cref{alg:estimate:size}, we uniformly sample $\gamma = O(2^i \log^2 n)$ vertices from $\Vsparsealg^{(t)}$ and return $\widetilde{s}^{(t)} = \frac{\card{\Vsparsealg^{(t)}}}{\gamma} \cdot \hat{\gamma}$, where $\hat{\gamma}$ is the number of samples belonging to $S$ at time $t$.
For each sample, the indicator that denotes a sample is in $S$ has expectation $\card S/\card{\Vsparsealg^{(t)}}$.
Put $\xi=\card{\Vsparsealg^{(t)}}/(1000\cdot2^iL)$.
By \Cref{prop:chernoff}, with high probability,
\[
 |\widetilde s^{(t)}-\card S|\le .08\card S+\xi,
\]
which implies both stated bounds.

Note that in the usage of \Cref{alg:estimate:size}, we do sparsity tests on the samples.
For the actual tests, a successful $5\zeta$-sparsity test returns no on every $4.5\zeta$-dense vertex and only on $5\zeta$-dense vertices.
Apply the same concentration bound to these two fixed sets and to their complements to obtain
\[
 0.9\card{\bigcup_{j\in S_i(3)}\Vsparsej^{(t)}(4.5\zeta)}-\xi
 \le\bar u_i\le
 1.1\card{\bigcup_{j\in S_i(3)}\Vsparsej^{(t)}(5\zeta)}+\xi,
\]
\[
 0.9\card{\Vsparsej^{(t)}\setminus\Vsparsej^{(t)}(5\zeta)}-\xi
 \le\bar u_j\le
 1.1\card{\Vsparsej^{(t)}\setminus\Vsparsej^{(t)}(4.5\zeta)}+\xi.
\]
Here $\bar u_j$ is the estimate in Line~\ref{line:negli} when $j\ne i$.
These inequalities hold for every set of successful test answers, since the tested counts lie between the corresponding fixed-set counts.
\end{proof}

\begin{remark}
\Cref{lem:est-size} also applies when $S$ is the set of degree-zero vertices in $\Vsparsealg$.
The same concentration argument shows that $C\log^2 n$ samples, for a sufficiently large constant $C$, estimate their fraction within additive error $1/100$ with high probability, as required by \Cref{alg:check:n:singleton}.
\end{remark}

We next show that when our algorithm decides to form cluster on a level $i$, it eliminates all dense vertices in that level (and also all levels in $S_i(3)$).
Note that although we estimate the size of $\Vsparsei(5\zeta)$ for every $h_i$ counted changes, we do not afford to maintain the exact degree bucketing list that contains all vertices in $\Vsparsei(5\zeta)$ as mentioned earlier.
Fortunately, due to the validity of $\Vsparsealgi$ in \Cref{thm:Vsingleton}, we form clusters on all vertices in $\Vsparsei(5\zeta)$.
\begin{lemma}
\label{lem:eliminate:level}
After a full scan of $\Vsparsealgi$, including newly appended entries, by \Cref{alg:check-and-try-cluster}, we have $\Vsparsej^{(t)}(5\zeta)=\emptyset$ for every $j\in S_i(3)$.
Every $5\zeta$-dense vertex $v$ in these levels belongs to a cluster. 
If fewer than $h_i$ changes counted by $c_i$ have occurred since the start of this completed scan, a vertex that is now $2\zeta$-dense with degree in $[2^i,2^{i+1})$ cannot be in $\Vsparsealg$.
\end{lemma}
\begin{proof}
Consider a $5\zeta$-dense vertex $v$ with degree in $[2^{i-1},2^{i+2})$ at the end of the completed scan.
By \Cref{def:sdd,lem:vertex-pair-agreement}, its positive $5\zeta$-agreeing neighbors are $10.1\zeta$-dense.
Indeed, their pairwise agreement tolerance is $10\zeta/(1-5\zeta)$, their degrees are at least $(1-5\zeta)(\deg(v)+1)-1$, and each has at most $\mathord{(5\zeta+\frac{5\zeta}{1-5\zeta})(\deg(v)+1)}$ neighbors outside this agreeing set.
For $\deg(v)<10/\eps$, integrality instead makes $N[v]$ an isolated exact clique.
These vertices pass the $12\zeta$ test and lie above the degree threshold.
Apply \Cref{thm:Vsingleton}, whose required lists include $\Vsparsealgi$ because the size bucket of $K_v$ differs from $i$ by at most two.

Suppose the second statement of \Cref{thm:Vsingleton} holds.
Then at least $40\eps\card{K_v}$ vertices of $K_v$ are singletons
in $\Vsparsealgi$ at the end of the scan.
Since $v$ is $5\zeta$-dense, fewer than
$5\zeta\deg(v)<40\eps\card{K_v}$ vertices of $K_v$ fail
$5\zeta$-agreement with $v$.
Thus at least one of these listed singletons, say $u$,
is in $5\zeta$-agreement with $v$.
By the preceding argument, $u$ is $10.1\zeta$-dense and lies
above the degree threshold, so the scan adds $u$ to a cluster.
If $u$ later returns to the list, its new entry is also scanned.
Since the graph is fixed during the scan, the same argument
applies at every visit.
Hence $u$ cannot remain a singleton at the end of the full scan, which is a contradiction.
Therefore, the first statement of \Cref{thm:Vsingleton} holds: some cluster $K$ contains at least $(1-91\eps)\card{K_v}$ vertices of $K_v$.
If $v\notin K$, then $v\in K_v\setminus K$, so this statement places $v$ itself in $\Vsparsealgi$.
Since $v$ is $5\zeta$-dense and lies above the degree threshold, the same scan argument gives a contradiction.
Thus $v\in K$, proving the assertions at the end of the scan.

After at most $b$ graph updates, a vertex that is $\theta$-dense at an endpoint of degree $d>b$ is $\left(\theta+\frac{(1+\theta)b}{d-b}\right)\text{-dense}$ throughout its lifetime in the interval, since degrees and neighborhood differences change by at most $b$, and at most $b$ neighbors appear.
Under the assumed bound of fewer than $h_i$ counted changes, either $b=0$ or $b<\zeta2^i/10$.
Hence a current $2\zeta$-dense vertex of degree at least $2^i$ was $3\zeta$-dense throughout the interval and, if present during the scan, belonged to the scanned degree window.
It is added to a cluster by the scan or its later insertion, and any refreshment processes it again.

It remains to exclude a later removal by identification.
The scan conclusion and \Cref{def:dense:core,lem:num:alg:add} give $\card{K^{(t_0)}}\ge(1-91\eps)(1-\eps)\deg(v)/(1+3\eps)$ for the containing cluster, while later additions satisfy the same bound by their dense-core guarantee.
Since degrees change by at most $b$, at a first such removal we have $\card{K^{(t_0)}}\ge\deg(v)/2$.
If a new cluster $K'$, formed from $w$, identifies $v$, then $\card{K'}\ge\deg(v)/4$ and $\card{N(v)\cap K'}>\card{K'}/4$.
Among these neighbors, the number failing agreement with $v$ or $w$, or lying outside $N[w]$, is at most
\[
 3\zeta\deg(v)+\frac{\eps}{20}\deg(w)
 +\frac{\eps}{1-\eps}\card{K'}<2\eps\card{K'}<\card{K'}/4,
\]
where $\deg(w)\le\card{K'}/(1-\eps)$ by \Cref{lem:dense-vertex-clique-forming}.
A common agreeing neighbor therefore exists, so $v,w$ agree at tolerance $(3\zeta+\eps/20)/(1-\eps/20)<\eps/5$.
The formation guarantee already includes $v$ in $K'$, contradicting its removal into $\Vsparsealg$.
Thus $v$ is not in $\Vsparsealg$ after the scan completes.
\end{proof}

Given the above lemma, we now show that at each time $t$, each level $i$ is either non-dominant, or non-emergent, or negligible.
Let level $i$ be dominant, emergent, and non-negligible at time $t$, let $a$ be the size of the dense part of level $i$, and let $b$ be the current value of its counter $c_i$.
If $a>2b$, level $i$ should already pass the tests in the last check and be eliminated.
Otherwise, if $a\le2b$, level $i$ may fail the tests, but we still do a partial scan.
Then since level $i$ is important now, the number of higher-degree vertices in its list during the last check is less than our budget $\widetilde O(2^i)$, and all dense vertices in level $i$ are captured by the partial scan.
In the following, we use the notion of ``returned time $t$'', which denotes the time at which the algorithm has finished processing the $t$-th update.
\begin{lemma}
\label{lem:main:alg}
For each returned time $t \in [T]$ and level $i \in [L]$, with high probability, at least one of the following holds.
\begin{itemize}
    \item \textbf{Non-dominant:} $\card{\Vsparsei^{(t)}(2\zeta)} \le \frac{1}{10 \cdot 2^i \log n}\card{\Vsparsealg^{(t)}}$;
    \item \textbf{Non-emergent:} $\card{\Vsparsei^{(t)}(2\zeta)} \le 10 \card{\bigcup_{j\in S_i(3)}\VsparseCj^{(t)}}$;
    \item \textbf{Negligible:} $\card{\Vsparsei^{(t)}(2\zeta)} \le \frac{1}{10 \log n}\card{\bigcup_{j'\in S_{j}(3)}\VsparseCjp^{(t)}}$\\ for some $j \in [L]$ with $j \ge i-4$.
\end{itemize}
\end{lemma}
\begin{proof}
We assume for contradiction that at some returned time $t$, all the three conditions above are not matched.
Let $\bar t$ be the start of the most recent completed level-$i$ check, put $a=\card{\Vsparsei^{(t)}(2\zeta)}>0$, and let $b$ be the current value of $c_i$.
By Lines~\ref{line:level-counter-increments} and~\ref{line:level-counter-reset} of \Cref{alg:deletion-clique-forming}, $b$ counts all graph updates and changes in membership of $\Vsparsealg$ since $\bar t$, including those during the check.
Such a check exists, since otherwise fewer than $h_i\le2^i$ insertions have occurred and no vertex has degree at least $2^i$.
We have $b<h_i$, and if $b>0$, then $b<\zeta2^i/10$.

By \Cref{lem:eliminate:level}, all but at most $b$ vertices of $\Vsparsei^{(t)}(2\zeta)$ belonged to the $4.5\zeta$-dense set in $S_i(3)$ at $\bar t$.
Moreover, at most $b$ vertices have left $\Vsparsealg$ since $\bar t$.
Since we assume that level $i$ is dominant, emergent, and non-negligible at time $t$, these observations give
\[
 \card{\Vsparsealg^{(\bar t)}}<10\cdot2^iLa+b,
 \qquad
 \card{\bigcup_{j\in S_i(3)}\Vsparsej^{(\bar t)}(4.5\zeta)}\ge a-b.
\]
We also need bounds on the vertices that were $4.5\zeta$-sparse at $\bar t$.
Call these the ``old sparse vertices'', with their bucket indices taken at $\bar t$.
Again by \Cref{lem:eliminate:level}, an old sparse vertex in bucket $j\ge i-4$ remains $\zeta$-sparse in $S_j(3)$ unless it leaves $\Vsparsealg$.
Thus each such vertex is either counted in the corresponding sparse population at time $t$ or is among the at most $b$ departing vertices.
Our assumptions on level $i$ therefore bound the number of old sparse vertices by less than $a/10+b$ in bucket $i$, less than $10La+b$ in each bucket $j\ge i-4$, and less than $10L^2a+b$ in their union.
The last bound has only one additive $b$, since each departing vertex is counted once across the union.

\paragraph{Case 1:} $a\ge2b$.
By \Cref{lem:est-size} and these bounds,
\[
 \bar u_i\ge 0.4a,\qquad
 \frac{\card{\Vsparsealg^{(\bar t)}}}{100\cdot2^iL}<0.11a,\qquad
 \bar u'_i<0.7a,\qquad
 \bar u'_j<12La.
\]
All three comparisons pass, so \Cref{alg:check-and-try-cluster} runs with $\mathrm{FLAG}_i=0$ and the full scan is complete.

\paragraph{Case 2:} $a<2b$.
\Cref{alg:check-and-try-cluster} attempts the partial scan even if a comparison fails.
We bound the number of unsuccessful visits by separating entries already present at $\bar t$ from entries added later.
An unsuccessful visit to an entry present at $\bar t$ must concern an old sparse vertex in a bucket $j\ge i-4$.
There are fewer than $10L^2a+b$ such vertices.
Each list entry contributes at most one unsuccessful visit: during this fixed-graph scan, an unsuccessfully visited entry stays in the list without being visited again, and adding an existing entry has no effect.
Every remaining unsuccessful visit is therefore charged to a distinct entry into $\Vsparsealg$ after $\bar t$, including any reentry of a vertex present at $\bar t$.
There are at most $b$ such entries, because $b$ counts all changes in membership since $\bar t$.
Hence the total number of unsuccessful visits is less than
\[
 (10L^2a+b)+b
 =10L^2a+2b
 <22L^2b
 <2.2\zeta2^iL^2
 <2^iL^2.
\]
The partial scan therefore completes before exhausting its allowance.

In both cases, \Cref{lem:eliminate:level} gives $\Vsparsei^{(t)}(2\zeta)=\emptyset$, which is a contradiction.
\end{proof}

Before proving our main result, we bound the number of degree-zero vertices in $\Vsparsealg$.
\begin{lemma}
\label{lem:degree-zero-cleanup}
With high probability, at every returned time $t\in[T]$, at most
$\card{\Vsparsealg^{(t)}}/10$ vertices in $\Vsparsealg^{(t)}$
have degree zero.
Moreover, each traversal in \Cref{alg:check:n:singleton} removes
at least a $1/40$ fraction of the vertices in $\Vsparsealg$
present at its start.
\end{lemma}
\begin{proof}
When $\Vsparsealg$ is nonempty, apply \Cref{alg:estimate:size}
to its subset of degree-zero vertices with fresh $C\log n$
samples, and divide the output by $\card{\Vsparsealg}$.
By the concentration argument in the proof of
\Cref{lem:est-size}, for sufficiently large $C$, this estimates
the fraction of degree-zero vertices to additive error $1/100$
with high probability.
Condition on these estimates being accurate.

Whenever \Cref{alg:check:n:singleton} starts a traversal,
the estimated fraction is at least $1/20$, so the true
fraction is at least $1/20-1/100=1/25>1/40$.
The graph does not change during the traversal, and all
these degree-zero vertices are removed from $\Vsparsealg$.

When the algorithm returns, either $\Vsparsealg$ is empty
or the estimated fraction is less than $1/20$.
In the latter case, the true fraction is less than
$1/20+1/100=3/50<1/10$.
After a traversal no degree-zero vertex remains, and the level checks that follow do not change the graph and add only vertices of positive degree.
\end{proof}

Finally, we are able to show our main theorem for the correctness of $\Vsparsealg$.
\begin{restatable}{theorem}{DeletionCliqueFormation}
    \label{thm:Vsingleton-property-guarantees-clique-formation}
    For each returned time $t\in[T]$ and each level $i\in [L]$, \Cref{alg:deletion-clique-forming} ensures that
    \begin{itemize}
        \item either $\card{\VsparseIi}\leq 10 \card{\bigcup_{j\in S_i(3)}\VsparseCj^{(t)}}$.
        \item or $\sum_{v\in \VsparseIi}\deg(v) \leq \frac{1}{C_0 \log{n}} \cdot \sum_{j\in [L]} \sum_{v\in \VsparseCj} \deg(v)$ for some absolute constant $C_0$, which may be chosen as $1/20$.
    \end{itemize}
    In other words, \Cref{alg:deletion-clique-forming} ensures the same guarantees on $\Vsparsealg$ as prescribed by \Cref{def:partial-correct-new-sdd}.
\end{restatable}
\begin{proof}
Every vertex in $\VsparseIi^{(t)}$ is $\zeta$-dense and hence $2\zeta$-dense, so \Cref{lem:main:alg} applies.
Note that if level $i$ is \textbf{non-emergent}, it directly satisfies the first condition in \Cref{thm:Vsingleton-property-guarantees-clique-formation}.
Since $\card{\Vsparsei}=\card{\VsparseIi}+\card{\VsparseCi}$, the correct vertices in $S_i(3)$ then constitute at least $1/11$ of $\Vsparsei$, as required in \Cref{def:partial-correct-new-sdd}.

For a \textbf{negligible} level, choose $j\ge i-4$ from \Cref{lem:main:alg}.
Vertices in $S_j(3)$ have degree at least $2^{j-1}$, giving
\[
 \sum_{v\in\VsparseIi}\deg(v)
 \le\frac{2^{i+1}}{10\log n}
       \card{\bigcup_{j'\in S_j(3)}\VsparseCjp}
 \le\frac{6.4}{\log n}\sum_{v\in\VsparseC}\deg(v).
\]
For a \textbf{non-dominant} level,
\[
 \sum_{v\in\VsparseIi}\deg(v)
 \le\frac{\card{\Vsparsealg}}{5\log n}.
\]
It remains to bound $\card{\Vsparsealg}$ by the number of correct vertices.

Finally, let $S_D$, $S_E$, $S_N$ be \textbf{non-dominant} levels, \textbf{non-emergent} levels, and \textbf{negligible} levels, respectively.
Without loss of generality, we assume they are disjoint sets since otherwise we can assign a level to any of them.
Summing the three bounds in \Cref{lem:main:alg}, using $\sum_i2^{-i}\le2$ and the fact that a bucket occurs in at most three sets $S_i(3)$, gives
\[
 \begin{aligned}
 \sum_{i\in S_D}\card{\VsparseIi}&\le\frac{\card{\Vsparsealg}}{5\log n},\\
 \sum_{i\in S_E}\card{\VsparseIi}&\le30\card{\VsparseC},\qquad
 \sum_{i\in S_N}\card{\VsparseIi}\le\card{\VsparseC}.
 \end{aligned}
\]
By \Cref{lem:degree-zero-cleanup}, at most $\card{\Vsparsealg}/10$ additional vertices have degree zero.
Therefore,
\[
 \card{\Vsparsealg}
 \le\frac{\card{\Vsparsealg}}{5\log n}
       +32\card{\VsparseC}+\frac{\card{\Vsparsealg}}{10},
\]
and hence $\card{\Vsparsealg}\le60\card{\VsparseC}$.
All vertices in $\VsparseC$ have positive degree, so the non-dominant bound is at most $12\sum_{v\in\VsparseC}\deg(v)/\log n$.
Both cases satisfy the second condition with $C_0=1/20$.
\end{proof}

Combining \Cref{thm:cor:cluster} (correctness of each cluster) and \Cref{thm:Vsingleton-property-guarantees-clique-formation} (correctness of $\Vsparsealg$), we complete the proof of correctness and show that our maintained partition satisfies \Cref{def:partial-correct-new-sdd}.
If the small-graph offline solution is reported instead, its own approximation guarantee applies.

\section{Analysis of Time Efficiency}
\label{sec:time-efficiency}
In this section, we analyze the amortized update time of our overall algorithms.
We gradually build the time efficiency proof with the following steps.
\begin{itemize}
    \item We first establish the efficient tracking of the statistics (e.g., $\inc(K)$ and $\del(K)$) maintained by \Cref{alg:vertex-dynamic-alg}.
    \item We then compute the time needed for each call of the subroutine. Essentially, the average time to conduct operations on each vertex is $\polylog(n)$, for instance, for a cluster $K$ of size $|K|$, it takes $|K|\polylog(n)$ time for it to form.
    \item Finally, we amortize the time complexity for the calls of these subroutines:
    \begin{itemize}
         \item We first charge the time of all cluster formations and refreshments in our algorithm to all data stream updates, establishing a $\polylog(n)$ amortized time.
        \item We then demonstrate that the time for maintaining $\Vsparsealg$ in \Cref{alg:deletion-clique-forming} is bounded by $\polylog(n)$ times the total time for cluster formations.
    \end{itemize}
\end{itemize}

The main challenge is the charging argument for cluster formations.
When a new cluster $K$ is formed, we must charge the time for its formation to a sufficiently large group of data stream updates such that the amortized time is $\polylog(n)$. 
For example, suppose that there is a cluster $K'$ and $\eps|K'|$ data stream additions occur in $K'$, so our algorithm refreshes $K'$ and forms $K$, then we can charge the time to form $K$ to the $\eps|K'|$ data stream updates that trigger the refreshment.
However, the algorithmic dynamic might be substantially more complicated, since vertices in $K$ may have been previously removed from an older cluster due to the identification step in our algorithm. 
Charging may be problematic in this case, since we cannot directly find $\eps|K|$ data stream updates to charge to, and if we trace back to the clusters that previously identified the vertices in $K$, it could lead to an infinite charging chain or double-charging.

To resolve this issue, we design a forward-charging budgeting scheme. 
We add a budget for each data stream update and transfer the budget to the clusters with which it is associated, e.g., we transfer the budget of $v$ to $K$ if $v$ is considered as a data stream addition to $K$.
Then during each cluster formation, the budgets are passed on to the vertices the cluster identifies, so we avoid dating back to historical events via this forward-charging method.
We establish an essential technical lemma, called the ``minted budget inequality'', which states that whenever a cluster is formed, its vertices have a enough total budget to pay for its own formation and pass on to the identified vertices (with size $O(\eps|K|)$).
This directly implies the correctness of our budgeting and charging method.
Please refer to the detailed explanation in \Cref{sec:charging:budgeting}.

\subsection{Amortizing the time efficiency for auxiliary control flows in \texorpdfstring{\Cref{alg:vertex-dynamic-alg}}{algorithm}}
Apart from the calls for \Cref{alg:new-ac} and \Cref{alg:refresh-ac}, \Cref{alg:vertex-dynamic-alg} tracks several statistics, including $c(K)$, $\inc(K)$, and $\del(K)$.
By \Cref{prop:data-strcuture-stats}, the tracking of $c(K)$ and $\inc(K)$ can be accomplished in $O(\log{n})$ (worst-case) time per counter change.
We need to argue about $\del(K)$ since it needs to track the number of vertices \emph{ever joined} the cluster. The implementation is as follows. 
\begin{tbox}
\begin{itemize}
    \item We use a doubly-linked list to connect each vertex $v$ with each maintained cluster it ever joined, with one record per vertex and cluster. Let this set be $\calK_{v} = \{K_1, K_2, \cdots\}$;
    \item If $v$ is deleted by the data stream, we notify all the clusters, and delete $v$ from the corresponding lists.
    \item If any cluster $K_i$ is refreshed or discarded (emptied), we notify $\calK$ and remove $K_i$ from $\calK$, charging the removal of its records to their creation.
\end{itemize}
\end{tbox}
We remark that the approach is general and applies to all formed clusters, i.e., we also track the cluster formed by \Cref{alg:deletion-clique-forming}.
The efficiency for tracking $\del(K)$ is given as follows.
\begin{lemma}
\label{lem:track-del-counter}
With high probability, for any vertex $v \in V$, $\card{\calK_{v}} = O(\log{n})$. Thus when a vertex is deleted from the graph, all corresponding $\del(K)$ counters can be updated in $O(\log^2 n)$ time.
\end{lemma}
\begin{proof}
This is a direct application of \Cref{lem:two:clique}: list the distinct clusters in $\calK_v$ in the order in which $v$ was first added to them.
The initial size of each cluster is more than $3/2$ times the initial size of the preceding cluster.
Adding $v$ to the same cluster again creates no new record, and refreshing a cluster removes its record.
Therefore, since no cluster can be larger than $n$ size, the vertex can stay in the counters of at most $O(\log{n})$ clusters simultaneously.
Combining this with \Cref{prop:data-strcuture-stats} leads to the desired statement.
\end{proof}

We can then conclude the time complexity to maintain the counters in \Cref{alg:vertex-dynamic-alg}.

\begin{lemma}
\label{lem:main-alg-tracking-stats}
With high probability, \Cref{alg:vertex-dynamic-alg} maintains the counters $c(K)$, $\inc(K)$, and $\del(K)$ in $O(\log^{2}{n})$ time per direct stream update; counter changes caused by internal operations take $O(\log^{2}{n})$ time each and are charged to those operations.
\end{lemma}

\subsection{Time complexity of each subroutine in \texorpdfstring{\Cref{alg:vertex-dynamic-alg}}{algorithm}}
\label{subsec:time-complexity-each-update}
We first establish the time complexity of each subroutine before charging the running time to the respective vertex insertion and deletions induced by the data stream.

\paragraph{Subroutines for cluster formation and refreshment.} We provide running time bounds for the algorithms \Cref{alg:new-ac} (cluster formation) and \Cref{alg:refresh-ac} (cluster refreshment) as follows.

\begin{lemma}
\label{lem:time-complexity-cluster-forming}
Let $K$ be the cluster formed by a single run of \Cref{alg:new-ac}. The time complexity for this single run of \Cref{alg:new-ac} is $O(\card{K}\log^{3}n)$ for constant $\eps$.
\end{lemma}
\begin{proof}
By the guarantee in \Cref{lem:dense-vertex-clique-forming}, the formation of the cluster takes $O\left(\frac{\card{K}\cdot \log^{3}{n}}{\eps^2}\right)$ time, which is $O(\card{K}\log^{3}n)$ for choice of constant $\eps$. After $K$ is formed, we obtain the identification set $S$ with time $O(|K|\log^{2}{n})$ (by \Cref{lem:dense-vertex-clique-forming-generalize} and the choice of $\delta=1/2$), and processing the identified vertices and their membership records takes $O(|K|\log{n}\cdot \frac{\log^{2}(n)}{\eps^2})$ time (\Cref{prop:data-strcuture-stats,lem:track-del-counter}), which is again $O(\card{K}\log^{3}n)$ for constant $\eps$.
\end{proof}


\begin{lemma}
\label{lem:time-complexity-cluster-refresh}
Let $K$ be the cluster that triggers \Cref{alg:refresh-ac}.
Apart from new cluster formations, merges, and removal of records charged to earlier operations, the time complexity for this single run of \Cref{alg:refresh-ac} is $O(\card{K}\log^{3}{n})$ for constant $\eps$.
\end{lemma}
\begin{proof}
   By \Cref{lem:sparse-test}, line~\ref{line:sparsity-test-dismantle} takes $O\left(\card{K}\frac{\log^{3}n}{\eps^2}\right)$ time to check the sparsity for each vertex. 
   We do not count the time for new cluster formation and merging vertices to clusters here (see \Cref{sec:charging:budgeting}).
   Finally, the addition of the vertices to $\Vsparsealg$ and $\Vsparsealgi$ takes $O(\card{K} \log^{2}{n})$ time. 
\end{proof}

\paragraph{Subroutines used by \Cref{alg:deletion-clique-forming}.} We similarly provide the relatively straightforward analysis for the time complexity needed for \algref{estimate:size} (size estimation) and \algref{check:n:singleton} (removing zero-degree vertices).

\begin{lemma}
\label{lem:time-complexity-size-estimation}
A single call on \algref{estimate:size} runs in $O(\gamma \log^{3}{n})$ time for constant choices of $\eps$. Here, $\gamma$ is the parameter used in \algref{estimate:size}.
\end{lemma}
\begin{proof}
    We only sample $\gamma$ vertices from the set $\Vsparsealg$, and we use \algref{estimate:size} in two ways: checking whether the vertex belongs to a degree bucket and checking whether the vertex is at least $5\zeta$ sparse. Both operations can be accomplished in $O(\log^{3}{(n)})$ time for each vertex for constant choice of $\eps$ due to \Cref{lem:sparse-test}, which leads to the desired time complexity.
\end{proof}

\begin{lemma}
\label{lem:time-complexity-merge-to-cluster}
Each degree-zero sampling test in \algref{check:n:singleton} takes $\polylog(n)$ time, and each traversal at step $t$ takes $\card{\Vsparsealg^{(t)}}\polylog(n)$ time, excluding the level checks required by the counters.
\end{lemma}
\begin{proof}
The test uses $O(\log n)$ samples and degree queries.
Each traversal checks every vertex in $\Vsparsealg$ and removes each degree-zero vertex from $\Vsparsealg$ and all $L$ lists, taking $\polylog(n)$ time per vertex.
The level checks required by the counters are charged separately in \Cref{lem:vs:time}.
\end{proof}

\subsection{Amortizing the time efficiency for cluster formation and refreshment}
\label{sec:charging:budgeting}

We amortize the time complexity for cluster formation and refreshment to \emph{past updates in the graph}, including \emph{insertion} and \emph{deletion} of the vertices of the graph.
At a high level, every time a new cluster $K$ is formed we must find a sufficiently large group of data stream updates (e.g., $\Omega(|K|)$ updates) to charge for its computational cost. 
The easy case occurs when $K$ is formed or refreshed due to the accumulation of data stream additions or removals (e.g., $\inc(K)$ or $\del(K)$ reaching the $\eps|K|$ threshold), or when a large portion of the vertices in $K$ never joins a cluster before, allowing us to cleanly charge the work directly to the corresponding data stream updates. 

However, a fundamental challenge arises with algorithmic cascades: vertices forming $K$ may have been previously removed from an older cluster due to algorithmic identification, meaning that no new data stream updates with respect to vertices in $K$ are available to charge. 
We could attempt to trace these algorithmically removed vertices back to the original insertions that are responsible for their removal, e.g., if vertices in $K$ were previously identified and moved to $\Vsparsealg$ by $K'$, then we charge the time of forming $K$ to that of forming $K'$. 
However, it is hard to construct a well-defined charging mapping, e.g., it could happen that vertices in $K$ were previously moved to $\Vsparsealg$ by $K'$, while vertices in $K'$ were previously moved to $\Vsparsealg$ by an earlier cluster $K''$, and we risk double-charging the same data stream updates across an infinite chain of cluster fragmentations. 

\paragraph{Our budgeting method.}
To resolve this issue without tracking complex historical lineages, we design a forward-charging budgeting scheme. 
We add a budget $\Psi$ for each data stream update and transfer the budget to the clusters with which it is associated, e.g., we transfer a part of $\Psi$ to $K$ if a vertex is a data stream addition to $K$.
Then we show that every newly-formed cluster      has a total budget of at least $\Omega(\eps|K|\Phi)$.
Therefore, it pools the budget to pay for its own formation work, while using the leftover budget to mint new $\Phi$ budgets for the vertices it removes during the identification step. 
The key observation is that a cluster at most identifies $O(\eps|K|)$ vertices due to \Cref{lem:core}, so the cost is safely charged as long as we find a larger total budget.
This process is captured by the following minted budget inequality:
\[\underbrace{C_1\eps|K|\Phi}_{\text{Unused Budgets}} \ge \underbrace{|K| \cdot \frac{\polylog(n)}{\poly(\eps)}}_{\text{Time for algorithmic operations}} + \underbrace{C_2\eps|K|\Phi}_{\text{Budgets being passed on}}.\]
Then the inequality holds if $C_1>C_2$ and $\Phi > \frac{\polylog(n)}{(C_1-C_2)\poly(\eps)}$.
In the following analysis, we show that when a cluster is formed, a large portion of its vertices comes from $\Vsparsealg$, so they receive a budget $\Phi$ from previous data stream updates, either due to identification or a refreshment of an old cluster.
The charging scheme is summarized as follows:

\begin{tbox}
\begin{itemize}
\item Budgeting and charging scheme:
\begin{itemize}
    \item When the data stream inserts or deletes a vertex $v$, add a budget $\Psi$ on $v$ and if $v$ joins a cluster $K$, transfer the budget to the cluster (specified in Line~\ref{line:budget:transfer}).
    \item When a cluster is newly formed or refreshed, the budgets on the corresponding vertices are charged to perform the algorithmic operations. Then, the budgets are passed on to those vertices that are moved to $\Vsparsealg$ due to the cluster formation or refreshment.
\end{itemize}
\item We prove these core invariants in the following analysis:
\begin{enumerate}
    \item When a cluster $K$ is formed or refreshed, at least $\Omega(\eps)$-fraction of the vertices in $K$ have an unused budget $\Phi$.
    \item The total amount of budgets $K$ has is larger than the amount of usage, including the time for algorithmic operations and the budgets that pass on to other vertices.
\end{enumerate}
\end{itemize}
\end{tbox}


If both invariants hold, it is instant that we have enough budget during each formation to pay for the algorithmic operations and the budgets that pass on to others.
Therefore, the charging framework implies a $\Psi$ amortized time, as we add a budget $\Psi$ on each data stream update:
\begin{lemma}
\label{lem:time-charge-framework}
Suppose that the invariants hold in the algorithm, then the amortized time for forming and refreshing clusters is at most $\Psi$.
\end{lemma}
\begin{proof}
All work of formations and refreshments is paid from budgets. New budget is created only at stream updates, $\Psi$ per update, and every other step only moves existing budget as follows. Transfers move part of an update's budget to a cluster and when a cluster is refreshed, its transferred budget is handed to its remaining vertices (see \obsref{obs:transfer}). During formation, the budget $\Phi$ given to each of the fewer than
$2.01\eps\card{K}$ identified vertices comes from the at least
$2.5\eps\Phi\card{K}$ collected from vertices of $K$ (see
\Cref{lem:inv:1,lem:inv:2}), a merge spends the added vertex's budget,
and a formation from a vertex of degree below $10/\eps$ spends that
vertex's budget. By the two invariants no budget becomes negative, so after $m$ updates the total time of formations and refreshments is at most $\Psi m$.
\end{proof}

Given \Cref{lem:time-charge-framework}, our goal is to show that $\Psi = O\left(\frac{\polylog(n)}{\poly(\eps)}\right)$, so that we achieve the efficient update time as desired.
We next formalize the budgeting and charging scheme.

\begin{tbox}
\label{box:charging}

\textbf{Initialization:} When the data stream inserts or deletes a vertex $v$, we add a budget $\Psi$ on $v$.
For an insertion, $\Phi$ is the spendable budget of $v$, while the remaining $\Psi-\Phi$ is reserved for transfers to a cluster $K$ whenever $v$ joins $K$ and increases $\inc(K)$.
Spending the vertex budget does not reduce this reserve. 
\vspace{1ex}

\textbf{Budget transferring:} \label{line:budget:transfer}
Transfer the budget on $v$ to a cluster $K$ under the following cases:
\begin{itemize}
\item If $v$ joins $K$ and increases $\inc(K)$ ($v$ is inserted to the graph after $K$ is formed), transfer a total budget of $\frac{2\Phi}{\eps}$ to $K$ from the insertion reserve. Note that this event could happen after $v$ is inserted, e.g., $v$ joins $\Vsparsealg$ or some other cluster $K'$ at first and then joins $K$ later. Adding the same vertex to the same cluster again does not increment $\inc(K)$ again.
\item If $v$ is deleted from the graph at $t$, for each existing cluster $K$ at $t$ that $v$ joins before, transfer a total budget of $\frac{2\Phi}{\eps}$ to $K$.
\end{itemize}

\textbf{Budget passing on:} When a cluster $K$ is formed or refreshed, use its available budgets to charge the operations (see \Cref{lem:inv:1} and \obsref{obs:transfer}), and pass on the budget of a cluster $K$ under the following cases:
\begin{itemize}
    \item If $K$ is refreshed, add a budget $\Phi$ on each $v \in K$ when the old cluster is removed, before processing its vertices, 
    i.e., $v$ moves to $\Vsparsealg$ or a new cluster.
    \item If $K$ is newly-formed, use the budget as follows:
    \begin{enumerate}
        \item \textbf{Algorithmic operations:} Charge the time for cluster formation/refreshment and membership bookkeeping to the budgets collected at formation.
        \item \textbf{Vertex being identified:} For each vertex $v$ that is removed from a cluster by identification and added to $\Vsparsealg$, add a budget $\Phi$ on $v$. A vertex already outside all maintained clusters retains its existing budget.
    \end{enumerate}
    Finally, zero-out the spendable budget for each vertex in $K$. A merge consumes the added vertex's budget in the same way.
\end{itemize}

\end{tbox}

We first show that when a cluster is refreshed, it obtains $\Phi |K|$ budgets since it is associated with enough data stream updates.
\begin{observation}
\obslab{obs:transfer}
During the refreshment of a cluster $K$, it has a total budget of at least $\Phi |K|$, i.e., each remaining vertex gets a budget $\Phi$.
\end{observation}
\begin{proof}
At least $\eps\card{K^{(t_0)}}$ counted data stream additions or removals trigger the refreshment.
Their transfers contribute at least $2\Phi\card{K^{(t_0)}}$.
By \Cref{lem:num:alg:add}, the remaining cluster has size at most $2\card{K^{(t_0)}}$, so there is budget $\Phi$ for every remaining vertex.
A one-vertex cluster cannot receive a merge by \Cref{lem:dense-vertex-clique-forming}.
\end{proof}

We next show that any vertex in $\Vsparsealg$ has a budget $\Phi$, so it suffices to prove that a large fraction of vertices in a newly-formed cluster come from $\Vsparsealg$.
\begin{observation}
\obslab{obs:budget:vs}
All vertices in $\Vsparsealg$ have an unused budget of $\Phi$.
\end{observation}
\begin{proof}
A vertex that has never joined a cluster retains the budget $\Phi$ assigned at insertion.
A vertex removed from a cluster receives $\Phi$ either from refreshment, by \obsref{obs:transfer}, or from the cluster that identifies it.
Removing a degree-zero vertex from $\Vsparsealg$, or restoring it to the lists, does not spend this budget.
\end{proof}

The next corollary extends \Cref{thm:Vsingleton} to any almost-clique, at the moment just before a new cluster is formed and we use it to prove the first invariant.

\begin{corollary}
\label{cor:generic:free}
Immediately before a new cluster is formed, let $K$ be any
$\eps$-almost-clique of size at least two.
At least one of the following holds:
\begin{enumerate}
    \item There is a maintained cluster $K'$ containing at least a
    $(1-91\eps)$-fraction of $K$, and every vertex in $K\setminus K'$
    lies outside all maintained clusters.
    \item At least a $40\eps$-fraction of $K$ lies outside all
    maintained clusters.
\end{enumerate}
\end{corollary}

\begin{proof}
For an $\eps$-almost-clique $K$, \Cref{def:sdd} gives
$|N[u]|\le\card K/(1-\eps)$ and at most
$\eps\card K/(1-\eps)$ non-neighbors per vertex.
These are stronger than the bounds used in \Cref{thm:Vsingleton},
so the part of its proof establishing the two alternatives applies to $K$.
Immediately before formation, all clusters whose counters reached a
refreshment threshold have been removed, and previous formations and
identifications are complete.
The same argument therefore shows that the relevant vertices lie outside
all maintained clusters; we do not need the final part of that proof
establishing list membership.
In particular, these vertices may include the vertex being inserted or
vertices awaiting processing during refreshment, even if they have not
yet been added to $\Vsparsealg$.
\end{proof}

With \obsref{obs:transfer} and \obsref{obs:budget:vs}, we are able to show
the first invariant.
The proof is based on \Cref{cor:generic:free}: either enough vertices
of the new cluster have unused budgets, or almost all of it lies in an
existing cluster.
In the latter case, we show that the vertex from which the new cluster
is formed should instead merge into the existing cluster.

\begin{lemma}[Invariant I]
\label{lem:inv:1}
When a cluster $K$ is formed in our algorithm from a vertex of degree
at least $10/\eps$, at least $2.5\eps\card K$ vertices in $K$ have
an unused budget $\Phi$.
For a smaller-degree vertex, its budget instead pays for the formation.
\end{lemma}

\begin{proof}
Note that any newly-formed cluster $K$ in our algorithm is an
$\eps$-almost-clique.
Let $v$ be the vertex from which $K$ is formed, and first suppose
that $\deg(v)\ge10/\eps$.
It suffices to show that, immediately before $K$ is formed, at least
$2.5\eps\card K$ of its vertices lie outside all maintained clusters.
Indeed, each such vertex has an unused budget $\Phi$ by
\obsref{obs:transfer} and \obsref{obs:budget:vs}, including $v$ and vertices
awaiting processing during refreshment.

If the second conclusion of \Cref{cor:generic:free} holds, our proof
is done.
For the first conclusion, if $\card{K\setminus K'}\ge2.5\eps\card K$,
then all these vertices lie outside all maintained clusters and have
unused budgets.
Thus, it suffices to consider the case where
$\card{K\setminus K'}<2.5\eps\card K$, so $K'$ contains more than
$(1-2.5\eps)\card K$ vertices of $K$.
The vertex $v$ lies outside $K'$.
We show that $v$ should have been merged into $K'$, contradicting the
formation of $K$.

Let $t_0$ be the time when $K'$ was formed.
We first bound its size at formation, using the same analysis as in
\Cref{thm:Vsingleton}.
For any $u\in K\cap K'$, \Cref{def:sdd,lem:num:alg:add} give
\[
 (1-12\eps)\card{(K')^{(t_0)}}
 \le |N[u]|
 \le \frac{\card K}{1-\eps},
\]
and hence $\card{(K')^{(t_0)}}\le1.11\card K$.
Write $S=K\cap K'$.
The bound from the proof of \Cref{lem:tol:non-neighbor}, which does
not require Case I, shows that every $x\in K'\setminus K$ has at most
$15\eps\card{(K')^{(t_0)}}$ non-neighbors in $K'$.
Consequently,
\[
 \card{N(x)\cap S}
 \ge \card S-15\eps\card{(K')^{(t_0)}}
 > (1-19.15\eps)\card K
 > \card K/2.
\]
Since $K$ is an $\eps$-almost-clique, the total number of positive
edges leaving $K$ is at most $\eps\card K^2/(1-\eps)$.
Counting the edges from $K'\setminus K$ to $S$ therefore gives
\[
 \card{K'\setminus K}
 \le \frac{2\eps}{1-\eps}\card K.
\]

We can now verify the merge conditions for $v$.
By \Cref{lem:dense-vertex-clique-forming},
$\card{N(v)\setminus K}<\eps\deg(v)/20$ and
$\card K\le(1+0.152\eps)\deg(v)$.
Together with the bounds on $K\setminus K'$ and $K'\setminus K$,
this gives
\[
 \frac{\card{N(v)\setminus K'}}{\deg(v)}
 < \eps/20+2.5\eps(1+0.152\eps)
 < 3\eps,
\]
\[
 \frac{\card{K'}}{\deg(v)}
 \le \left(1+\frac{2\eps}{1-\eps}\right)(1+0.152\eps)
 < 1+4\eps.
\]
Thus, \Cref{lem:dense-vertex-clique-forming} would merge $v$ into $K'$
instead of forming a new cluster, a contradiction.

For $\deg(v)<10/\eps$, the same subroutine forms an isolated exact
clique of size at most $10/\eps+1$ and identifies no outside vertex,
so $v$'s budget suffices.
\end{proof}

Now, we establish the second invariant by carefully choosing the
value of $\Phi$.

\begin{lemma}[Invariant II]
\label{lem:inv:2}
Let $\Phi=O(\log^3n/\eps^3)$, with a sufficiently large constant.
Then the total budget of $K$ ($2.5\eps\Phi\card K$ when formation
starts from a vertex of degree at least $10/\eps$) pays for the
algorithmic operations and the budgets passed on to other vertices.
\end{lemma}

\begin{proof}
We begin by computing the total number of algorithmic operations
charged to a cluster $K$.
Throughout this proof, $\card K$ denotes its size at formation.
By \Cref{lem:time-complexity-cluster-forming}, the time complexity
for cluster formation, including the identification step, is
$O(\card K\log^3n/\eps^2)$.
By \Cref{lem:num:alg:add}, at most $3\eps\card K$ vertices outside
$K$ at its formation are first added to $K$ before its refreshment.
These additions, together with the creation and eventual removal
of their membership records, can be charged to the time of cluster
formation.
Here, we note that a vertex might leave and re-enter $K$ several
times, which is not counted as a new addition in the preceding bound.
However, we do not need to charge all these operations to $K$.
A repeated addition of a vertex outside all maintained clusters uses
the budget $\Phi$ supplied by its preceding removal from a cluster;
an addition during the formation of another cluster is charged to
that formation.
Finally, by \Cref{lem:time-complexity-cluster-refresh}, refreshing $K$
takes $O(\card K\log^3n/\eps^2)$ time.
Thus, the average charge on each vertex of the initial cluster is
$O(\log^3n/\eps^2)$.

In addition, when $K$ is formed, it identifies all vertices outside
$K$ that connect to at least an $\alpha=1/2$ fraction of its vertices,
and it may also report vertices with more than $\card K/4$ neighbors
in $K$.
Since $K$ is an $(\eps/2)$-almost-clique, \Cref{lem:core} bounds the
number of all reported vertices by
\[
 \frac{2\eps}{1-\eps/2}\card K
 < 2.01\eps\card K.
\]
Thus, we only need to pass on a budget $\Phi$ to at most
$2.01\eps\card K$ vertices.
By \Cref{lem:inv:1}, when formation starts from a vertex of degree
at least $10/\eps$, at least $2.5\eps\card K$ vertices have an unused
budget $\Phi$.
Therefore,
\[
 2.5\eps\Phi\card K
 \ge
 O\left(\frac{\card K\log^3n}{\eps^2}\right)
 +2.01\eps\Phi\card K
\]
for $\Phi=O(\log^3n/\eps^3)$ with a sufficiently large constant.
For a smaller-degree vertex, the new cluster is an isolated exact
clique of size at most $10/\eps+1$ and identifies no outside vertex,
so that vertex's single budget $\Phi$ pays for the same operations.
\end{proof}

\begin{lemma}
$\Psi = O\left(\frac{\log^4n}{\eps^4}\right)$ is sufficient for the budget transferring step in Line~\ref{line:budget:transfer}.
\end{lemma}
\begin{proof}
A deletion transfers budget to at most $O(\log n)$ clusters by \Cref{lem:track-del-counter}.
For insertions, fix a vertex $v$ and list the distinct clusters $K_1,\ldots,K_d$ whose insertion counters increase when $v$ is first added, in that order.
Let $t_i$ be the formation time of $K_i$; each $K_i$ was formed before $v$ was inserted.

At its addition to $K_{i+1}$, $v$ has at least $(1-12\eps)\card{K_{i+1}^{(t_{i+1})}}$ positive neighbors in $K_{i+1}^{(t_{i+1})}$ by the proof of \Cref{lem:num:alg:add}.
These vertices and their edges to $v$ already existed when $v$ was added to $K_i$.
At that earlier time, at most $7\eps\card{K_{i+1}^{(t_{i+1})}}$ of them were outside $K_{i+1}$, by \Cref{lem:dense:change} applied immediately before the later addition.
The others lie outside the dense core containing $v$ in $K_i$, so
\[
 (1-19\eps)\card{K_{i+1}^{(t_{i+1})}}
 \le\frac{4\eps(1+4\eps)}{1-4\eps}\card{K_i^{(t_i)}}
 <\frac{1-19\eps}{2}\card{K_i^{(t_i)}}.
\]
Here the factor $1+4\eps$ also covers an addition triggering immediate refreshment.
Such a merge has degree at least $10/\eps$, and its dense core and \Cref{lem:num:alg:add} imply $\card{K_i^{(t_i)}}>1/\eps$.
Thus the extra vertex changes the size bound from $1+3\eps$ to $1+4\eps$; the proof of \Cref{lem:num:alg:add} retains its $12\eps$ bound.

The initial cluster sizes therefore decrease by a factor greater than two, so $d=O(\log n)$.
Adding $v$ to the same cluster again does not increase its insertion counter.
Each update consequently needs $O(\log n)\cdot2\Phi/\eps$, in addition to its initial vertex budget, which is the stated bound on $\Psi$.
\end{proof}

As a corollary of \Cref{lem:time-charge-framework}, we have the following result.
\begin{lemma}
\label{lem:cluster:time}
The amortized time for cluster formation and refreshment in \Cref{alg:vertex-dynamic-alg} and \Cref{alg:deletion-clique-forming} is $O(\polylog(n))$ for a constant $\eps$.
\end{lemma}
The same budgeting argument holds for every finite prefix of internal operations, since all charged budgets come from past updates.
If $\ell$ counts all additions to clusters, including formation and repeated additions, then after $m$ stream updates we have $\ell\le m\polylog(n)$ whenever the local subroutine guarantees hold.
This bound does not depend on the number of level checks.

\subsection{Amortizing the time efficiency for \texorpdfstring{\Cref{alg:deletion-clique-forming}}{algorithm}}

In this section, we show that the time required to maintain $\Vsparsealg$ and $\Vsparsealgi$ in \Cref{alg:deletion-clique-forming} is upper bounded by the number of operations that add a vertex to a cluster up to $\polylog(n)$ factors.
Therefore, since we showed that such operations can be charged to the data stream inputs, the amortized update time of \Cref{alg:deletion-clique-forming} is also $\polylog(n)$.

\begin{lemma}
\label{lem:vs:time}
Let $\ell$ be the total number of operations that we add a vertex to a cluster (including forming new clusters and repeated additions) and let $m$ be the total number of data stream updates, then the total update time of \Cref{alg:deletion-clique-forming} is $(\ell+m) \cdot \polylog(n)$.
As a result, the amortized update time of \Cref{alg:deletion-clique-forming} is $\polylog(n)$.
\end{lemma}
\begin{proof}
\textbf{Time for adding and removing vertices from $\Vsparsealg$:}
An entry is caused by a data stream insertion, removal from a cluster, or identification of a singleton previously removed by \Cref{alg:check:n:singleton}.
Charge the first two to the insertion or preceding addition to a cluster.
Charge the last to the identifying cluster, which identifies only $O(\eps\card K)$ vertices by \Cref{lem:core}.
Repeated identification of a vertex already in $\Vsparsealg$ creates no entry.
A departure is caused by an addition to a cluster, a deletion, or removal of a degree-zero vertex; the last is charged to its preceding entry.
Hence there are $O(m+\ell)$ changes counted by each $c_i$ in \Cref{alg:deletion-clique-forming}.
Updating all $L$ counters takes $O(L)$ time per counted change.

\textbf{Time for running \Cref{alg:check:n:singleton}:}
By \Cref{lem:degree-zero-cleanup}, every traversal removes at least a $1/40$ fraction of the vertices present at its start, so its work is charged to their preceding entries.
The tests cost $\polylog(n)$ per output or traversal.

\textbf{Time for checking sizes:}
The counter $c_i$ is reset at the start of each check and counts changes during that check toward the next one.
Thus each level-$i$ check requires at least $h_i$ new counted changes, giving $O((m+\ell)/h_i)$ checks.
Each uses $2^i\polylog(n)$ work for estimation and at most $2^iL^2$ unsuccessful visits in a partial scan of \Cref{alg:check-and-try-cluster}.
Since $2^i/h_i=O(1/\zeta)$, this totals $(m+\ell)\polylog(n)$ over all levels.

\textbf{Time for maintaining vertices with lower degrees:}
Each removal below the threshold $2^i/10$ is charged to the creation of that list entry.
Entries arise when a vertex enters $\Vsparsealg$ or is identified by a new cluster, and both were charged above.
By \Cref{alg:check-and-try-cluster}, a scan visits each entry at most once unless it is removed and created again, and it does not copy the entire list before starting.

\textbf{Time for maintaining vertices with higher degrees:}
Consider a full scan of \Cref{alg:check-and-try-cluster} whose comparisons pass, and let $D$ count the $5\zeta$-dense vertices initially in the three buckets $S_i(3)$.
With $\xi=\card{\Vsparsealg}/(1000\cdot2^iL)$, \Cref{lem:est-size} and the dominance comparison give
\[
 \bar u_i\le1.1D+\xi,\qquad \xi\le 0.1\bar u_i,
 \qquad \bar u_i\le(11/9)D.
\]
If bucket $j\ge i-4$ initially contains $P$ vertices that are $5\zeta$-sparse, the other two comparisons give
\[
 0.9P-\xi\le\bar u'_j\le100L\bar u_i,
 \qquad P=O(LD).
\]
Every unsuccessful visit to an initial vertex above the threshold belongs to one of these sparse sets, because a $5\zeta$-dense vertex passes the $12\zeta$ test.
There are at most $L$ such buckets, hence $O(L^2D)$ such visits.
By \Cref{lem:eliminate:level}, at least $D$ vertices are added to clusters during the scan, paying for this work.
Successful visits in either kind of scan are charged to additions to clusters, and visits to newly created entries are charged to their creation.
By \Cref{lem:cluster:time}, $\ell\le m\polylog(n)$ independently of how often checks run, so the total time is $m\polylog(n)$.
\end{proof}

Combining \Cref{lem:cluster:time} and \Cref{lem:vs:time}, we have our main theorem for update time.
\begin{theorem}
The entire algorithm runs in $\polylog(n)$ amortized time.
\end{theorem}

Assume $m=\poly(n)$. By the preceding time analysis, there are at most
$\poly(n)$ randomized subroutine calls before the first failure or
termination. Using fresh randomness with sufficiently small
inverse-polynomial failure probability for each call, a union bound
over the possible positions of the first failure shows that all calls
succeed with high probability. Hence the correctness and time guarantees
hold throughout the update sequence, proving
\Cref{thm:master-theorem-CC}.

\section*{AI Disclosure}

The initial draft of this paper was finished before July 2026.
AI is not used in discovering the research direction, constructing the algorithms, and forming the analysis.
Later, the authors used GPT Astra and Claude Opus 5.5 to check the initial draft, simplify proofs, and assist in improving the presentation.
GPT Astra was used to assist in the proof of the adversarially robust result.
All mathematical results and proofs were independently verified by human authors. 
The manuscript was written entirely by the authors, who take complete responsibility for the correctness of this work.

\allowdisplaybreaks

\FloatBarrier

\bibliographystyle{alpha}
\bibliography{references}

\clearpage
\appendix

\section{A Lower Bound for Maintaining the Strict Sparse-dense Decomposition}
\label{sec:original-decomp-lb}
In this section, we prove a lower bound showing that no dynamic algorithm can maintain the sparse-dense decomposition structure as in \Cref{def:sdd} with $o(\sqrt{n})$ amortized update time.
This highlights the difficulty for algorithms to efficiently maintain decompositions, which justifies the necessity to separate ``decomposition'' from ``clustering'' to use the more relaxed notion as in \Cref{def:partial-correct-new-sdd}.

\begin{lemma}
\label{lem:vanilla-sdd-lower-bound}
For any constant $\eps\in (0,1/4)$ and the corresponding constant $\eta_0$ as defined in \Cref{def:sdd},
there exists a dynamically changing graph with at most $n$ vertices and a polynomial-length sequence of vertex updates, such that any dynamic algorithm must suffer an expected amortized update time of $\Omega(\sqrt{n})$ per vertex update if it explicitly maintains a valid standard $\eps$-sparse-dense decomposition (as in \Cref{def:sdd}) at all time steps with probability at least $99/100$.
\end{lemma}
\begin{proof}
Let $k$ be a parameter such that $k= \lfloor\sqrt{n/2}\rfloor$. We will consider a ``dense set'' of vertices $V_0$ of size $N = k^2$. We construct a fixed worst-case sequence consisting of an initialization phase followed by an alternating update cycle repeated $M = n$ times. At a high level, we will use the insertion and deletion of a set of at most $k$ vertices to force updates on all vertices in the dense set of $k^2$ size, which will lead to the lower bound. The formal argument is given as follows.

\paragraph{Initialization Phase (Construction of a graph $G_0$):}
We sequentially insert $N$ vertices partitioned into $k$ disjoint cliques $C_1, C_2, \dots, C_k$, each of size exactly $k$. A vertex insertion arrives with all its edges connecting to previously inserted vertices in its respective clique. This requires $N = \Theta(n)$ vertex insertions. Let the resulting graph be $G_0 = (V_0, E_0)$.

Note that in the state of $G_0$, all vertices in $C_1, C_2, \cdots, C_k$ should be categorized as \emph{almost-cliques}. 
Indeed, for
any vertex $v \in C_i$, we have degree $\deg_{G_0}(v) = k - 1$. For any neighbor $u \in C_i$, their closed neighborhoods satisfy $N[v] \Delta N[u] = \emptyset$, meaning $|N[v] \Delta N[u]| = 0 < \eta_0 \eps \cdot \max\{\deg(v)+1, \deg(u)+1\}$ for sufficiently large $n$.
As such, $v$ has no neighbor that disagrees with $v$, which means $v$ cannot be in $\Vsparse$.


\paragraph{Alternating Phases:}
For $M$ iterations, the data stream alternates between the following two steps. In what follows, we let $\epstilde = \eta_0 \cdot \eps$ for the concise of the notation.
\begin{enumerate}
    \item \textbf{Insertion Step (Transition to $G_1$):} We sequentially insert a set $S$ of $s = \lceil4\eps k\rceil$ vertices. Each vertex $x \in S$ is inserted with edges connecting to all $N=k^2$ vertices in $V_0$, but no edges to other vertices in $S$. This requires $\Theta(\sqrt{n})$ vertex insertions. Let the resulting graph be $G_1$.
    \item \textbf{Deletion Step (Transition to $G_0$):} We sequentially delete all $s$ vertices in $S$. This requires $s$ vertex deletions and reverts the the graph to $G_0$.
\end{enumerate}

\paragraph{Analysis of State $G_1$:}
In $G_1$, the degree of any $v \in V_0$ is $\deg_{G_1}(v) = k - 1 + s$, and the degree of any $x \in S$ is $\deg_{G_1}(x) = N = k^2$. 
We claim that all vertices in $V_0$ are sparse in $G_1$. 
Suppose an $\eps$-almost-clique $K$ contains a vertex $v\in V_0$.
If $K$ also contains a vertex in $S$, then $\Delta(K)\geq k^2+1$, so $v$ must have at least $(1-2\eps)(k^2+1)-1$ neighbors in $K$, contradicting $\deg(v)=k-1+s=O(k)$ for sufficiently large $k$.
Otherwise $K\subseteq V_0$, so $\Delta(K)=k+s$, whereas $v$ has at least $s$ neighbors outside $K$.
Since $s\geq4\eps k>\eps(k+s)$ for sufficiently large $k$ and $\eps<1/4$, this contradicts the exterior-degree condition.
Therefore, no vertex of $V_0$ can belong to an almost-clique. 
Moreover, putting every vertex of $G_1$ into $\Vsparse$ gives a valid standard $\eps$-sparse-dense decomposition. Every $v\in V_0$ has the $s$ neighbors $x\in S$, and $\card{N[v]\sym N[x]}=k^2-k+s-1>\epstilde\max\{\card{N[v]},\card{N[x]}\}=\epstilde(k^2+1)$. Since $\epstilde\le\eps$ and $s\le 4\eps k+1$, we have $\epstilde\deg(v)=\epstilde(k-1+s)\le\eps k(1+4\eps)<4\eps k\le s$, so $v$ is $\epstilde$-sparse. Every $x\in S$ has only such neighbors, so $x$ is $\epstilde$-sparse too.
This gives us the following observations:
\begin{itemize}
    \item when the graph is $G_0$, no $v\in V_0$ should be in $\Vsparse$;
    \item when the graph is $G_1$, all vertices $v\in V_0$ should be in $\Vsparse$.
\end{itemize}

\paragraph{Expected Amortized Update Time:}
During each of the $M$ alternating phases, the data stream makes exactly $2s = \Theta(\sqrt{n})$ vertex updates. Let $\mathcal{E}_t$ be the event that the algorithm maintains a valid standard $\eps$-SDD at time step $t$, and the algorithm ensures $\Pr[\mathcal{E}_t] \ge 0.99$. 
Let $t_1$ be the end of the Insertion Step (State $G_1$) and $t_2$ be the end of the Deletion Step (State $G_0$). By the union bound, $\Pr[\mathcal{E}_{t_1} \cap \mathcal{E}_{t_2}] \ge 1 - (0.01 + 0.01) = 0.98$. 
Conditioned on this event occurring, the explicit decomposition must transition from placing all $N$ vertices of $V_0$ in $V_{\text{sparse}}$ at $t_1$ to placing them in dense clusters at $t_2$. Thus, the algorithm is forced to explicitly reassign the state of at least $N = \Theta(n)$ vertices. 
Therefore, the expected computation time the algorithm spends explicitly updating the assignments per phase is at least $0.98 \times \Omega(n) = \Omega(n)$. 

Repeating the above process for $M=n$ times, the total expected number of operations is at least $\Omega(n^2)$. On the other hand, the total number of data stream update is at most
\begin{align*}
    n + n\cdot(2\sqrt{n}) = O(n^{1.5}).
\end{align*}
Therefore, the amortized running time has to be at least $\Omega(\sqrt{n})$, as desired.
\end{proof}

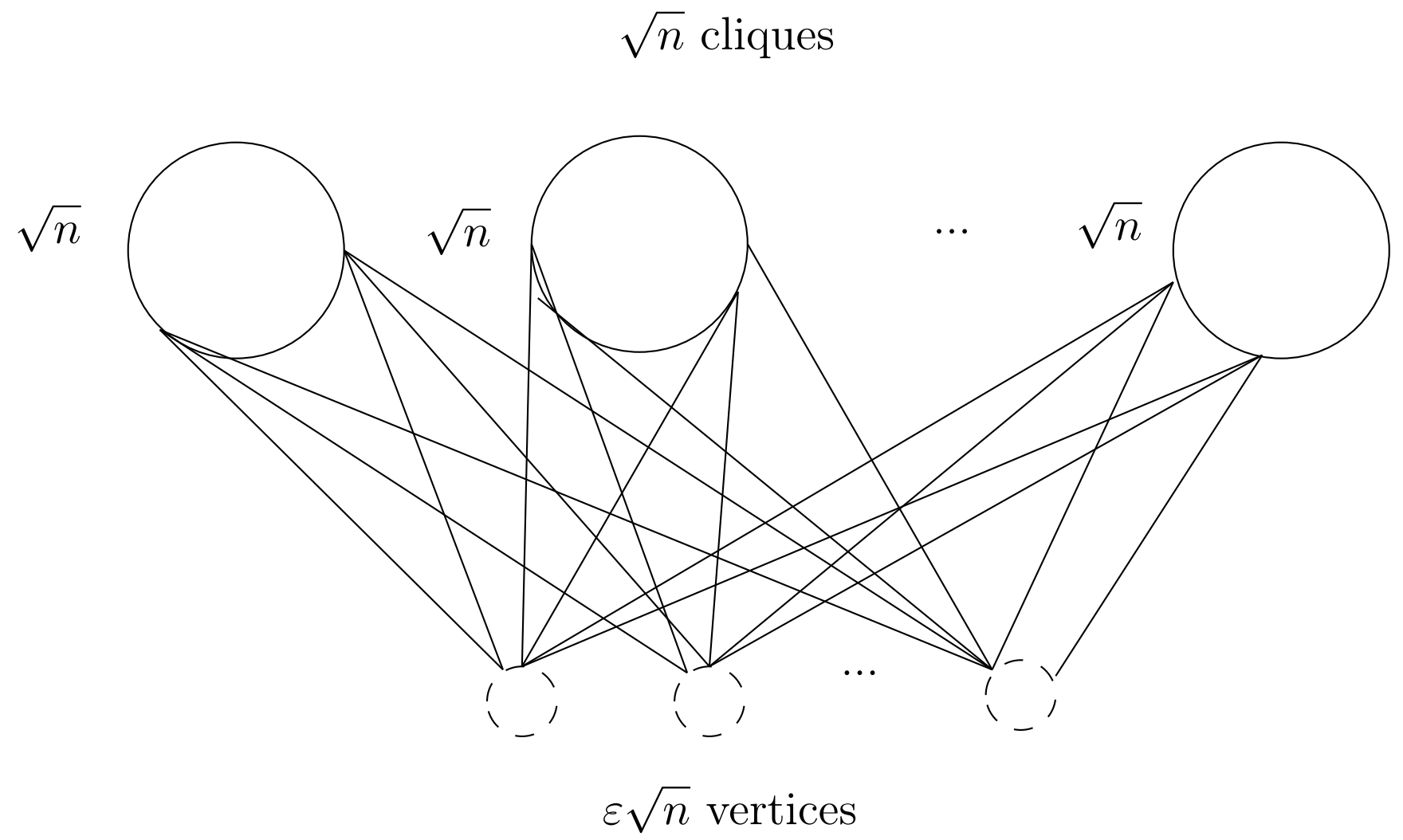
\begin{figure}[htbp]
  \centering
  \begin{tikzpicture}[x=0.75pt,y=0.75pt,yscale=-1,xscale=1]

\draw   (128,113) .. controls (128,94.22) and (143.22,79) .. (162,79) .. controls (180.78,79) and (196,94.22) .. (196,113) .. controls (196,131.78) and (180.78,147) .. (162,147) .. controls (143.22,147) and (128,131.78) .. (128,113) -- cycle ;
\draw   (255,111) .. controls (255,92.22) and (270.22,77) .. (289,77) .. controls (307.78,77) and (323,92.22) .. (323,111) .. controls (323,129.78) and (307.78,145) .. (289,145) .. controls (270.22,145) and (255,129.78) .. (255,111) -- cycle ;
\draw   (457,113) .. controls (457,94.22) and (472.22,79) .. (491,79) .. controls (509.78,79) and (525,94.22) .. (525,113) .. controls (525,131.78) and (509.78,147) .. (491,147) .. controls (472.22,147) and (457,131.78) .. (457,113) -- cycle ;
\draw  [dash pattern={on 4.5pt off 4.5pt}] (241,255) .. controls (241,248.92) and (245.92,244) .. (252,244) .. controls (258.08,244) and (263,248.92) .. (263,255) .. controls (263,261.08) and (258.08,266) .. (252,266) .. controls (245.92,266) and (241,261.08) .. (241,255) -- cycle ;
\draw  [dash pattern={on 4.5pt off 4.5pt}] (300,255) .. controls (300,248.92) and (304.92,244) .. (311,244) .. controls (317.08,244) and (322,248.92) .. (322,255) .. controls (322,261.08) and (317.08,266) .. (311,266) .. controls (304.92,266) and (300,261.08) .. (300,255) -- cycle ;
\draw  [dash pattern={on 4.5pt off 4.5pt}] (398,253) .. controls (398,246.92) and (402.92,242) .. (409,242) .. controls (415.08,242) and (420,246.92) .. (420,253) .. controls (420,259.08) and (415.08,264) .. (409,264) .. controls (402.92,264) and (398,259.08) .. (398,253) -- cycle ;
\draw    (138,138) -- (246,245) ;
\draw    (196,113) -- (246,245) ;
\draw    (255,111) -- (252,244) ;
\draw    (320,126) -- (252,244) ;
\draw    (457,123) -- (252,244) ;
\draw    (485,146) -- (252,244) ;
\draw    (138,138) -- (304,246) ;
\draw    (196,113) -- (311,244) ;
\draw    (255,111) -- (304,246) ;
\draw    (320,126) -- (311,244) ;
\draw    (457,123) -- (311,244) ;
\draw    (485,146) -- (311,244) ;
\draw    (138,138) -- (400,245) ;
\draw    (196,113) -- (400,245) ;
\draw    (257,128) -- (400,245) ;
\draw    (323,111) -- (400,245) ;
\draw    (457,123) -- (400,245) ;
\draw    (485,146) -- (420,247) ;

\draw (380,105) node [anchor=north west][inner sep=0.75pt]   [align=left] {...};
\draw (91,97) node [anchor=north west][inner sep=0.75pt]   [align=left] {$\displaystyle \sqrt{n}$};
\draw (220,98) node [anchor=north west][inner sep=0.75pt]   [align=left] {$\displaystyle \sqrt{n}$};
\draw (425,96) node [anchor=north west][inner sep=0.75pt]   [align=left] {$\displaystyle \sqrt{n}$};
\draw (281,36) node [anchor=north west][inner sep=0.75pt]   [align=left] {$\displaystyle \sqrt{n}$ cliques};
\draw (351,244) node [anchor=north west][inner sep=0.75pt]   [align=left] {...};
\draw (276,280) node [anchor=north west][inner sep=0.75pt]   [align=left] {$\displaystyle \varepsilon \sqrt{n}$ vertices};
\end{tikzpicture}
  \caption{The picture for the proof in \Cref{lem:vanilla-sdd-lower-bound}. $\eps \sqrt{n}$ vertices are inserted and removed from the graph, which makes the $\sqrt{n}$ cliques switching from forming and dismantling, forcing high time complexity.}
  \label{fig:counter-example-for-SDD}
\end{figure}

A conceptual picture of the above proof is shown as in \Cref{fig:counter-example-for-SDD}.

\section{Robustness against a white-box adversary}
\label{sec:white-box}

In this section, we prove that our algorithm is robust against a white-box adversary, who is allowed to inspect the entire maintained state and all random coins drawn so far before choosing the next vertex insertion or deletion, including its incident edges.
Each update is fully processed before the next update is given.
Random coins are drawn independently when needed and may be revealed to the adversary immediately afterward.
Coins not yet drawn are not revealed.

\begin{lemma}
\label{lem:fresh}
Consider a call to a randomized subroutine of \Cref{lem:sparse-test,lem:dense-vertex-clique-forming,lem:dense-vertex-clique-forming-generalize,lem:est-size}, including the degree-zero estimate in \Cref{alg:check:n:singleton}.
If the hypotheses of its lemma hold for the current graph and input, then the call fails with at most its stated probability, even conditioned on all coins drawn before it.
\end{lemma}
\begin{proof}
Once the earlier coins are fixed, the current graph and the input to the call are fixed.
The call draws fresh independent coins, so the proof of its lemma applies unchanged.
\end{proof}

\begin{theorem}[White-box robustness]
\label{thm:white-box}
For the correlation clustering problem on a signed complete graph $G=(V,E^+\cup E^-)$, there exists a randomized fully dynamic vertex-update algorithm that, given the positive subgraph $G^+=(V,E^+)$ and a sequence of at most $T=\operatorname{poly}(n)$ vertex insertions and deletions chosen by any white-box adversary, together with all incident positive edges, maintains an $O(1)$-approximation to the optimal correlation clustering with high probability in $\operatorname{polylog}(n)$ amortized update time.
Both guarantees hold on the same event throughout the update sequence.
\end{theorem}

\begin{proof}
Use the algorithm of \Cref{thm:master-theorem-CC} with fresh randomness in every call.
Future insertions and deletions may depend on all previously revealed coins, but they never need to view these coins as random during that time.
In other words, let $t$ be the time where the randomized guarantee concerns the arguments in arguments in \Cref{sec:cor:cluster,sec:cor:vs,sec:time-efficiency} happens,
any subsequent arguments in \Cref{sec:cor:cluster,sec:cor:vs,sec:time-efficiency} for $\geq (t+1)$ time never need to \emph{reuse} randomness of time $t$ (although they will reply on the high-probability \emph{conditioning} at time $t$). 
In particular, degree-change bounds and refreshment rules hold regardless of how updates are chosen, and no earlier sample must remain representative of a later graph.

Fix an adversary and the polynomial upper bound $T$ on the number of updates.
The bound for every execution prefix without a failure in \Cref{sec:time-efficiency} gives a fixed polynomial $M$ bounding the number of randomized calls up to and including the first failure.
Set each call's failure probability to at most $n^{-c}/M$, for any fixed constant $c>0$.
Let $H_j$ be the entire history before call $j$ draws its fresh coins.
For every such history $h$ with no earlier failure, the input satisfies the required conditions and \Cref{lem:fresh} gives
\[
 \Pr(\text{call }j\text{ fails}\mid H_j=h)\le n^{-c}/M.
\]
Consequently,
\[
 \Pr(\text{some call fails})
 \le\sum_{j=1}^{M}\Pr(\text{call }j\text{ is the first failure})
 \le n^{-c}.
\]
On the complementary event, induction over the updates gives correctness and the amortized time bound throughout the execution.

For the small-graph offline algorithm, amplify its constant success probability by fresh independent runs and return the least-cost clustering, and it still preserves the polylogarithmic time bound.
In all procedures, sample sizes and corresponding count thresholds are amplified together, including both $500\log n$ and $400\log n$ in \Cref{lem:dense-vertex-clique-forming-generalize}.
Since $M$ is polynomial in $n$ and $T=\operatorname{poly}(n)$, the amplification preserves all polylogarithmic time bounds.
\end{proof}

\end{document}

%% file: preamble.tex
\pdfoutput=1
\usepackage[T1]{fontenc}
\usepackage[protrusion=true,expansion=true]{microtype}
\usepackage{amsmath,amssymb,amsfonts,amsthm}
\usepackage{subcaption}
\usepackage{graphicx}
\usepackage{fullpage}
\usepackage[backref=page,linktocpage=true]{hyperref}
\usepackage{color}
\usepackage{wrapfig}
\usepackage{tikz}
\usetikzlibrary{decorations.pathreplacing}
\usepackage{algorithm}
\usepackage[noend]{algpseudocode}
\usepackage[framemethod=tikz]{mdframed}
\usepackage{xspace}
\usepackage{pgfplots}
\usepackage{framed}
\usepackage{thmtools}
\usepackage{thm-restate}
\usepackage{tabu}
\usepackage{fancyhdr}
\pgfplotsset{compat=1.5}
\usepackage{enumitem}

\usepackage{placeins}
\usepackage{afterpage}
\usepackage{booktabs}

\usepackage{subcaption}
\usetikzlibrary{shapes.geometric, positioning, calc}
\usetikzlibrary{arrows.meta,fit,backgrounds}

\newtheorem{theorem}{Theorem}
\newtheorem{corollary}[theorem]{Corollary}
\newtheorem{lemma}{Lemma}[section]
\newtheorem{proposition}{Proposition}[section]
\newtheorem{remark}{Remark}[section]
\newtheorem{claim}[lemma]{Claim}

\newtheorem{observation}[lemma]{Observation}

\theoremstyle{definition}
\newtheorem{problem}{Problem}
\newtheorem{definition}{Definition}

\newenvironment{proofof}[1]{\begin{trivlist} \item {\bf Proof
#1:~~}}
  {\qed\end{trivlist}}

\newcommand{\namedref}[2]{\hyperref[#2]{#1~\ref*{#2}}}

\newcommand{\claimlab}[1]{\label{claim:#1}}
\newcommand{\claimref}[1]{\namedref{Claim}{claim:#1}}

\renewcommand{\algref}[1]{\namedref{Algorithm}{alg:#1}}

\newcommand{\obslab}[1]{\label{obs:#1}}
\newcommand{\obsref}[1]{\namedref{Observation}{obs:#1}}

\newcommand{\eps}{\varepsilon}
\newcommand{\epstilde}{\widetilde{\varepsilon}}

\def \calA    {\mdef{\mathcal{A}}}

\def \calK    {\mdef{\mathcal{K}}}

\def \calU    {\mdef{\mathcal{U}}}

\newcommand{\Etilde}{\ensuremath{\widetilde{E}}\xspace}

\newcommand{\myqed}[1]{\let\qed\relax \hspace*{\fill} #1 \ensuremath{\square}}

\newcommand{\SDD}{\ensuremath{\texttt{SDD}}\xspace}

\newenvironment{tbox}{\begin{tcolorbox}[
		enlarge top by=5pt,
		enlarge bottom by=5pt,
		 breakable,
		 boxsep=0pt,
                  left=4pt,
                  right=4pt,
                  top=10pt,
                  arc=0pt,
                  boxrule=1pt,toprule=1pt,
                  colback=white
                  ]
	}
{\end{tcolorbox}}

\newcommand{\mdef}[1]{{\ensuremath{#1}}\xspace}  

\DeclareMathOperator*{\polylog}{polylog}
\DeclareMathOperator*{\poly}{poly}

\newcommand{\set}[1]{\mdef{\left\{#1\right\}}}                        

\newcommand{\ignore}[1]{}

\makeatletter
\renewcommand*{\@fnsymbol}[1]{\textcolor{mahogany}{\ensuremath{\ifcase#1\or *\or \dagger\or \ddagger\or
 \mathsection\or \triangledown\or \mathparagraph\or \|\or **\or \dagger\dagger
   \or \ddagger\ddagger \else\@ctrerr\fi}}}
\makeatother

\providecommand{\email}[1]{\href{mailto:#1}{\nolinkurl{#1}\xspace}}

\definecolor{mahogany}{rgb}{0.75, 0.25, 0.0}
\definecolor{darkblue}{rgb}{0.0, 0.0, 0.55}
\definecolor{darkpastelgreen}{rgb}{0.01, 0.75, 0.24}
\definecolor{darkgreen}{rgb}{0.0, 0.2, 0.13}
\definecolor{darkgoldenrod}{rgb}{0.72, 0.53, 0.04}
\definecolor{darkred}{rgb}{0.55, 0.0, 0.0}
\definecolor{forestgreenweb}{rgb}{0.13, 0.55, 0.13}
\definecolor{greencss}{rgb}{0.0, 0.5, 0.0}
\definecolor{bleudefrance}{rgb}{0.19, 0.55, 0.91}

\hypersetup{
     colorlinks   = true,
     citecolor    = darkblue,
	 linkcolor	  = darkred
}

\fancypagestyle{pg}
{
\lhead{}
\rhead{}
\cfoot{--\ \thepage\ --}

}

\newcommand{\paren}[1]{\ensuremath{\left(#1\right)}\xspace}
\newcommand{\card}[1]{\left\vert{#1}\right\vert}

\newcommand{\Vsparse}{V_{\textnormal{sparse}}}

\newcommand{\Vsparsei}{\widetilde{V}_{\textnormal{singleton},(i)}}
\newcommand{\Vsparsej}{\widetilde{V}_{\textnormal{singleton},(j)}}

\newcommand{\Vsparsealgi}{\widetilde{V}_{\textnormal{list},(i)}}
\newcommand{\Vsparsealgj}{\widetilde{V}_{\textnormal{list},(j)}}
\newcommand{\Vsparsealg}{\widetilde{V}_{\textnormal{singleton}}}

\newcommand{\VsparseC}{\widetilde{V}_{\textnormal{singleton,correct}}}
\newcommand{\VsparseCj}{\widetilde{V}_{\textnormal{singleton,correct},(j)}}
\newcommand{\VsparseCjp}{\widetilde{V}_{\textnormal{singleton,correct},(j')}}
\newcommand{\VsparseI}{\widetilde{V}_{\textnormal{singleton,incorrect}}}
\newcommand{\VsparseCi}{\widetilde{V}_{\textnormal{singleton,correct},(i)}}
\newcommand{\VsparseIi}{\widetilde{V}_{\textnormal{singleton,incorrect},(i)}}

\newcommand{\Vsingleton}{\widetilde{V}_{\textnormal{singleton}}}
\newcommand{\VsingletonC}{\widetilde{V}_{\textnormal{singleton, correct}}}
\newcommand{\VsingletonW}{\widetilde{V}_{\textnormal{singleton, incorrect}}}

\newcommand{\nsparse}{\ensuremath{n_{\textnormal{singleton}}}\xspace}

\newcommand{\sym}{\,\triangle\,}

\newcommand{\bracket}[1]{\left[#1\right]}
\newcommand{\expect}[1]{\mathbb{E}\bracket{#1}}

\newcommand{\SDDtol}{\ensuremath{\texttt{SDD}^{\textnormal{tol}}}\xspace}
\newcommand{\SDDpar}{\ensuremath{\texttt{SDD}^{\textnormal{partial}}}\xspace}

\newcommand{\OPT}{\ensuremath{\mbox{\sc opt}}\xspace}

\newcommand{\ALG}{\ensuremath{\mbox{\sc alg}}\xspace}

\newcommand{\cost}{\ensuremath{\textsf{\textnormal{cost}}}\xspace}

\newcommand{\degp}[1]{\ensuremath{\deg}^{+}(#1)\xspace}

\newcommand{\charge}[2]{\ensuremath{\textnormal{\textsf{charge}}(#1,#2)}\xspace}

\newcommand{\pl}{\ensuremath{(+)}\xspace}
\newcommand{\nl}{\ensuremath{(-)}\xspace}

\newcommand{\spcharge}[2]{\ensuremath{\textnormal{\textsf{sparse-charge}}(#1,#2)}\xspace}

\newcommand{\discharge}[2]{\ensuremath{\textnormal{\textsf{clique-disagreement-charge}}(#1,#2)}\xspace}

\newcommand{\occharge}[2]{\ensuremath{\textnormal{\textsf{outside-clique-charge}}(#1,#2)}\xspace}
\newcommand{\iccharge}[2]{\ensuremath{\textnormal{\textsf{inside-clique-charge}}(#1,#2)}\xspace}

\newcommand{\inc}{\ensuremath{\texttt{inc}}\xspace}

\newcommand{\del}{\ensuremath{\texttt{del}}\xspace}

\newcommand{\chargeset}[1]{\ensuremath{\textnormal{{ChargeSet}}(#1)}\xspace}

\newcommand{\edgecost}[2]{\ensuremath{\textsf{cost}^{#1}(#2)}\xspace}

\newcommand{\degtu}[2]{\ensuremath{\deg^{(#1)}(#2)}}

\usepackage{tcolorbox}
\tcbuselibrary{skins,breakable}
\tcbset{enhanced jigsaw}

\newif\ifnotes\notesfalse
\ifnotes
\newcommand{\samson}[1]{\textcolor{red}{{\bf (Samson:} {#1}{\bf ) }} \marginpar{\tiny\bf
             \begin{minipage}[t]{0.5in}
               \raggedright S:
            \end{minipage}}}
\else
\newcommand{\samson}[1]{}
\fi

\newcommand{\chen}[1]{\textcolor{RURed}{[\textbf{Chen:} #1]}}
\newcommand{\prat}[1]{\textcolor{bleudefrance}{[\textbf{P:} #1]}}
\newcommand{\shenghao}[1]{\textcolor{brown}{\textbf{[Shenghao:}~#1\textbf{]}}}

\usepackage[normalem]{ulem}
\usepackage[compact]{titlesec}

\definecolor{DarkRed}{rgb}{0.5,0.1,0.1}
\definecolor{RURed}{rgb}{0.95,0.1,0.1}
\definecolor{DarkBlue}{rgb}{0.1,0.1,0.5}

\newtheorem{mdresult}{Result}

\definecolor{Red}{rgb}{0.9,0,0}

\usepackage{nameref}
\usepackage[nameinlink]{cleveref}

\crefname{property}{property}{Property}
\creflabelformat{property}{(#1)#2#3}

\crefname{equation}{eq}{Eq}
\creflabelformat{equation}{(#1)#2#3}

\crefname{thm}{theorem}{Theorem}
\creflabelformat{Theorem}{(#1)#2#3}

\crefname{algocf}{Algorithm}{Algorithm}
\creflabelformat{algocf}{#1#2#3}

\crefname{claim}{claim}{claim}
\Crefname{claim}{Claim}{Claim}

\theoremstyle{definition}
\newtheorem{mdalg}{Algorithm}
\newenvironment{Algorithm}{\begin{tbox}\begin{mdalg}}{\end{mdalg}\end{tbox}}

\newcommand{\Kdense}{\ensuremath{K^{\textnormal{tolerant}}}\xspace}
\newcommand{\Ktolerate}{\ensuremath{K^{\textnormal{tolerant}}}\xspace}
\newcommand{\Ktolerant}{\ensuremath{K^{\textnormal{tolerant}}}\xspace}
\newcommand{\Kisolate}{\ensuremath{K^{\textnormal{isolate}}}\xspace}
\newcommand{\Vcharge}{\ensuremath{V^{\textnormal{charge}}}\xspace}
\newcommand{\Echarge}{\ensuremath{E^{\textnormal{charge}}}\xspace}

%% file: references.bib
@inproceedings{AADGW22,
  author = {Vikrant Ashvinkumar and Sepehr Assadi and Chengyuan Deng and Jie Gao and Chen Wang},
  title = {{Evaluating stability in massive social networks: Efficient streaming algorithms for structural balance}},
  editor = {Nicole Megow and Adam~D. Smith},
  booktitle = {Approximation, Randomization, and Combinatorial Optimization. Algorithms and Techniques, {APPROX/RANDOM} 2023, September 11-13, 2023, Atlanta, Georgia, {USA}},
  volume = {275},
  series = {LIPIcs},
  pages = {58:1--58:23},
  publisher = {Schloss Dagstuhl - Leibniz-Zentrum f{\"{u}}r Informatik},
  year = {2023},
}

@article{ailon2008aggregating,
  author = {Nir Ailon and Moses Charikar and Alantha Newman},
  title = {{{Aggregating Inconsistent Information: Ranking and Clustering}}},
  journal = {Journal of the ACM (JACM)},
  volume = {55},
  number = {5},
  pages = {1--27},
  year = {2008},
}

@inproceedings{AhmadianE0M20,
  author = {Sara Ahmadian and Alessandro Epasto and Ravi Kumar and Mohammad Mahdian},
  title = {{Fair correlation clustering}},
  editor = {Silvia Chiappa and Roberto Calandra},
  booktitle = {The 23rd International Conference on Artificial Intelligence and Statistics, {AISTATS} 2020, 26-28 August 2020, Online [Palermo, Sicily, Italy]},
  volume = {108},
  series = {Proceedings of Machine Learning Research},
  pages = {4195--4205},
  publisher = {{PMLR}},
  year = {2020},
}

@inproceedings{AgrawalHKMT09,
  author = {Rakesh Agrawal and Alan Halverson and Krishnaram Kenthapadi and Nina Mishra and Panayiotis Tsaparas},
  title = {{Generating labels from clicks}},
  booktitle = {Proceedings of the Second International Conference on Web Search and Web Data Mining, {WSDM}},
  pages = {172--181},
  year = {2009},
}

@inproceedings{ArasuRS09,
  author = {Arvind Arasu and Christopher R{\'{e}} and Dan Suciu},
  title = {{Large-scale deduplication with constraints using dedupalog}},
  booktitle = {Proceedings of the 25th International Conference on Data Engineering, {ICDE}},
  pages = {952--963},
  year = {2009},
}

@inproceedings{AssadiSW23,
  author = {Sepehr Assadi and Vihan Shah and Chen Wang},
  title = {{Streaming algorithms and lower bounds for estimating correlation clustering cost}},
  editor = {Alice Oh and Tristan Naumann and Amir Globerson and Kate Saenko and Moritz Hardt and Sergey Levine},
  booktitle = {Advances in Neural Information Processing Systems 36: Annual Conference on Neural Information Processing Systems 2023, NeurIPS 2023, New Orleans, LA, USA, December 10 - 16, 2023},
  year = {2023},
}

@inproceedings{Assadi022,
  author = {Sepehr Assadi and Chen Wang},
  title = {{Sublinear time and space algorithms for correlation clustering via sparse-dense decompositions}},
  editor = {Mark Braverman},
  booktitle = {13th Innovations in Theoretical Computer Science Conference, {ITCS} 2022, January 31 - February 3, 2022, Berkeley, CA, {USA}},
  volume = {215},
  series = {LIPIcs},
  pages = {10:1--10:20},
  publisher = {Schloss Dagstuhl - Leibniz-Zentrum f{\"{u}}r Informatik},
  year = {2022},
}

@article{bansal2004correlation,
  author = {Nikhil Bansal and Avrim Blum and Shuchi Chawla},
  title = {{{Correlation Clustering}}},
  journal = {Machine learning},
  volume = {56},
  number = {1},
  pages = {89--113},
  year = {2004},
}

@article{BCCGM24,
  author = {Soheil Behnezhad and Moses Charikar and Vincent Cohen{-}Addad and Alma Ghafari and Weiyun Ma},
  title = {{Fully dynamic correlation clustering: Breaking 3-approximation}},
  journal = {CoRR},
  volume = {abs/2404.06797},
  year = {2024},
}

@inproceedings{ChuzhoyKhanna19,
  author    = {Julia Chuzhoy and Sanjeev Khanna},
  title     = {A New Algorithm for Decremental Single-Source Shortest Paths
               with Applications to Vertex-Capacitated Flow and Cut Problems},
  booktitle = {Proceedings of the 51st Annual ACM SIGACT Symposium on Theory of Computing},
  pages     = {389--400},
  publisher = {ACM},
  year      = {2019},
  doi       = {10.1145/3313276.3316320},
  url       = {https://doi.org/10.1145/3313276.3316320}
}

@inproceedings{Wajc20,
  author    = {David Wajc},
  title     = {Rounding Dynamic Matchings Against an Adaptive Adversary},
  booktitle = {Proceedings of the 52nd Annual ACM SIGACT Symposium on Theory of Computing},
  pages     = {194--207},
  publisher = {ACM},
  year      = {2020},
  doi       = {10.1145/3357713.3384258},
  url       = {https://doi.org/10.1145/3357713.3384258}
}

@inproceedings{BernsteinEtAl22,
  author    = {Bernstein, Aaron and van den Brand, Jan and
               Probst Gutenberg, Maximilian and Nanongkai, Danupon and
               Saranurak, Thatchaphol and Sidford, Aaron and Sun, He},
  title     = {Fully-Dynamic Graph Sparsifiers Against an Adaptive Adversary},
  booktitle = {49th International Colloquium on Automata, Languages, and Programming (ICALP 2022)},
  pages     = {20:1--20:20},
  series    = {Leibniz International Proceedings in Informatics (LIPIcs)},
  volume    = {229},
  publisher = {Schloss Dagstuhl -- Leibniz-Zentrum f{\"u}r Informatik},
  year      = {2022},
  doi       = {10.4230/LIPIcs.ICALP.2022.20},
  url       = {https://drops.dagstuhl.de/entities/document/10.4230/LIPIcs.ICALP.2022.20}
}

@inproceedings{AjtaiBJSSWZ22,
  author    = {Mikl{\'o}s Ajtai and Vladimir Braverman and T. S. Jayram and
               Sandeep Silwal and Alec Sun and David P. Woodruff and Samson Zhou},
  title     = {The White-Box Adversarial Data Stream Model},
  booktitle = {Proceedings of the 41st ACM SIGMOD-SIGACT-SIGAI Symposium on Principles of Database Systems},
  pages     = {15--27},
  publisher = {ACM},
  year      = {2022},
  doi       = {10.1145/3517804.3526228},
  url       = {https://doi.org/10.1145/3517804.3526228}
}

@inproceedings{Ben-EliezerJWY20,
  author    = {Omri Ben-Eliezer and Rajesh Jayaram and David P. Woodruff and Eylon Yogev},
  title     = {A Framework for Adversarially Robust Streaming Algorithms},
  booktitle = {Proceedings of the 39th ACM SIGMOD-SIGACT-SIGAI Symposium on Principles of Database Systems},
  pages     = {63--80},
  publisher = {ACM},
  year      = {2020},
  doi       = {10.1145/3375395.3387658},
  url       = {https://doi.org/10.1145/3375395.3387658}
}

@inproceedings{HassidimKMMS20,
  author    = {Avinatan Hassidim and Haim Kaplan and Yishay Mansour and
               Yossi Matias and Uri Stemmer},
  title     = {Adversarially Robust Streaming Algorithms via Differential Privacy},
  booktitle = {Advances in Neural Information Processing Systems},
  volume    = {33},
  year      = {2020},
  url       = {https://papers.nips.cc/paper_files/paper/2020/hash/0172d289da48c48de8c5ebf3de9f7ee1-Abstract.html}
}

@inproceedings{BehnezhadCMTFOCS22,
  author = {Soheil Behnezhad and Moses Charikar and Weiyun Ma and Li{-}Yang Tan},
  title = {{Almost 3-approximate correlation clustering in constant rounds}},
  booktitle = {63rd {IEEE} Annual Symposium on Foundations of Computer Science, {FOCS} 2022, Denver, CO, USA, October 31 - November 3, 2022},
  pages = {720--731},
  publisher = {{IEEE}},
  year = {2022},
}

@inproceedings{BehnezhadCMT23,
  author = {Soheil Behnezhad and Moses Charikar and Weiyun Ma and Li{-}Yang Tan},
  title = {{Single-pass streaming algorithms for correlation clustering}},
  editor = {Nikhil Bansal and Viswanath Nagarajan},
  booktitle = {Proceedings of the 2023 {ACM-SIAM} Symposium on Discrete Algorithms, {SODA} 2023, Florence, Italy, January 22-25, 2023},
  pages = {819--849},
  publisher = {{SIAM}},
  year = {2023},
}

@inproceedings{BehnezhadDHSS19,
  author = {Soheil Behnezhad and Mahsa Derakhshan and MohammadTaghi Hajiaghayi and Cliff Stein and Madhu Sudan},
  title = {{Fully dynamic maximal independent set with polylogarithmic update time}},
  editor = {David Zuckerman},
  booktitle = {60th {IEEE} Annual Symposium on Foundations of Computer Science, {FOCS} 2019, Baltimore, Maryland, USA, November 9-12, 2019},
  pages = {382--405},
  publisher = {{IEEE} Computer Society},
  year = {2019},
}

@inproceedings{braverman2025fully,
  author = {Vladimir Braverman and Prathamesh Dharangutte and Shreyas Pai and Vihan Shah and Chen Wang},
  title = {{Fully dynamic adversarially robust correlation clustering in polylogarithmic update time}},
  booktitle = {International Conference on Artificial Intelligence and Statistics},
  pages = {1477--1485},
  publisher = {PMLR},
  year = {2025},
}

@inproceedings{BhaskaraDV18,
  author = {Aditya Bhaskara and Samira Daruki and Suresh Venkatasubramanian},
  title = {{Sublinear algorithms for {MAXCUT} and correlation clustering}},
  booktitle = {45th International Colloquium on Automata, Languages, and Programming, {ICALP} 2018, July 9-13, 2018, Prague, Czech Republic},
  pages = {16:1--16:14},
  year = {2018},
}

@article{BonchiGK13,
  author = {Francesco Bonchi and David Garc{\'{\i}}a{-}Soriano and Konstantin Kutzkov},
  title = {{Local correlation clustering}},
  journal = {CoRR},
  volume = {abs/1312.5105},
  year = {2013},
}

@inproceedings{Cohen-AddadLLNFOCS2023,
  author = {Vincent Cohen-Addad and Euiwoong Lee and Shi Li and Alantha Newman},
  title = {{Handling correlated rounding error via preclustering: A 1.73-approximation for correlation clustering}},
  booktitle = {2023 IEEE 64th Annual Symposium on Foundations of Computer Science (FOCS)},
  pages = {1082--1104},
  year = {2023},
}

@inproceedings{Cohen-AddadLNFOCS2022,
  author = {Vincent Cohen-Addad and Euiwoong Lee and Alantha Newman},
  title = {{Correlation clustering with sherali-adams}},
  booktitle = {2022 IEEE 63rd Annual Symposium on Foundations of Computer Science (FOCS)},
  pages = {651--661},
  year = {2022},
}

@inproceedings{Cohen-AddadLPTYZSTOC2024,
  author = {Vincent Cohen-Addad and David~Rasmussen Lolck and Marcin Pilipczuk and Mikkel Thorup and Shuyi Yan and Hanwen Zhang},
  title = {{Combinatorial correlation clustering}},
  booktitle = {Proceedings of the 56th Annual ACM Symposium on Theory of Computing},
  series = {STOC 2024},
  pages = {1617–1628},
  address = {New York, NY, USA},
  publisher = {Association for Computing Machinery},
  year = {2024},
}

@inproceedings{CaoCL0NV24,
  author = {Nairen Cao and Vincent Cohen{-}Addad and Euiwoong Lee and Shi Li and Alantha Newman and Lukas Vogl},
  title = {{Understanding the cluster linear program for correlation clustering}},
  editor = {Bojan Mohar and Igor Shinkar and Ryan O'Donnell},
  booktitle = {Proceedings of the 56th Annual {ACM} Symposium on Theory of Computing, {STOC} 2024, Vancouver, BC, Canada, June 24-28, 2024},
  pages = {1605--1616},
  publisher = {{ACM}},
  year = {2024},
}

@inproceedings{CCLLLNTVYZdynamicCC,
  author       = {Nairen Cao and
                  Vincent Cohen{-}Addad and
                  Euiwoong Lee and
                  Shi Li and
                  David Rasmussen Lolck and
                  Alantha Newman and
                  Mikkel Thorup and
                  Lukas Vogl and
                  Shuyi Yan and
                  Hanwen Zhang},
  editor       = {Sayan Bhattacharya and
                  Danupon Nanongkai and
                  Michael Benedikt and
                  Gabriele Puppis},
  title        = {Static to Dynamic Correlation Clustering},
  booktitle    = {53rd International Colloquium on Automata, Languages, and Programming,
                  {ICALP} 2026, Royal Holloway, University of London, Egham, United
                  Kingdom, July 7-10, 2026},
  series       = {LIPIcs},
  volume       = {374},
  pages        = {48:1--48:23},
  publisher    = {Schloss Dagstuhl - Leibniz-Zentrum f{\"{u}}r Informatik},
  year         = {2026},
  url          = {https://doi.org/10.4230/LIPIcs.ICALP.2026.48},
  doi          = {10.4230/LIPICS.ICALP.2026.48},
  bibsource    = {dblp computer science bibliography, https://dblp.org}
}

@inproceedings{Cohen-AddadFLMN22,
  author = {Vincent Cohen{-}Addad and Chenglin Fan and Silvio Lattanzi and Slobodan Mitrovic and Ashkan Norouzi{-}Fard and Nikos Parotsidis and Jakub Tarnawski},
  title = {{Near-optimal correlation clustering with privacy}},
  editor = {Sanmi Koyejo and S.~Mohamed and A.~Agarwal and Danielle Belgrave and K.~Cho and A.~Oh},
  booktitle = {Advances in Neural Information Processing Systems 35: Annual Conference on Neural Information Processing Systems 2022, NeurIPS 2022, New Orleans, LA, USA, November 28 - December 9, 2022},
  year = {2022},
}

@inproceedings{CambusKLPU24,
  author = {M{\'{e}}lanie Cambus and Fabian Kuhn and Etna Lindy and Shreyas Pai and Jara Uitto},
  title = {{A {(3} + {\(\varepsilon\)})-approximate correlation clustering algorithm in dynamic streams}},
  editor = {David~P. Woodruff},
  booktitle = {Proceedings of the 2024 {ACM-SIAM} Symposium on Discrete Algorithms, {SODA} 2024, Alexandria, VA, USA, January 7-10, 2024},
  pages = {2861--2880},
  publisher = {{SIAM}},
  year = {2024},
}

@inproceedings{ChakrabartiKP08,
  author = {Deepayan Chakrabarti and Ravi Kumar and Kunal Punera},
  title = {{A graph-theoretic approach to webpage segmentation}},
  booktitle = {Proceedings of the 17th International Conference on World Wide Web, {WWW}},
  pages = {377--386},
  year = {2008},
}

@inproceedings{Cohen-AddadLMNP21,
  author = {Vincent Cohen{-}Addad and Silvio Lattanzi and Slobodan Mitrovic and Ashkan Norouzi{-}Fard and Nikos Parotsidis and Jakub Tarnawski},
  title = {{Correlation clustering in constant many parallel rounds}},
  editor = {Marina Meila and Tong Zhang},
  booktitle = {Proceedings of the 38th International Conference on Machine Learning, {ICML} 2021, 18-24 July 2021, Virtual Event},
  volume = {139},
  series = {Proceedings of Machine Learning Research},
  pages = {2069--2078},
  publisher = {{PMLR}},
  year = {2021},
}

@inproceedings{Cohen-AddadLMP24,
  author = {Vincent Cohen{-}Addad and Silvio Lattanzi and Andreas Maggiori and Nikos Parotsidis},
  title = {{Dynamic correlation clustering in sublinear update time}},
  booktitle = {Forty-first International Conference on Machine Learning, {ICML} 2024, Vienna, Austria, July 21-27, 2024},
  publisher = {OpenReview.net},
  year = {2024},
}

@inproceedings{ChenSX12,
  author = {Yudong Chen and Sujay Sanghavi and Huan Xu},
  title = {{Clustering sparse graphs}},
  booktitle = {Advances in Neural Information Processing Systems 25: 26th Annual Conference on Neural Information Processing Systems},
  pages = {2213--2221},
  year = {2012},
}

@inproceedings{ChechikZ19,
  author = {Shiri Chechik and Tianyi Zhang},
  title = {{Fully dynamic maximal independent set in expected poly-log update time}},
  editor = {David Zuckerman},
  booktitle = {60th {IEEE} Annual Symposium on Foundations of Computer Science, {FOCS} 2019, Baltimore, Maryland, USA, November 9-12, 2019},
  pages = {370--381},
  publisher = {{IEEE} Computer Society},
  year = {2019},
}

@inproceedings{DalirrooyfardMM24,
  author = {Mina Dalirrooyfard and Konstantin Makarychev and Slobodan Mitrovic},
  title = {{Pruned pivot: Correlation clustering algorithm for dynamic, parallel, and local computation models}},
  booktitle = {Forty-first International Conference on Machine Learning, {ICML} 2024, Vienna, Austria, July 21-27, 2024},
  publisher = {OpenReview.net},
  year = {2024},
}

@inproceedings{dalirrooyfard2025sparse,
  author = {Mina Dalirrooyfard and Konstantin Makarychev and Slobodan Mitrovi{\'c}},
  title = {{Sparse-pivot: Dynamic correlation clustering for node insertions}},
  booktitle = {Proceedings of the 42nd International Conference on Machine Learning, {ICML} 2025},
  series = {Proceedings of Machine Learning Research},
  year = {2025},
}

@article{GiotisG06,
  author = {Ioannis Giotis and Venkatesan Guruswami},
  title = {{Correlation clustering with a fixed number of clusters}},
  journal = {Theory Comput.},
  volume = {2},
  number = {13},
  pages = {249--266},
  year = {2006},
}

@article{levorato2017evaluating,
  author = {Mario Levorato and Rosa Figueiredo and Yuri Frota and L{\'u}cia Drummond},
  title = {{Evaluating balancing on social networks through the efficient solution of correlation clustering problems}},
  journal = {EURO Journal on Computational Optimization},
  volume = {5},
  number = {4},
  pages = {467--498},
  year = {2017},
}

@article{ShiDELM21,
  author = {Jessica Shi and Laxman Dhulipala and David Eisenstat and Jakub Lacki and Vahab~S. Mirrokni},
  title = {{Scalable community detection via parallel correlation clustering}},
  journal = {Proc. {VLDB} Endow.},
  volume = {14},
  number = {11},
  pages = {2305--2313},
  year = {2021},
}

@inproceedings{Veldt22,
  author = {Nate Veldt},
  title = {{Correlation clustering via strong triadic closure labeling: Fast approximation algorithms and practical lower bounds}},
  booktitle = {International Conference on Machine Learning, {ICML}},
  pages = {22060--22083},
  year = {2022},
}

@inproceedings{VeldtGW18,
  author = {Nate Veldt and David~F. Gleich and Anthony Wirth},
  title = {{A correlation clustering framework for community detection}},
  booktitle = {Proceedings of the 2018 World Wide Web Conference on World Wide Web, {WWW}},
  pages = {439--448},
  year = {2018},
}

@inproceedings{VeldtWG17,
  author = {Nate Veldt and Anthony~Ian Wirth and David~F. Gleich},
  title = {{Correlation clustering with low-rank matrices}},
  booktitle = {Proceedings of the 26th International Conference on World Wide Web, {WWW}},
  pages = {1025--1034},
  year = {2017},
}

@inproceedings{zhang2008correlation,
  author = {Zhenya Zhang and Hongmei Cheng and Wanli Chen and Shuguang Zhang and Qiansheng Fang},
  title = {{Correlation clustering based on genetic algorithm for documents clustering}},
  booktitle = {2008 IEEE Congress on Evolutionary Computation (IEEE World Congress on Computational Intelligence)},
  pages = {3193--3198},
  publisher = {IEEE},
  year = {2008},
}

@inproceedings{CaoCL0LNTVY025,
  author = {Nairen Cao and Vincent Cohen{-}Addad and Euiwoong Lee and Shi Li and David Rasmussen Lolck and Alantha Newman and Mikkel Thorup and Lukas Vogl and Shuyi Yan and Hanwen Zhang},
  editor = {Michal Kouck{\'{y}} and Nikhil Bansal},
  title = {Solving the Correlation Cluster {LP} in Sublinear Time},
  booktitle = {Proceedings of the 57th Annual {ACM} Symposium on Theory of Computing, {STOC} 2025, Prague, Czechia, June 23-27, 2025},
  pages = {1154--1165},
  publisher = {{ACM}},
  year = {2025},
  url = {https://doi.org/10.1145/3717823.3718181},
  doi = {10.1145/3717823.3718181},
  bibsource = {dblp computer science bibliography, https://dblp.org}
}

@article{SorianoS26,
  author = {David Garc{\'{\i}}a{-}Soriano and Antoine Schohn},
  title = {Approximate Dual Separation for the Cluster {LP:} a 1.387 approximation for Correlation Clustering},
  journal = {CoRR},
  volume = {abs/2607.27829},
  year = {2026},
  url = {https://doi.org/10.48550/arXiv.2607.27829},
  doi = {10.48550/ARXIV.2607.27829},
  eprinttype = {arXiv},
  eprint = {2607.27829},
  bibsource = {dblp computer science bibliography, https://dblp.org}
}
